\documentclass[a4paper,11pt]{article}
\pdfoutput=1 

\usepackage{jheppub} 
\usepackage{bbm,bm,graphicx,mathtools,color,slashed,hyperref}
\usepackage{amssymb,amsmath}
\usepackage{dsfont}
\usepackage{braket}
\usepackage{ascmac}
\usepackage{subfig}
\usepackage{here}
\usepackage{multirow}
\usepackage{float}

\newcommand{\im}{\mathrm{i}}

\newcommand{\rme}{\mathrm{e}}

\title{
From Self-Dual to Physical  $\mathbb{C}P^{N-1}$:
 Anomalies, Boundary Stokes Phenomenon, and Global Structure of $\theta$-vacua 
}

\author[1,2]{Yui Hayashi,}
\affiliation[1]{Department of Physics, The University of Tokyo, 7-3-1 Hongo, Bunkyo-ku, Tokyo 113-0033, Japan}
\affiliation[2]{Yukawa Institute for Theoretical Physics,
Kyoto University, Kyoto, 606-8502, Japan}
\emailAdd{yui.hayashi@yukawa.kyoto-u.ac.jp}

\author[3]{ Mithat \"{U}nsal}
\affiliation[3]{Department of Physics, North Carolina State University, Raleigh, NC 27607, USA}
\emailAdd{unsal.mithat@gmail.com}

\abstract{
We introduce a two-coupling generalization of  $\mathbb{C}P^{N-1}$ model that continuously interpolates between the self-dual ($\epsilon=0$)  and the physical ($\epsilon=g$) theories   as a  useful  nonperturbative  tool. 
At $\epsilon \neq g$, this model possesses a chiral imbalance, which may be viewed as a real topological deformation (imaginary-$\theta$). We demonstrate that exact quantum equivalence between first- and second-order formulations strictly requires a topological counterterm sourced by a bosonic chiral anomaly. Solving this deformed theory at large $N$ yields two primary results. First,  we analytically determine the nonperturbative vacuum structure of the self-dual theory, a self-dual vacuum with a dynamically generated field-strength condensate. 
Second, we resolve a fundamental paradox where saddles with   $(\theta + 2\pi n) \sim \mathcal{O}(N)$ ($n$ is branch number) spuriously yield lower energy densities than the physical ground state.  Because the effective action possesses an essential singularity at $F=0$, we show that the Lefschetz thimble analysis must be generalized to include boundary thimbles. A boundary Stokes phenomenon renders the problematic saddles topologically inactive, fully restoring the validity of the large-$N$ expansion for strongly coupled theories.
}

\begin{document}

\maketitle

\section{Introduction and summary}

In standard asymptotically free quantum field theories (QFTs) that dynamically generate a strong scale --such as $SU(N)$ Yang-Mills theory in 4d and $\mathbb{C}P^{N-1}$ models in 2d-- there is no inherent dimensionless expansion parameter outside of the large-$N$ limit \cite{tHooft:1973alw, DAdda:1978vbw, DAdda:1978dle, DAdda:1982lsk, Witten:1978bc}.   Consequently, there is a crucial need to develop novel analytical techniques capable of probing the non-perturbative dynamics of these models directly in infinite volume, serving as a complement to adiabatic continuity approaches \cite{Dunne:2012ae, Dunne:2012zk}.

In many such QFTs, regardless of properties such as supersymmetry or dimensionality, the theory in Euclidean space possesses a self-dual and anti-self-dual sector. This paper rests on the premise that one can gain significant analytical control in these theories by treating these sectors asymmetrically, i.e, introduce an  asymmetry as a control parameter.  The Minkowski space realization of self-dual and anti-self-dual sectors correspond to positive and negative helicity sectors in 4d gauge theories, and left/right-moving chiral sectors in 2d gauge theories. In the path integral formulation, one can restrict the functional integration entirely to self-dual configurations. In Minkowski space, for gauge theories, this restriction corresponds to the Siegel-Chalmers theory, also known as self-dual Yang-Mills theory (SDYM) \cite{Chalmers:1996rq}, which strictly reduces the theory to a purely first-order formalism.

The self-dual restriction is demonstrably simpler in perturbation theory, as evidenced in the context of perturbative QCD and maximally helicity violating (MHV) amplitudes, and exhibits remarkable features such as one-loop exactness \cite{Bardeen:1995gk, Witten:2003nn, Krasnov:2016emc, Monteiro:2022nqt, Bittleston:2020hfv}.
However, the non-perturbative analysis of such self-dual theories remains far less established (for early attempts, see \cite{Frenkel:2006fy, Frenkel:2008vz, Frenkel:2007ux, Losev:2017qrj}).

\paragraph{\texorpdfstring{$\epsilon$}{epsilon}-deformation of self-dual theory vs. real topological deformation of physical theory:}
A central motivation for our approach is that self-dual theories can be systematically deformed, establishing a continuous analytical pathway from the idealized self-dual limit back to the fully symmetric, physical theory~\cite{Witten:2003nn,Domurcukgul:2025xgf}. Remarkably, this deformation of the self-dual sector is exactly equivalent to deforming the physical theory by a real topological term (an imaginary $\theta$-angle), which physically functions as a chemical potential for topological charge~\cite{Bhanot:1984rx, Azcoiti:2003qe}: \footnote{The same correspondence also holds between $\epsilon$-deformation of SDYM theory (interpolating between self-dual and physical Yang-Mills)~\cite{Witten:2003nn,Domurcukgul:2025xgf} and a specific imaginary theta angle deformation of pure Yang-Mills theory~\cite{Panagopoulos:2011rb, DElia:2012pvq, DElia:2013uaf, Bonati:2015sqt}.}
\begin{align}
    \epsilon\text{-deformation of } S_{\text{selfdual}} = \text{real topological deformation of } S_{\text{physical}}
    \label{eq:intro_correspondence}
\end{align}
Consequently, this exact correspondence provides a novel framework for probing the non-perturbative structure of a broad class of QFTs by unifying two traditionally distinct theoretical toolboxes. On one hand, it connects with the scattering amplitudes literature, which heavily exploits the perturbative simplicity and geometric elegance of first-order self-dual formulations. On the other hand, it ties directly into the non-perturbative techniques of lattice gauge theory, where imaginary $\theta$-angles are routinely utilized to circumvent the Euclidean sign problem and map out phase diagrams via analytic continuation. Bridging these two domains and perspectives allows us to resolve long-standing non-perturbative puzzles at both the self-dual and physical limits.

Dialing the  $\epsilon$-deformation, first-order formalism of the $\mathbb{C}P^{N-1}$ model smoothly interpolates between the exact self-dual restriction (at $\epsilon \to 0$) and the standard physical theory (at $\epsilon = g$). The pure self-dual theory has only one coupling $g$, entering in the coefficient of the topological term. The coefficient exhibits exact one-loop running in perturbation theory with a strong scale $\Lambda_g$. A crucial feature of the interpolating theory is the presence of the deformation parameter $\epsilon$, which also exhibits asymptotic freedom and generates its own renormalization group invariant scale, $\Lambda_\epsilon$.  The theory possesses a renormalization group invariant which is, at one-loop order (ignoring the effects of Jacobians momentarily), the ratio of the strong scales  $\Lambda_\epsilon/\Lambda_g$.
 Notably, the instanton and anti-instanton fugacities become asymmetric in the deformed model; as $\epsilon \to 0$, local anti-instantons acquire infinite action and completely decouple, leaving a purely holomorphic, instanton-dominated vacuum.
\begin{figure}[t]
    \centering
    \includegraphics[width=0.7 \linewidth]{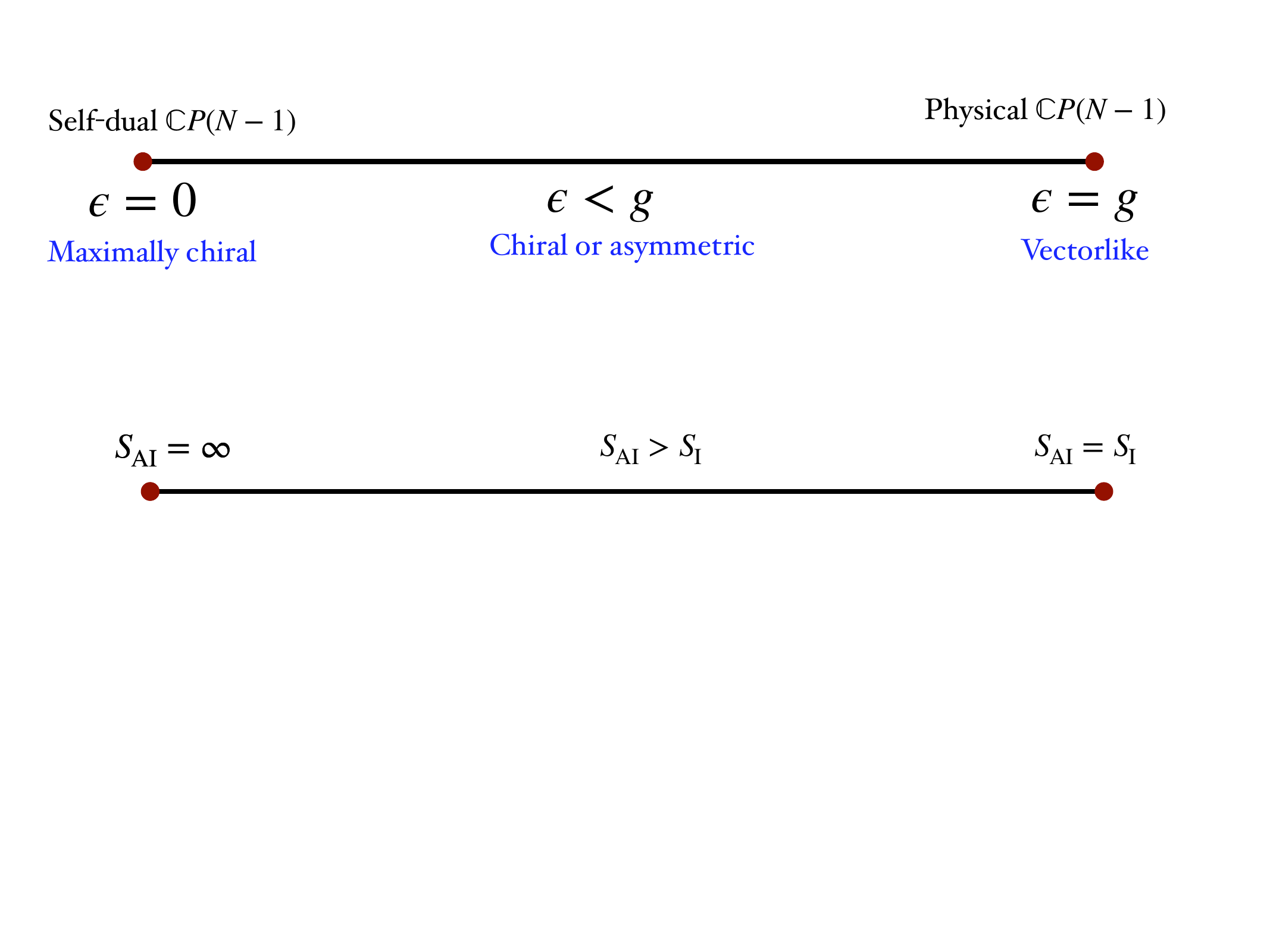}
    \vspace{-3cm}
    \caption{  The spectrum of the $\epsilon$-deformed $\mathbb{C}P^{N-1}$ model, illustrating the role of $\epsilon$ as a chiral dial. At the exact physical point ($\epsilon = g$), the theory is  vectorlike, and the instanton and anti-instanton actions are equal. Deforming the theory into the regime $\epsilon < g$ introduces a chiral asymmetry, which exponentially penalizes anti-instantons. In the strict limit $\epsilon \to 0$, the theory becomes maximally chiral; local anti-instantons acquire infinite action  and decouple entirely. The theory reduces to the  self-dual version of $\mathbb{C}P^{N-1}$ model.
    }
    \label{fig:deform}
\end{figure}

In this paper, we present two important aspects of the $\epsilon$-deformed $\mathbb{C}P^{N-1}$ model: (1) necessity of the topological counterterm for the correspondence (\ref{eq:intro_correspondence}) and (2) large-$N$ analysis: $F$-condensate, $\theta$ dependence, and boundary Stokes phenomenon.

\paragraph{(1) Necessity of the topological counterterm}

We first find that it is essential to properly incorporate the topological counterterm arising from the bosonic version of the chiral anomaly for the correspondence (\ref{eq:intro_correspondence}).
For example, the topological counterterm is necessary to obtain the self-dual limit from the topological deformation of $S_{\text{physical}}$.

This counterterm is tied up with the inherent non-vectorlike structure of our deformation. 
In the self-dual theory, the first-order sigma model field $\phi$ and the auxiliary field $h$ naturally reorganize into a ``bosonic Dirac spinor", $\Psi= (\phi, h)^T$  mapping the kinetic action directly onto the structure of a  Dirac operator. 
The asymmetric zero-mode spectrum of the functional measure plays a prominent role. 
Integration over these functional spaces generates an anomalous logarithmic Jacobian governed by the Atiyah-Singer index theorem.
This is nothing but the chiral anomaly for the bosonic spinor.
In this way, to translate the first-order formalism to the second-order canonical formalism, or vice versa, the anomalous topological term emerges.  
We will show that   this topological counterterm is essential to get a finite result in the self-dual limit in the large-$N$ calculation.



\paragraph{(2) Large-\texorpdfstring{$N$}{N} analysis: \texorpdfstring{$F$}{F}-condensate, $\theta$ dependence, and boundary Stokes phenomenon}

By applying large-$N$ saddle-point techniques to the $\epsilon$-deformed theory, we address two fundamental questions regarding the $\mathbb{C}P^{N-1}$ model: one concerning the non-perturbative vacuum of the self-dual theory, and the other concerning the global structure of the $\theta$-angle dependence, on which a potential inconsistency has been pointed out \cite{Sugeno:2025exv} even in the full physical theory.

\begin{itemize}
\item{\bf $F$-Condensate in the Self-Dual Vacuum:} 
Different from the self-dual ${\cal N}=(2,2)$  theory which is believed to be   a  logarithmic conformal field theory  (log-CFT)\cite{Frenkel:2006fy, Frenkel:2008vz, Frenkel:2007ux}, our large-$N$ analysis reveals a different vacuum structure for purely bosonic self-dual theory. 
The theory dynamically generates a macroscopic field strength condensate $F_* \simeq \Lambda_g^2$, and simultaneously, the Lagrange multiplier field develops $M_* \simeq -\Lambda_g^2$, governed by the dynamical scale $\Lambda_g$ which breaks scale invariance.
The dynamics locks the $M$-field to the field strength, the lowest state in the spectrum obeys $F + M = 0$, thus reducing the lowest Landau levels to exact zeromodes.
The low-energy effective theory consists of such Landau levels, rather than log-CFT.

    \item \textbf{Boundary Stokes Phenomenon and  global aspects of $\theta$ dependence:} 
    Standard saddle-point approximations in the $\mathbb{C}P^{N-1}$ model incorporating a $\theta$-angle suffer from the problem of ``unphysical" saddles \cite{Sugeno:2025exv}, which lead to erroneous vacuum energy densities and contradictory results to the classical works \cite{DAdda:1978vbw, DAdda:1978dle, DAdda:1982lsk, Witten:1978bc} (See \cite{Vicari:2008jw, DelDebbio:2002xa} for a general review, including numerical works).\footnote{Early on, Witten anticipated on physical grounds that the number of metastable $\theta$-vacua must truncate at   $\mathcal{O}(N)$ \cite{Witten:1980sp} rather than being exactly $N$, a behavior later demonstrated semi-classically on $\mathbb{R}^3 \times S^1$ \cite{Aitken:2018mbb}. However, Ref.~\cite{Sugeno:2025exv} was the first to sharply expose the formal mathematical paradox this lack of truncation causes in the strongly coupled continuum theory.}   
    More precisely, the saddle-point approximation is applied to the fixed-$n$ sector of the partition function $Z(\theta) =  \sum_{n \in \mathbb{Z}}\tilde{Z}(\theta+2\pi n)$ from the Poisson-sum representation of the Dirac quantization.
    On the $\tilde{Z}(\theta+2\pi n)$ integral, a saddle that is physical and relevant for some range of $\theta + 2 \pi n$, starts to give pathological contributions in some other range, despite the fact that a standard Stokes phenomenon (which occurs due to competition between different saddles) does not take place, as pointed out in \cite{Sugeno:2025exv}. 
    This is a severe problem because the full partition function is the sum over $n$, which always involves such pathological contributions.

We resolve this breakdown by carefully executing a Lefschetz thimble analysis in the presence of  {\it boundaries,} which accounts for the singularity at zero field strength ($F=0$).\footnote{In Lefschetz thimble analysis with boundaries, the boundary points are placed on the same footing as saddle points. They possess their own thimbles, and they enter the Stokes phenomenon in a non-trivial way, competing with standard saddles. Due to the non-analytic structure of large-$N$ effective action, the boundary Stokes phenomenon occurs
in the $\mathbb CP^{N-1}$ model.} We demonstrate that this singularity acts as a boundary endpoint on the integration contour, triggering a {\it boundary Stokes phenomenon}. As the parameters $\epsilon$ and $\theta$ are varied, the integration contour shifts from a combination of the saddle thimble and the endpoint thimble to purely the endpoint thimble,  rendering the unphysical saddles irrelevant/inactive and salvaging the large-$N$ expansion.
We also highlight that the saddle is always irrelevant for $\theta/N \sim O(1)$ in the physical limit $\epsilon = g$, whereas the saddle is relevant for a certain region of $\theta/N$ in the $\epsilon$-deformed theory.
This immediate irrelevance does not contradict the classic works \cite{DAdda:1978vbw, DAdda:1978dle, DAdda:1982lsk, Witten:1978bc} (see Section \ref{sec:weak_field_maintext} and Appendix \ref{sec:scaling_subtleties}).
To our knowledge, this is the first application of the boundary Stokes phenomenon to resolve vacuum structures in a quantum field theory path integral.
\end{itemize}

This paper is organized as follows.
In Section \ref{sec:formulation}, we introduce the interpolating model between the self-dual and physical $\mathbb{C}P^{N-1}$ model.
We discuss the correspondence (\ref{eq:intro_correspondence}) between the first-order and second-order formulations and the necessity of the topological counterterm.
After examining the formulations, Section \ref{sec:saddle-point-general} introduces the large-$N$ saddle-point analysis of this interpolating model and presents the $F$-condensate vacuum, and then we face the same unphysical saddle problem as Ref.~\cite{Sugeno:2025exv}.
The resolution of this problem, the boundary Stokes phenomenon, is addressed in Section \ref{sec:Stokesph}.
Finally, in Section \ref{sec:amplitude_fixedQ}, we briefly explore the connection from the large-$N$ analysis to the exact self-dual theory.
Section \ref{sec:discussion} provides a discussion.
Appendix \ref{sec:zeromodes} details the topological counterterm by bosonic chiral anomaly.
Appendices \ref{sec:endpoint_vs_negativeF_appendix}, \ref{sec:scaling_subtleties}, \ref{app:weak_field} serve as supplementary materials for Section \ref{sec:Stokesph}.
Appendix \ref{app:boundary_stokes} gives a brief introduction to the boundary Stokes phenomenon.

\section{Self-dual \texorpdfstring{$\mathbb{C}P^{N-1}$}{CP(N-1)} vs. topological deformation of \texorpdfstring{$\mathbb{C}P^{N-1}$}{CP(N-1)} }
\label{sec:formulation}

In this section, we introduce a topological deformation of the $\mathbb{C}P^{N-1}$ model that interpolates between the self-dual restriction and physical theory.

We first construct a self-dual version of the $\mathbb{C}P^{N-1}$ model, the 2D counterpart to the Chalmers-Siegel limit of Yang-Mills theory \cite{Chalmers:1996rq, Witten:2003nn}. This is achieved by restricting the path integral to self-dual configurations, which, as we show, reduces the theory to a purely first-order formalism (Section \ref{sec:firstorder_SDCPN}). 
Then, we rewrite the physical $\mathbb{C}P^{N-1}$ model in terms of the first-order formalism (Section \ref{sec:first-order-physical}).
An important finding here is that the topological counterterm has to be properly incorporated in order to switch between the first-order formalism and second-order formalism (Section \ref{sec:bosonic_chiral}).
By formulating both the self-dual and physical theories in the first-order formalism, we can construct an interpolating theory between these two theories (Section \ref{sec:epsilon-def-theory}).
This interpolation is described by the $\epsilon$-deformation of the self-dual theory in the first-order formalism, and it is equivalent to the imaginary-$\theta$ deformation of the physical theory in the second-order formalism.
The topological counterterm is once again essential for this identification.

The interpolating theory has two couplings, $\epsilon$ and $g$, which generate two independent renormalization group invariant scales, $\Lambda_\epsilon$ and $\Lambda_g$. The ratio of these two dynamically generated scales serves as the fundamental dimensionless parameter in our formalism (Section \ref{sec:RGeqs}).

\subsection{Self-dual \texorpdfstring{$\mathbb{C}P^{N-1}$}{CP(N-1)} and  the  first-order formalism}
\label{sec:firstorder_SDCPN}

We start with the partition function for the standard $\mathbb{C}P^{N-1}$ model. The bare action includes the standard second-order kinetic term and the topological $\theta$-angle:
\begin{align}
   S = \int d^2 x \frac{1}{g^2 } |D_\mu \bm{\phi}|^2 - \im \theta \int \frac{d^2 x}{2\pi} F, 
\end{align}
where $D_\mu = \partial_\mu - \im A_\mu$, $F = \epsilon_{\mu\nu}\partial_\mu A_\nu$. The gauge field is auxiliary and can be integrated out to give $A_\mu = -\im \bm{\phi}^\dagger \partial_\mu \bm{\phi}$. The field $\bm{\phi}$ obeys the constraint $|\bm{\phi}|^2=1$. The partition function of the theory is given by $Z= \int D\bm{\phi} \;  e^{-S}$. 

To construct the self-dual theory, we first restrict the path integral to purely self-dual fields. The self-duality equation in the $\mathbb{C}P^{N-1}$ model is given by $D_+ \bm{\phi}=0$, where $D_+ = D_1 + \im D_2$ is the chiral covariant derivative. To construct the self-dual formulation and the corresponding first-order formalism, we insert a Dirac delta functional $\delta(D_+ \bm{\phi}) \delta((D_+ \bm{\phi})^\dagger)$ into the path integral:  
\begin{align}
     Z_{\rm sd}= \int D\bm{\phi} \;  e^{-S} \;  \delta(D_+ \bm{\phi}) \; \delta((D_+ \bm{\phi})^\dagger).
\end{align}
We exponentiate this constraint by introducing a complex auxiliary field $h(x)$ that acts purely as a Lagrange multiplier: 
 \begin{align}
   \delta(D_+ \bm{\phi}) \delta((D_+ \bm{\phi})^\dagger) = \int Dh D h^{\dagger} \;  \exp\left[ \int \left( \im h^\dagger D_+ \bm{\phi} + \im (D_+ \bm{\phi})^\dagger h \right) \right].
\end{align}
Using the exact algebraic identity relating the standard kinetic term to the chiral covariant derivative, we have $|D_\mu \bm{\phi}|^2 = |D_+ \bm{\phi}|^2 + F$. Substituting this into the original action gives:
\begin{align}
   S_{\rm sd} = \int d^2 x \left[ \frac{1}{g^2_{\text{sd}} } |D_+ \bm{\phi}|^2 + \frac{1}{g^2_{\text{sd}} } F - \im h^\dagger D_+ \bm{\phi} - \im (D_+ \bm{\phi})^\dagger h \right] - \im \theta \int \frac{d^2 x}{2\pi} F.
\end{align}
We added the subscript ``sd'' to $g^2$ as a reminder that it is the coupling of the self-dual theory. We will describe the relation between $g^2_{\text{sd}}$ and  $g^2$ momentarily.

Because $h$ is simply an integration variable, we are free to perform a local field redefinition. By this manipulation, we can exactly cancel the $|D_+ \bm{\phi}|^2$ term by shifting  $h$ as follows:
\begin{align}
   h \longrightarrow h - \frac{\im}{2g^2_{\text{sd}} } D_+ \bm{\phi}.
\end{align}
We are left with the exact path integral for the self-dual limit: 
\begin{align}
Z_{{\rm sd}} &= \int  D \bm{\phi} D h \exp[- S_{\rm sd}], \\
   S_{\rm sd} &= \int d^2 x \left[ - \im h^\dagger D_+ \bm{\phi} - \im (D_+ \bm{\phi})^\dagger h  \right] + \left(\frac{1}{g^2_{\text{sd}} }   - \im \frac{\theta}{2 \pi}  \right) \int d^2 x F.
   \label{SDaction}
\end{align}
The kinetic terms here are strictly in the first-order formalism, acting as the exact two-dimensional counterpart to the Chalmers-Siegel formulation. The perturbation theory of this limit is completely dictated by the first term. The second term is purely topological and does not enter into the perturbation theory around the perturbative vacuum. In this formalism, the standard second-order kinetic energy is completely eliminated in favor of a purely first-order derivative coupling, and the topological structure is naturally packaged into a complexified coupling parameter.

Varying this first-order action yields a set of classical equations of motion that are intricately linked through the composite gauge connection. Variation with respect to the auxiliary Lagrange multipliers $h$ and $h^\dagger$ and   $\bm{\phi}$ and $\bm{\phi}^\dagger$  results in the following equations:
\begin{align}
    D_+ \bm{\phi} &= 0, \\
    D_- h &= 0.
\end{align}
While these equations describe a purely self-dual (holomorphic) physical field $\bm{\phi}$ and a purely anti-self-dual (anti-holomorphic) auxiliary field $h$, their dynamics are strongly coupled. Because the gauge field in the $\mathbb{C}P^{N-1}$ model is a composite auxiliary field completely determined by the scalar configuration, the anti-holomorphic field $h$ must propagate strictly in the background gauge geometry dictated by the self-dual physical field $\bm{\phi}$. 

Note that in the self-dual limit, anti-self-dual configurations are not permitted. Therefore, the path integral must reduce to a sum over instanton sectors. Later on, we will determine the vacuum structure of both the standard 
$\mathbb {C}P^{N-1}$ model, its self-dual limit as well as the interpolating theories by using the large-$N$ analysis.

\subsection{Physical theory in the first-order formalism}
\label{sec:first-order-physical}

As a starting point for constructing the interpolating theory between the self-dual $\mathbb {C}P^{N-1}$ model and physical one, we rewrite the physical theory in terms of the first-order formalism $(\phi, h)$.
At the classical level, it can be achieved by the deformation $\Delta S  = \int d^2x \;   g^2 |h|^2$.
At the quantum level, however, this turns out to be insufficient, and we address this issue in the next subsection.
\medskip

\noindent
{\bf Naive integration of the auxiliary field:} Let us first consider the naive version of the self-dual theory without the counterterm (at $\theta=0$ for convenience) and deform it with  $\Delta S  = \int d^2x \;   g^2 |h|^2$. 
\begin{equation}
    S_{\text{bare}} = \int d^2x \left( g^2 |h|^2 - \im h^\dagger D_+ \phi - \im (D_+ \phi)^\dagger h \right) + \frac{1}{g^2} \int d^2x \, F.
\end{equation}
Classically, one can integrate out $h$ by shifting the field:
\begin{equation}
    h \to h' = h - \frac{\im}{g^2} D_+ \phi
\end{equation}
This perfectly diagonalizes the action:
\begin{equation}
    S_{\text{bare}} = \int d^2x \frac{1}{g^2} |D_+\phi|^2 + \int d^2x \, g^2 |h'|^2 + \frac{1}{g^2} \int d^2x \, F
\end{equation}
Using the geometric identity $|D_+\phi|^2 = |D_\mu\phi|^2 - F$, and performing the   $h'$  integration naively,  
\begin{equation}
    S_{\text{bare}} \underbrace{\longrightarrow}_{\rm naive \;  int.} \int d^2x \frac{1}{g^2} |D_\mu\phi|^2
\end{equation}
where $-\frac{1}{g^2} \, F$  generated by the identity exactly cancels the explicit  $\frac{1}{g^2} \, F$   topological term. 

This procedure is same as using classical  equations of motions, $g^2 h=\im D_+ \phi, \;  g^2 h^{\dagger}=\im 
(D_+ \phi)^\dagger  $.  Therefore, 
classically, this perfectly recovers the standard second-order $\mathbb{C}P^{N-1}$ kinetic action.    

The above procedure is correct  classically as well as to all orders in perturbation theory  \cite{Witten:2003nn, Bittleston:2024efo, Domurcukgul:2025xgf}. However, it is not correct nonperturbatively in the full quantum theory.  

\medskip

\noindent
{\bf Necessity of the topological counterterm:}
In this integration procedure, we need to recall that the fields in the first-order formalism are chiral.
With a careful prescription to the integral measure, we will obtain 
\begin{equation}
    S_{\text{bare}} \underbrace{\longrightarrow}_{\rm full~int.}     S_{\text{quantum}} = \int d^2x \frac{1}{g^2} |D_\mu\phi|^2 - \frac{N \log g^2}{2\pi} \int d^2x \, F, \label{eq:int_out_withanomaly}
\end{equation}
where the new $\log g^2$ topological term is generated via the bosonic version of the chiral anomaly, which is derived in the next subsection.

Then, the correct first-order formalism of the physical $\mathbb {C}P^{N-1}$ model should incorporate the topological counterterm $\Delta S_{\text{counter}}$, which is chosen to cancel the residual topological term in (\ref{eq:int_out_withanomaly}).
The action is given by,
\begin{align}
   S&=    S_{\text{bare}}  +  \Delta S_{\text{counter}} \cr 
    &= \int d^2x \left( g^2 |h|^2 - \im h^\dagger D_+ \phi - \im (D_+ \phi)^\dagger h \right) + 
    \left( \frac{1}{g^2}  + \frac{N \log g^2}{2\pi} \right)  \int d^2x \, F.
    \label{withCounter}
\end{align}
This motivates us to define the $\epsilon$-deformed self-dual theory (with  $\Delta S  = \int d^2x \;   \epsilon^2 |h|^2$) interpolating between the self-dual theory (\ref{SDaction}) and the physical theory (\ref{withCounter}), with the following identification of the coupling:
\begin{align}
  \frac{1}{g^2_{\text{sd}} }   = \frac{1}{g^2 }+ \frac{N}{2\pi} \log (g^2) .
    \label{sdcoupling}
\end{align}

\subsection{Bosonic chiral anomaly and topological counterterm}
\label{sec:bosonic_chiral}

Let us now address how to properly integrate out $h$ (\ref{eq:int_out_withanomaly}).
As promised, we must treat the chiral structure, especially the zeromodes, with great care in integrating out $h$.
As a result, the anomalous $\log g^2$ topological term is generated through the chiral anomaly of the bosonic spinor $(\phi,h)$.
While we focus here only on (\ref{eq:int_out_withanomaly}), it should be noted that the same mechanism applies more generally to the $\epsilon$-deformed theory introduced later. 
For a detailed discussion with a careful treatment of the zeromodes in the $\epsilon$-deformed theory, see Appendix \ref{sec:zeromodes}.

We define $\mathcal{H}_+$ as the space to which the positive-chirality component $\phi$ belongs, and $\mathcal{H}_-$ as the space to which the negative-chirality component $h$ belongs.
The chiral Dirac-like operator $D_+$ relates these two spaces:
\begin{align}
 &\phi \in    \mathcal{H}_+, \\
 &h \in \mathcal{H}_-, \cr 
 &D_+ :  \mathcal{H}_+ \rightarrow  \mathcal{H}_- \;  .
\end{align}
To rigorously define the path integral measure, we must expand both fields in the orthonormal eigenbasis of their respective Laplacians, $\Delta_+ = -D_- D_+$ (acting on $\mathcal{H}_+$) and $\Delta_- = -D_+ D_-$ (acting on $\mathcal{H}_-$). Standard spectral theory ensures that their non-zero eigenvalues match exactly.
However, the difference in the number of zeromodes is dictated by the Atiyah-Singer index theorem:
\begin{equation}
    \text{Index}(D_+) = n_+ - n_- = \frac{N}{2\pi} \int d^2x \, F
\end{equation}
where $n_+$ and $n_-$ are the dimensions of the respective kernels, and $N$ accounts for the number of field components.

When we perform the quantum integration over the shifted variable $h'$, we are evaluating a pure Gaussian integral over the target space $\mathcal{H}_-$:
\begin{equation}
    \int \mathcal{D}h' \exp\left( -g^2 \int d^2x \, |h'|^2 \right) = \left( \det_{\mathcal{H}_-} g^2 \right)^{-1} = (g^2)^{-\dim \mathcal{H}_-}
\end{equation}
In quantum field theory, $\dim \mathcal{H}_-$ must be regularized. Using heat kernel regularization with a UV cutoff scale $\Lambda_{\rm uv}$, which preserves the $U(1)$ gauge symmetry, the number of modes in each space is given by $N_+ = \text{Tr}_{\mathcal{H}_+} e^{-\Delta_+ / \Lambda_{\rm uv}^2}$ and $N_- = \text{Tr}_{\mathcal{H}_-} e^{-\Delta_- / \Lambda_{\rm uv}^2}$. Because the non-zero spectra cancel exactly, the difference is governed strictly by the index:
\begin{equation}
    N_- = N_+ - \text{Index}(D_+)
\end{equation}
Substituting this regularized dimension back into our Gaussian evaluation yields:
\begin{equation}
    \int \mathcal{D}h' e^{-g^2 \int |h'|^2} = (g^2)^{-N_+ + \text{Index}(D_+)} = (g^2)^{-N_+} \times \exp \left( \frac{N \log g^2}{2\pi} \int d^2x \, F \right)
    \label{Jac}
\end{equation}
The first factor, $(g^2)^{-N_+}$, is an infinite constant that can be absorbed into the normalization of the $\mathcal{D}\phi$ measure, because $\mathcal{D}\phi$ is effectively an $N_+$-dimensional integration.\footnote{Indeed, this absorption corresponds to the rescaling $\phi \mapsto g\phi$, which brings the kinetic term into the canonical form. For details, see Appendix \ref{sec:zeromodes}.} The second factor, however, is an anomalous topological Jacobian generated dynamically by the asymmetric functional measure. This derivation is the exact bosonic counterpart to Fujikawa's path-integral evaluation of the fermionic chiral anomaly~\cite{Fujikawa:1979ay, Fujikawa:1980rc}.

Because this anomalous Jacobian appears in the path integral as the exponential factor in \eqref{Jac}, it acts as a negative contribution to the effective action. Consequently, the true quantum effective action generated by integrating out $h'$ includes an unwanted topological shift:
\begin{equation}
    S_{\text{quantum}} = \int d^2x \frac{1}{g^2} |D_\mu\phi|^2 - \frac{N \log g^2}{2\pi} \int d^2x \, F
\end{equation}
This reveals a profound structural feature of the theory: if one attempts to write the standard $\mathbb{C}P^{N-1}$ model in the first-order ``bosonic spinor'' formalism, the path integral measure inherently generates a chiral anomaly. Therefore, to ensure the first-order formulation rigorously matches the standard physical theory, we must explicitly add a compensating topological counterterm to our initial bare action:
\begin{equation}
    \Delta S_{\text{counter}} = \frac{N \log g^2}{2\pi} \int d^2x \, F
\end{equation}
Therefore, to obtain physical $\mathbb CP^{N-1}$ from the self-dual theory, we must use action with this counterterm:
\begin{align}
   S&=    S_{\text{bare}}  +  \Delta S_{\text{counter}} \cr 
    &= \int d^2x \left( g^2 |h|^2 - \im h^\dagger D_+ \phi - \im (D_+ \phi)^\dagger h \right) + 
    \left( \frac{1}{g^2}  + \frac{N \log g^2}{2\pi} \right)  \int d^2x \, F, \tag{\ref{withCounter}}
\end{align}
Now, with the topological counterterm, if we integrate out the $h$ field, we reproduce the standard physical theory. 
This justifies that (\ref{withCounter}) is indeed the first-order formalism of the physical $\mathbb CP^{N-1}$ model, as promised in the previous subsection.

We note that the equivalent of this counterterm is also present in connecting SDYM theory to pure Yang-Mills theory, however, up to our knowledge, it is not explored in that literature. We will explore this in a subsequent work.

\subsection{From \texorpdfstring{$\epsilon$}{epsilon}-deformation of the self-dual theory to physical theory}
\label{sec:epsilon-def-theory}

To extrapolate to the standard $\mathbb{C}P^{N-1}$ model continuously and generate an interpolating theory, we introduce a kinetic deformation $\epsilon^2 |h|^2$  to the first-order action instead of $g^2 |h|^2$ as in \eqref{withCounter}. Starting with this $\epsilon$-deformed action, and including the necessary counterterm discussed above, we have
\begin{equation}
    S_{\text{def}} = \int d^2x \left[ \epsilon^2 |h|^2 - \im h^\dagger D_+ \phi - \im (D_+ \phi)^\dagger h \right] + \left( \frac{1}{g^2} - \im \frac{\theta}{2\pi} + \frac{N}{2\pi}\log(g^2) \right) \int d^2x \, F.
    \label{careful}
\end{equation}

This deformed theory indeed continuously interpolates between the self-dual and physical theories.
In the $\epsilon^2 \rightarrow 0$ limit, this action reduces to the self-dual action (\ref{SDaction}) with the identification of the couplings (\ref{sdcoupling}): $\frac{1}{g^2_{\text{sd}} }   = \frac{1}{g^2 }+ \frac{N}{2\pi} \log (g^2)$.
Also, in the $\epsilon^2 \rightarrow g^2$ limit, the deformed theory coincides with the first-order action of the physical theory (\ref{withCounter}).


As in the previous subsections, we can switch from the first-order formalism to the second-order formalism by integrating out the auxiliary field $h$.
Carefully accounting for the path integral measure ---specifically absorbing the factor $\epsilon^{-2N_{+}}$ into the $\mathcal{D}\phi$ integration measure--- we obtain the second-order formulation of our interpolating theory $S_{\text{def}}$, 
\begin{equation}
  S = \int d^2x \frac{1}{\epsilon^2} |D_\mu \phi|^2 -    \left(  \frac{1}{\epsilon^2}  - \frac{1}{g^2} + \im \frac{\theta}{2\pi} - \frac{N}{2\pi}\log\left(\frac{g^2}{\epsilon^2}\right) \right) \int d^2x \, F.
  \label{topdef}
\end{equation}
This single expression elegantly captures the interpolation. Setting $\epsilon = g$ clearly recovers the standard physical theory. 
Conversely, in the limit $\epsilon \rightarrow 0$, the kinetic penalty for any non-holomorphic configuration becomes infinite because integrating out $h$ in \eqref{careful} induces $\frac{1}{\epsilon^2} |D_+ \bm{\phi}|^2$. Functionally, this sharply localizes the path integral onto the moduli space of exact self-dual solutions ($D_{+}\phi = 0$), and the theory reduces to the self-dual sigma model.

Notice that the difference between this deformed action and the standard sigma model action is purely a real topological term: 
\begin{align}
    \Delta S_{\rm top} \equiv  S_{\rm def}  - S = - \left( \frac{1}{\epsilon^2} - \frac{1}{g^2} - \frac{N}{2\pi} \log\left(\frac{g^2}{\epsilon^2} \right) \right)  
    \int {d^2 x} F 
\end{align}
Because this deformation is topological, it leaves standard perturbation theory invariant; however, it profoundly alters the non-perturbative dynamics of the model. Specifically, in the deformed theory, the actions and fugacities of instantons ($Q=1$) and anti-instantons ($Q=-1$) become highly asymmetric:
\begin{align}
S_{inst} &= \frac{2\pi}{g^{2}} - \im\theta, & I_{inst} &= \exp\left(-\frac{2\pi}{g^{2}} + \im\theta\right), \\
S_{anti} &= 2\pi\left(\frac{2}{\epsilon^{2}} - \frac{1}{g^{2}}\right) + \im \theta, & I_{anti} &= \exp\left[-2\pi\left(\frac{2}{\epsilon^{2}} - \frac{1}{g^{2}}\right) - \im \theta\right].
\end{align}
This asymmetry explicitly demonstrates the mechanics of the self-dual limit. The instanton fugacity is independent of the deformation parameter $\epsilon$.  However, the anti-instanton fugacity depends crucially upon it: as we take $\epsilon \rightarrow 0$, local anti-instantons acquire infinite action and completely decouple from the path integral ($I_{anti} \rightarrow 0$), leaving a purely holomorphic, instanton-dominated vacuum. Restoring the physical limit $\epsilon \rightarrow g$ perfectly recovers the standard, symmetric weights for both topological sectors.

Note that   \eqref{careful}  can be viewed as an $\epsilon$ deformation of the self-dual limit, while  \eqref{topdef} can be viewed as a deformation of the physical theory by a real topological term, leading to the result advocated in the Introduction: 
\begin{align}
    \epsilon\text{-deformation of } S_{\text{selfdual}} = \text{real topological deformation of } S_{\text{physical}}
    \label{eq:intro_correspondence2}
\end{align}
Since the second term in \eqref{topdef} is  topological, its coefficient  remains invariant under renormalization group. This gives us a constraint on the running of the coupling constant.

\subsection{Renormalization group of the kinetic and topological couplings}
\label{sec:RGeqs}

Equation \eqref{topdef} represents a two-coupling generalization of the standard sigma model, characterized by a topological coupling $g$ and a kinetic coupling $\epsilon$. Because only $\epsilon$ appears  in front of the local kinetic term (not $g$), it dictates the 
scattering,  and obeys standard renormalization group running, hence the name kinetic coupling. Conversely, a specific algebraic combination of $\epsilon$ and $g$ governs the purely topological term in the effective action. Since the coefficient of a topological term cannot depend on the continuous renormalization scale $\mu$, this combination must be a renormalization group (RG) invariant. 

When we introduce the $\epsilon$-deformation,  upon integrating out the auxiliary $h$  field,  $\frac{1}{\epsilon^2}$ emerges as the coefficient of the kinetic term $\frac{1}{\epsilon^2} |D_\mu \bm{\phi}|^2$ and  $ \frac{1}{\epsilon^2} + \frac{N}{2\pi} \log(\epsilon^2)$ combination appears in the coefficient of the topological term. \footnote{
Recall that in the standard $\mathbb{C}P^{N-1}$ model, $g$ acts as the canonical physical coupling. In the self-dual limit ($\epsilon^2 \rightarrow +0$), however, it disappears from the local kinetic sector and only appears in the algebraic combination $\frac{1}{g_{sd}^2} \equiv \frac{1}{g^2} + \frac{N}{2\pi}\log(g^2)$ in the coefficient of the topological term. The logarithmic shift $\frac{N}{2\pi}\log(g^2)$ is the direct consequence of the bosonic chiral anomaly in the first-order formalism, arising strictly from the non-invariance of the path integral measure. This precise relationship between the canonical physical coupling and the chiral/holomorphic coupling parallels the mechanism uncovered by Novikov, Shifman, Vainshtein, and Zakharov~\cite{Novikov:1983uc, Novikov:1985rd} and clarified by Arkani-Hamed and Murayama~\cite{Arkani-Hamed:1997qui} in their resolution of exact beta functions in supersymmetric gauge theories. In both cases, the universal all-orders correction arises from the geometric non-invariance of the measure.
}

In the leading large-$N$ limit, the running of the $\epsilon$ is one-loop exact and define the beta-functions 
\begin{align}
 \beta_{\epsilon^2} = -\frac{N}{2\pi} \epsilon^4   \label{eq:one-loop-epsilon-RG}
\end{align}
 Hence, we identify the first RG-invariant scale of the large-$N$ theory as 
\begin{align}
\Lambda_{\epsilon}^2 &= \mu^2 \exp\left( - \frac{4\pi}{N \epsilon^2} \right).
\label{phys-run}
\end{align}
The second RG-invariant is the coefficient of the topological term:
 
\begin{equation}
 \left( \frac{1}{\epsilon^2} + \frac{N}{2\pi} \log(\epsilon^2) \right) - \left( \frac{1}{g^2} + \frac{N}{2\pi} \log(g^2) \right) = \mathcal{C} \label{RGconstraint2}
\end{equation}
Because $\mathcal{C}$ is  independent of $\mu$, unlike $\epsilon$ and $g$, this relation  locks the running of the kinetic coupling $\epsilon^2(\mu)$ to the topological coupling $g^2(\mu)$. 

To extract the implications for the $\beta$-functions, we apply the RG operator $\mu \frac{d}{d\mu}$ to the invariant constraint. Defining the $\beta$-functions as $\beta_{g^2} = \mu \frac{d}{d\mu} g^2$ and $\beta_{\epsilon^2} = \mu \frac{d}{d\mu} \epsilon^2$, we obtain an exact, all-orders algebraic map between the two flows:
\begin{equation}
\frac{\beta_{\epsilon^2}}{\epsilon^4} \left( 1 - \frac{N \epsilon^2}{2\pi} \right) = \frac{\beta_{g^2}}{g^4} \left( 1 - \frac{N g^2}{2\pi} \right) \label{exactbeta}
\end{equation}
This relation mathematically enforces that the running of the kinetic and topological couplings are permanently locked to one another. Solving for $\beta_{g^2}$ isolates the physical beta function, revealing a highly non-trivial structure:
\begin{equation}
    \beta_{g^2} = -\frac{N}{2\pi} g^4 \left[ \frac{1 - \frac{N \epsilon^2}{2\pi}}{1 - \frac{N g^2}{2\pi}} \right],
\end{equation}
where we have used the one-loop expression of $\beta_{\epsilon^2}$ (\ref{eq:one-loop-epsilon-RG}).

Note that using the self-dual coupling $\frac{1}{g^2_{sd}} \equiv \frac{1}{g^2} + \frac{N}{2\pi} \log(g^2)$,  it is simple to see that 
\begin{align}
  \beta_{g^2_{sd}} = -\frac{N}{2\pi} g^4_{sd}   \left( 1 - \frac{N \epsilon^2}{2\pi}  \right).
\end{align}
In the self-dual limit ($\epsilon \to 0$) and physical limit ($\epsilon=g$), we have exact one-loop running at large-$N$:
\begin{align}
\text{Self-dual limit:} \qquad \beta_{g_{sd}^2} &= -\frac{N}{2\pi} g^4_{sd} \quad \text{(1-loop exact)}  \\ 
\text{Physical limit:} \qquad \beta_{g^2} &= -\frac{N}{2\pi} g^4 \quad \text{(1-loop exact, large-N)}.
\label{top-run}
\end{align}
Note that in the physical limit, this one-loop exactness holds only at large $N$, whereas in the self-dual limit, it holds also for finite $N$.
In the self-dual limit, we keep only the leading order in $\epsilon$, justifying the use of (\ref{eq:one-loop-epsilon-RG}).

In particular, the strong scale of the self-dual limit is subtly different from the physical theory:\footnote{The $1/N^2$ factor in the self-dual limit is added for the later convenience.
In the large-$N$ calculation in this paper, we keep $\lambda_g =Ng^2$ finite, which corresponds to keeping $\frac{1}{N g^2_{\text{sd}}} + \frac{1}{2\pi} \log(N)$ finite. Note that this is not equivalent to keeping $N g^2_{\text{sd}}$ finite.
}
\begin{align}
 \text{Self-dual limit:} \qquad \Lambda_{g}^2 &= \frac{1}{N^2}\mu^2 \exp\left( - \frac{4\pi}{N g_{sd}^2} \right) = \mu^2 \left(\frac{1}{g^4 N^2}\right) \exp\left( - \frac{4\pi}{N g^2} \right),  \\
 \text{Physical limit:} \qquad \Lambda_{g}^2 &= \mu^2 \exp\left( - \frac{4\pi}{N g^2} \right).
\end{align}

Note that in the self-dual limit, although the running of $g_{sd}^2$ is one-loop exact, the canonical coupling $g^2$ receives perturbative loop corrections at all orders,
\begin{equation}
    \beta_{g^2} = -\frac{N}{2\pi}g^4 \frac{1}{1 - \frac{Ng^2}{2\pi}}.
\end{equation}
This is a direct consequence of the topological counterterm arising from the bosonic chiral anomaly. The emergence of an all-orders geometric denominator $(1 - Ng^2/2\pi)^{-1}$ is strikingly analogous to the exact Novikov--Shifman--Vainshtein--Zakharov (NSVZ) beta function in supersymmetric Yang--Mills theories~\cite{Novikov:1983uc, Novikov:1985rd}. 
In both frameworks, while the holomorphic or chiral coupling runs at one loop, the canonical physical coupling acquires an infinite tower of higher-loop corrections driven entirely by the Jacobian of the path integral measure. \footnote{
The structure of the denominator indicates that the exact large-$N$ beta function for $g^2$ possesses a singularity, diverging at exactly $Ng^2 = 2\pi$.
This singularity is superficial, however, as $g^2$ appears in the action only in the form of $\frac{1}{g^2}+ \frac{N}{2\pi} \log (g^2)$.
We can in principle derive the detailed structure of scale dependence of the coupling $N g^2$ in terms of   Lambert $W_{-1}$ and $W_0$, but we will not need this detailed information in this work. 
}

Finally, this formulation confirms the physical consistency of the deformation. If the interpolation is turned off by setting the invariant separation to zero ($\mathcal{C} = 0$), the physical and kinetic couplings become strictly equivalent ($\epsilon^2 = g^2$). In this undeformed limit, the numerator and denominator within the bracket exactly cancel out. The pole disappears, the infinite tower of higher-loop corrections identically vanishes, and the theory recovers the standard one-loop exact running of the standard large-$N$  $\mathbb{C}P^{N-1}$ model, $\beta_{g^2} = -\frac{N}{2\pi}g^4$. This shows that  the all-orders running of $g^2$ in deformed theory is  entirely driven by the topological imbalance between the two sectors.

\section{Saddle-point analysis at large \texorpdfstring{$N$}{N}}
\label{sec:saddle-point-general}

At large $N$, we can nonperturbatively study the interpolating theory (or the $\epsilon$-deformed theory) via the saddle-point analysis.
In this section, we first derive the large-$N$ effective action and its saddle-point equations.
Within the standard prescription for $\theta = 0$, we find a saddle for any $0<\epsilon<g$ and argue that the saddle point is continuously connected between $\epsilon = +0$ (almost self-dual limit) and $\epsilon=g$ (standard $\mathbb{C}P^{N-1}$) in Section \ref{sec:solving_equations}.

We then face a severe problem when $\frac{\theta+2\pi n}{N}$ is dialed, where $n$ is a label of sectors that must be summed up for imposing the Dirac quantization $\int F \in 2 \pi \mathbb{Z}$.
For example, even at $\theta =0$, the saddle at some sectors $\frac{n}{N} \sim O(1)$ seems more dominant than the standard saddle at $n=0$, i.e., as it stands, it has lower energy than the vacuum obtained via weak field analysis in classical works  
\cite{DAdda:1978vbw, DAdda:1978dle, DAdda:1982lsk, Witten:1978bc}.  
This is the problem that exists even in the physical $\mathbb{C}P^{N-1}$ model, as pointed out in Ref.~\cite{Sugeno:2025exv}.  
To resolve this problem, we need a careful analysis of the thimble structure, which is discussed in the next section.

\subsection{Large-\texorpdfstring{$N$}{N} Effective Action and Saddle Point Equations}

We now determine the large-$N$  effective action for the interpolating theory \eqref{careful} or equivalently \eqref{topdef}  for generic values of the $\epsilon$.  This will allow us to understand the non-perturbative ground state properties of both self-dual limit as well as the intricate global structure of the theta-vacua in the physical theory.  

To determine the large-$N$ effective action, we start with implementing the constraint $|\bm{\phi}(x)|^2 = 1$  via the auxiliary Lagrange multiplier field $M(x)$
\begin{align}
\prod_x \delta(|\bm{\phi}(x)|^2 - 1) = \prod_x \int_{ \im \mathbb{R}} dM(x) e^{-M(x)(|\bm{\phi}(x)|^2 - 1)} \equiv \int DM(x) e^{-\int d^2x \, M(x)(|\bm{\phi}(x)|^2 - 1)},
\end{align}
 We can write the standard partition function as (we will reinstate the topological term momentarily):
\begin{align}
Z = \int D\bm{\phi} DA_\mu DM \, \exp\left[ - \int d^2x \left( \frac{1}{\epsilon^2} |D_\mu \bm{\phi}|^2 + \frac{M}{\epsilon^2} (|\bm{\phi}|^2 - 1) \right)  \right].
\end{align}
Now, since $\bm{\phi}$ is quadratic and unconstrained, we can integrate it out exactly in the path integral to obtain an effective action in terms of $A_\mu$ and $M$:
\begin{align}
Z &= \int DA_\mu DM \, \left[ \det(-D_\mu^2 + M) \right]^{-N} \exp\left[ \frac{1}{\epsilon^2} \int d^2x \, M  \right] \cr
&= \int DA_\mu DM \, \exp\left[ -N \operatorname{tr} \log(-D_\mu^2 + M) + \frac{1}{\epsilon^2} \int d^2x \, M  \right].
\end{align}
For the topological term, we recall (\ref{topdef}):
\begin{align}
    S_{\mathrm{top}}=- \left[\frac{1}{\epsilon^2} - \frac{1}{g^2} + \im \frac{\theta}{2\pi} - \frac{N}{2\pi} \log \left( \frac{g^2}{\epsilon^2}\right) \right] \int d^2x~F =: -N\tau \int d^2x~F
\end{align}
including the counterterm. 

If we view the theory on $\mathbb{R}^2$ as the decompactification limit of the theory on some two-manifold $M_2$, we can replace the integral over the gauge potential $A$ with an integral over the field strength $F$, including a sum over all flux sectors via the Poisson resummation formula:
\begin{align}
\int DA \to \int DF \sum_{\nu \in \mathbb{Z}} \delta\left( \nu - \frac{1}{2\pi} \int F \right) = \int DF \sum_{n \in \mathbb{Z}} e^{i n \int F}. \label{eq:Poisson_summation}
\end{align}

In our analysis, we adopt the uniform ansatz of $(F,M)$ as a working assumption, restricting the path integral entirely to the constant modes. 
This ansatz is reasonable from the translational invariance.
Although the influence of nonzero momentum modes on the thimble analysis is highly non-trivial, we proceed by neglecting them to evaluate the effective action, assuming that the nonzero mode corrections do not change the conclusion drastically.

We evaluate the functional determinant by summing over Landau levels under the uniform $F$ background.
The Landau level spectrum is given by $\text{Spec}(-D^2) = \{\vert{}F\vert{} (2m+1) ; m=0,1,2,\dots\}$.
The degeneracy of the levels is given by $\operatorname{deg}(m) = \frac{F V_{\mathrm{2d}}}{2\pi}$, where $V_{\mathrm{2d}} = \int_{M_2} d^2x$ is the area of the two-dimensional space:
\begin{align}
\det(-D_\mu^2 + M) &= \prod_{m \ge 0} \left( 2F\left(m + \frac{1}{2}\right) + M \right)^{\operatorname{deg}(m)} \cr
&= \exp\left[ \sum_{m \ge 0} \operatorname{deg}(m) \log\left( 2F\left(m + \frac{1}{2}\right) + M \right) \right].
\end{align}
Using the proper-time integral representation for the logarithm (with a UV cutoff $\mu \to \infty$),
\begin{align}
\log\frac{\xi}{\mu^2} = - \int_{1/\mu^2}^\infty dt \frac{1}{t} e^{-\xi t}, \qquad \operatorname{Re}(\xi) > 0,
\end{align}
we can rewrite the determinant factor $\left[ \det(-D_\mu^2 + M) \right]^{-N}$ in the form:
\begin{align}
\left[ \det(-D_\mu^2 + M) \right]^{-N} &= \exp\left[ N \int_{M_2} d^2x \frac{F}{2\pi} \int_{1/\mu^2}^\infty dt \frac{1}{t} \sum_{m \ge 0} e^{-\left(2F\left(m + \frac{1}{2}\right) + M\right)t} \right] \cr
&= \exp\left[ N \int d^2x \int_{1/\mu^2}^\infty dt \frac{F}{4\pi t} e^{-Mt} \frac{1}{\sinh(Ft)} \right].
\end{align}

Now, restoring the complex topological term,  the partition function takes the form:
\begin{align}
Z(\theta) = \sum_{n \in \mathbb{Z}} \int DF DM \, \exp\left[ -N \int_{M_2} d^2x \, L_{\text{eff}}(M, F) \right],
\label{PF}
\end{align}
where the effective Lagrangian in terms of the background field strength $F$ and the dynamically generated mass gap $M$ takes the form:
\begin{align}
    L_{\rm eff}(M, F)  &= - \int_{1/\mu^2}^{\infty} dt \frac{F}{4 \pi t} e^{-Mt} \frac{1}{\sinh(Ft)} - \frac{M}{N\epsilon^2}  
    - {F} \tau_n \cr 
    \tau_n &:= \left[ \left( \frac{1}{\epsilon^2} - \frac{1}{g^2} \right) \frac{1}{N} + \im \frac{\theta + 2\pi n}{2\pi N} - \frac{1}{2\pi} \log  \left( \frac{g^2}{\epsilon^2} \right) \right] = \tau + \im \frac{2\pi n}{2 \pi N}  \cr 
     \tau_n &= \left[ \left( \frac{1}{ \lambda_\epsilon} - \frac{1}{\lambda_g} \right) 
     - \frac{1}{2\pi} \log  \left( \frac{\lambda_g}{ \lambda_\epsilon} \right)  + \im  \frac{\bar \theta_n}{2 \pi} \right] \equiv \tau_{r} + \im \tau_{i}
     \label{effLag}
\end{align}
For   $\epsilon=g$ where ${\rm Re}({\tau}_n)=0$, this reduces to the standard large-$N$ effective Lagrangian  $L_{\rm eff}(M, F)$  of the sigma model \cite{Rossi:2016uce, Aguado:2010ex, Lawrence:2012ua, Sugeno:2025exv, Campostrini:1991tw, Campostrini:1992ar}.  In the last line, we expressed  $\tau_n$ in terms of 't Hooft couplings and $\bar \theta_n$, both of which remain $O(N^0)$ in the large-$N$ limit. 

Note that in the large-$N$ limit, we keep $\tau_n$ fixed, i.e, 't Hooft couplings  $\epsilon^2 N$,  $g^2N$  and  $\frac{\theta + 2\pi n}{2\pi N}$  are $O(N^0)$ and  held fixed at some scale.
We can also trivially express  the term inside the log in terms of  't Hooft couplings, as $\log  \left( \frac{g^2N}{\epsilon^2N} \right)$.
Clearly, 
$\tau_n \equiv \tau_{r} + \im \tau_{i}$ has both the imaginary part (emanating from the topological theta angle and branch label) as well as a real part (which is sourced by the chiral asymmetry or deformation term) in the theory.

The $- \tau {F}$ term can be viewed as sourcing an $F$-condensate. Since $\tau  = \tau_r + \im \tau_i \in \mathbb C$, in the  Euclidean 
formulation, the $\tau_r$ sources a real magnetic field and   $\tau_i$ sources an imaginary magnetic field. In Minkowski formulation, this implies that  $\tau_r$ sources an imaginary  electric  field and   $\tau_i$ sources a real electric  field.  We can summarize this overall structure in a table. 

\begin{align}
\begin{array}{l|l|l}
\text{Formulation} & \tau_r & \tau_i \\
\hline
\text{Euclidean:} & B_r & B_i \\
\text{Minkowski:} & E_i & E_r \\
\text{Energy density: }  & \mathcal{E}_M< 0 & \mathcal{E}_M > 0
\end{array}
\label{eq:tau_sourcing}
\end{align}
By fixing $\tau$ in the UV,  we can probe both physical theory at arbitrary theta and level number (which realizes a genuine real electric field between   parallel plate capacitors  with charges  $ \pm \frac{\theta + 2\pi n}{2\pi N}$ at $\pm \infty$ in Minkowski formulation) and also determine the vacuum structure of the 
self-dual $\mathbb CP^{N-1}$ model (which realizes a Landau level problem  in a magnetic field in Euclidean formulation). Dialing $\tau \in \mathbb C$ allows us to probe the vacuum structure of the general theory.

\noindent 
{\bf Saddle point Equations:}
The vacuum is determined by demanding that the effective action is stationary with respect to variations in both the auxiliary field $M$ and the field strength $F$. 
 Taking $ {\partial L_{\rm eff}} / \partial M = 0$ yields the first saddle-point (Gap) equation:
\begin{align}
    I(M,F; \mu^2) := \int_{1/\mu^2}^{\infty} dt \frac{F}{4\pi} e^{-Mt} \left( \frac{1}{\sinh(Ft)} \right) = \frac{1}{N\epsilon^2} \label{eq:1st_saddle_point_eq}
\end{align}
Taking ${\partial L_{\rm eff}} / \partial F = 0$ yields the second saddle-point  equation:
\begin{align}
    J(M,F) := \int_{0}^{\infty} dt \frac{1}{4\pi t} e^{-Mt} \left( \frac{1}{\sinh(Ft)} - \frac{Ft \cosh(Ft)}{\sinh^2(Ft)} \right) = - \tau_n \label{eq:2nd_saddle_point_eq}
\end{align}

\subsection{Solving the saddle-point equation and continuity}
\label{sec:solving_equations}

Let us solve the saddle-point equations (\ref{eq:1st_saddle_point_eq}) and (\ref{eq:2nd_saddle_point_eq}).
We will see that the saddle point is continuously connected between $\epsilon = +0$ (almost self-dual limit) and $\epsilon=g$ (standard $\mathbb{C}P^{N-1}$).
For the almost self-dual case, we can determine the saddle point analytically.

\subsubsection{Structure of the saddle-point equations}

As we introduced ``chemical potential for the topological charge'' when $\epsilon < g$, we expect that the saddle-point value of $F$ satisfies $\operatorname{Re} F_*>0$.
Hence, we shall look for a (generally complex) saddle point, which is analytically continued from the positive real axis\footnote{As a possibility, a saddle point on a Riemann sheet analytically continued from the negative real axis may contribute.
In our situation ($\epsilon < g$), such a contribution is generally not dominant.
We will comment on this point later.
} $F>0$.

In this Riemann sheet, we can compute the integrals as follows.
For later convenience, we introduce
\begin{align}
    z:= M/F
\end{align}
For the first saddle-point equation, by extracting the asymptotic behavior for $\mu^2 \gg M, F$, we obtain:
\begin{align}
    I(M,F; \mu^2) &= \frac{F}{4\pi} \int_{1/\mu^2}^{\infty} dt \, e^{-Mt} \left( \frac{1}{\sinh(Ft)} \right) \notag \\
    &\simeq \frac{1}{4\pi} \left[ \log\left(\frac{\mu^2}{2F}\right) - \gamma - \psi\left(\frac{z+1}{2}\right) \right] ,
\end{align}
where $\gamma$ is the Euler-Mascheroni constant and $\psi(\cdot)$ is the digamma function.
For the second saddle-point equation, we have
\begin{align}
    J(M,F) &= \frac{1}{4\pi} \int_{0}^{\infty} \frac{dt}{t} \, e^{-Mt} \left( \frac{1}{\sinh(Ft)} - \frac{Ft \cosh(Ft)}{\sinh^2(Ft)} \right) \notag \\
    &= \frac{1}{4\pi} \left[ 2 \log \Gamma\left(\frac{z+1}{2}\right) - z \psi\left(\frac{z+1}{2}\right) + z - \log(2\pi) \right] .
\end{align}
Here, we notice the structure of the two saddle-point equations.
\begin{itemize}
    \item The second saddle-point equation, (\ref{eq:2nd_saddle_point_eq}) which can be written as $J(z)= -\tau_n$, determines the ratio $z_*=M_*/F_*$ of the saddle point $(M_*,F_*)$.
    \item The first saddle-point equation (\ref{eq:1st_saddle_point_eq}) yields
\begin{align}
    F_* = \frac{\mu^2}{2} \exp\left( -\frac{4\pi}{N\epsilon^2(\mu)} - \gamma - \psi\left(\frac{z_*+1}{2}\right) \right)
\end{align}
and $M_* = z_* F_*$.
\end{itemize}
Thus, the nontrivial task is to solve $J(M,F) = -\tau_n$ with respect to $z = M/F$.

Before solving the saddle-point equations, we calculate the vacuum energy density, which is given by
\begin{align}
     L_{\rm eff}(M, F)  &= - \int_{1/\mu^2}^{\infty} dt \frac{F}{4 \pi t} e^{-Mt} \frac{1}{\sinh(Ft)} - \frac{M}{N\epsilon^2}  
    - {F} \tau_n .
\end{align}
The integral is evaluated as follows:
\begin{align}
     \int_{1/\mu^2}^{\infty} dt \frac{F}{4 \pi t} e^{-Mt} \frac{1}{\sinh(Ft)} &= \frac{1}{4\pi} \left[ \mu^2 - M\log\left(\frac{\mu^2}{2F}\right) + M\gamma - F\log(2\pi) + 2F\log\Gamma\left(\frac{F+M}{2F}\right) \right] \label{eq:integral_vac_energy}
\end{align}
After some calculation using the saddle-point equations\footnote{
Using the first saddle point equation, we can eliminate $\log \mu^2$:
\begin{align}
    L_{\rm eff}(M_*, F_*) &=  -\frac{1}{4\pi} \left[ \mu^2 -M \psi\left(\frac{F+M}{2F}\right)  - F\log(2\pi) + 2F\log\Gamma\left(\frac{F+M}{2F}\right) \right] - F \tau_n
\end{align}
The expression has the same analytic form as the second saddle-point equation, and most terms are canceled: $L_{\rm eff}(M_*, F_*)  = \frac{1}{4\pi} (M_* - \mu^2)$.
}, we obtain
\begin{align}
     L_{\rm eff}(M_*, F_*)  &= \frac{1}{4\pi} (M_* - \mu^2).
\end{align}

\subsubsection{Solution at \texorpdfstring{$\theta + 2\pi n = 0$}{theta+2pi n=0} }
\label{sec:continuity_realtau}

We first look at the saddle that is continuously connected to the usual large-$N$ analysis at $\epsilon =g$.
To this end, we set $\theta =0$, and focus on the $n=0$ sector, which makes $\tau_n =\tau \in \mathbb{R}$.
Within our interest $0<\epsilon <g$, the topological term $\tau$ takes a value in $0< \tau <\infty$

\begin{figure}[t]
    \centering
    \includegraphics[width=0.6 \linewidth]{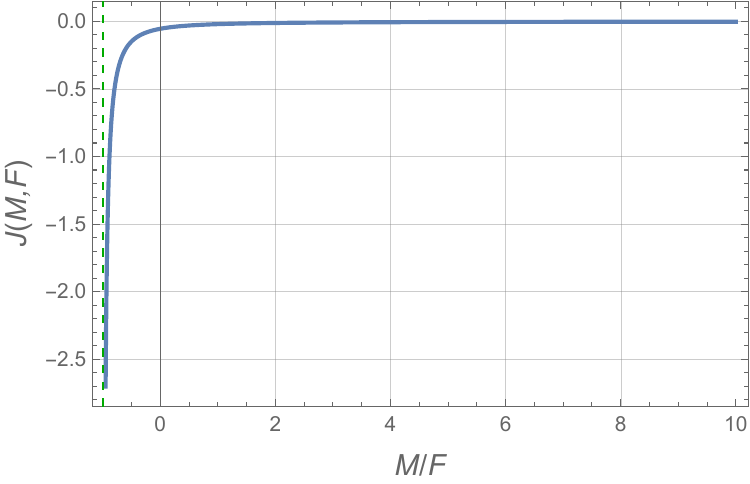}
    \caption{The 2nd saddle-point equation implies the one-to-one correspondence between $-1 < z < \infty \longleftrightarrow 0 < \tau <\infty$, where $z = M/F$.
    Note that $J(M,F) =\frac{1}{4\pi} \left[ \frac{M}{F} + 2\log\Gamma\left(\frac{F+M}{2F}\right) - \frac{M}{F}\psi\left(\frac{F+M}{2F}\right) - \log(2\pi) \right]$ is a function of $z$, and it is analytic for $-1<z<+\infty$ (across $z =0$).
    }
    \label{fig:plotJ2}
\end{figure}

From the 2nd saddle-point equation (at $\theta =0$ and $n=0$)
\begin{align}
    J(M,F) =  - \tau 
\end{align}
we can see a smooth one-to-one correspondence between $z=M/F$ and $\tau$:
\begin{align}
    -1 < z < \infty \longleftrightarrow 0 <\tau<\infty
\end{align}
This can be seen from Figure \ref{fig:plotJ2}, where $J(z)$ is plotted.

There are infinitely many other solutions in the region $z < -1$.
However, these solutions violate the stability condition of the fluctuation $M > - |F|$, and thus they are unphysical saddles.
As we will see later, this is also justified by the perspective of the contour deformation.

Note that we can recover the standard $\mathbb{C}P^{N-1}$ model (at $\epsilon^2 = g^2-0$, i.e., $\tau = +0$).
We can find a solution at large $z_*=M_*/F_* \rightarrow +\infty$, which is,
\begin{align}
    (M_*,F_*) = ( \mu^2 \rme^{-\gamma } \rme^{-\frac{4 \pi}{N\epsilon^2(\mu)}},+0) = ( \mu^2 \rme^{-\gamma } \rme^{-\frac{4 \pi}{Ng^2(\mu)}},+0)  ~~~~~~(\text{at }\epsilon^2 = g^2 -0) \label{eq:physical_saddle_location}
\end{align}
where we have used $\psi(z) \simeq \log z +O(z^{-1})$ for $|z| \rightarrow +\infty$.
This solution can also be obtained by solving the gap equation ($\partial L_{\mathrm{eff}}/\partial M = 0$) at $F=0$ as usual.

We can observe that the saddle point continuously moves in the region $0<\epsilon <g$.
At $\epsilon^2 = g^2-0$, i.e., $\tau = +0$, the saddle point is the (perturbed) one of the standard $\mathbb{C}P^{N-1}$ model.
As we increase $\tau: 0 \rightarrow +\infty$, the solution $z$ smoothly decreases as $z: +\infty \rightarrow -1$ shown in Figure \ref{fig:plotJ2}.
This solution is continuously connected to the saddle of the almost self-dual limit $\tau \rightarrow +\infty$, where $z = -1 +0$.

\begin{figure}[t]
    \centering
    \includegraphics[width=0.75 \linewidth]{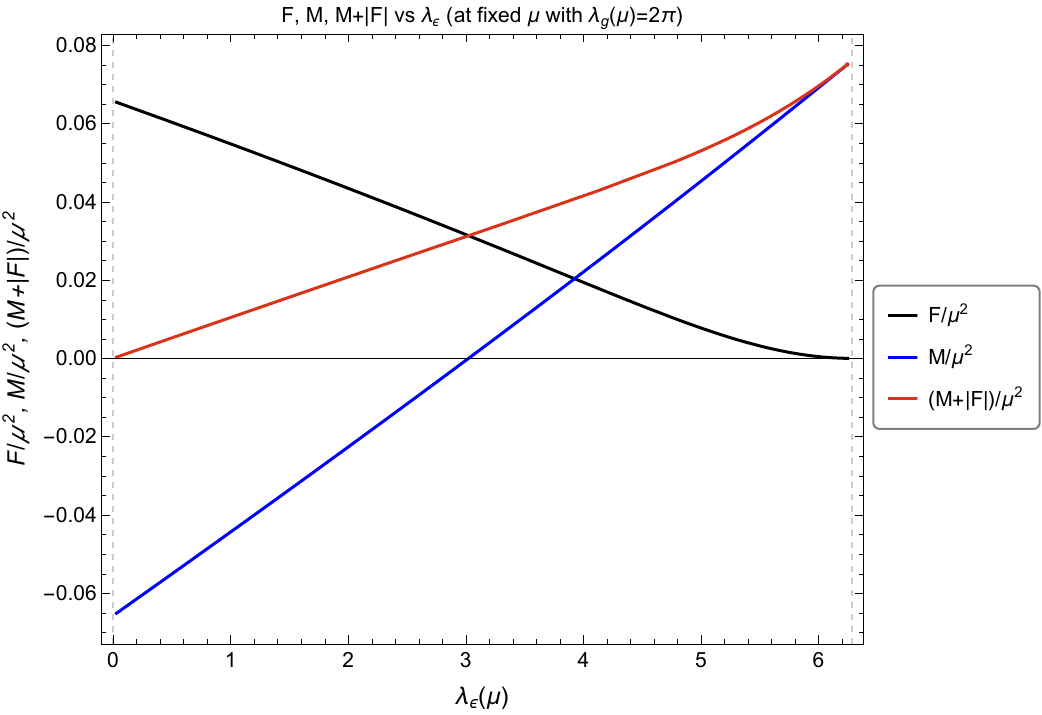}
    \caption{Plots of $M_*$, $F_*$, and $M_*+|F_*|$ as functions of $\lambda_\epsilon(\mu)=N\epsilon^2(\mu)$.
The fixed scale $\mu$ is chosen such that $\lambda_g(\mu) = Ng^2(\mu) = 2\pi$.
The self-dual theory corresponds to $\lambda_\epsilon = 0$, while the standard theory is recovered at $\lambda_\epsilon = \lambda_g = 2\pi$.
In addition to $M_*$ and $F_*$, we plot the mass gap of the fluctuation operator, $M_*+|F_*|$, which vanishes at the self-dual point.
    }
    \label{fig:MFvsepsilon}
\end{figure}

In Figure \ref{fig:MFvsepsilon}, we plot $M_*$, $F_*$, and $M_*+|F_*|$ as functions of $\lambda_\epsilon(\mu)=N\epsilon^2(\mu)$ at a fixed scale $\mu$.
This plot illustrates the continuity between the self-dual theory ($\lambda_\epsilon = 0$) and the standard $\mathbb{C}P^{N-1}$ model ($\lambda_\epsilon = \lambda_g$), where we define $\lambda_\epsilon = N\epsilon^2$ and $\lambda_g = Ng^2$.
This figure also demonstrates that the mass gap of the fluctuation operator, $M_*+|F_*|$, vanishes at the self-dual point.
Note that the stability bound is $M + |F| \geq 0$, and $M<0$ itself does not indicate instability.

\subsubsection{Analytic form at almost self-dual limit}
\label{sec:self-dual_limit_analytic}

We can derive the analytic form of $(M_*,F_*)$ at almost self-dual limit, where $\tau \gg 1$.
From Figure \ref{fig:plotJ2}, the ratio is located at $z = -1 +0$.
Let $\delta z$ be a deviation from $-1$: $z = -1 + \delta z$.
Then, the second saddle-point equation relates $\delta z$ and $\tau$:
\begin{align}
    J(M,F)&= \frac{1}{4\pi} \left[ 2 \log \Gamma\left(\frac{z+1}{2}\right) - z \psi\left(\frac{z+1}{2}\right) + z - \log(2\pi) \right] \notag \\
    &=  \frac{1}{4\pi} \left[ -\frac{2}{\delta z} - 2 \log(\delta z) + 1 - \gamma + \log\left(\frac{2}{\pi}\right) + O(\delta z) \right] = -\tau,
\end{align}
and we find
\begin{align}
  \delta z = \frac{1}{2\pi \tau} \left[ 1 - \frac{\log(2\pi \tau) + \frac{1}{2} \left( 1 - \gamma + \log\left(\frac{2}{\pi}\right) \right)}{2\pi \tau} \right] + O\left( \frac{\log^2 \tau}{\tau^3} \right).
\end{align}
Then, the first saddle-point equation gives the actual saddle-point variables:
\begin{align}
    F_* &= \frac{\mu^2}{2} \exp\left( -\frac{4\pi}{N\epsilon^2(\mu)} - \gamma - \psi\left(\frac{z_*+1}{2}\right) \right) \notag \\
    &= \frac{\mu^2}{2} \exp\left( -\frac{4\pi}{N\epsilon^2(\mu)} + \frac{2}{\delta z_*} + O(\delta z_*)\right) \notag \\
    &= \left[ 4 \pi \tau^2\rme^{4 \pi \tau + 1 - \gamma} \left(1+ O\left( \frac{\log \tau}{\tau} \right) \right)\right]  \mu^2 \rme^{-\frac{4\pi}{N\epsilon^2(\mu)}} , \\
    M_* &= z_* F_* = - \left[ 4 \pi \tau^2\rme^{4 \pi \tau + 1 - \gamma} \left(1+ O\left( \frac{\log \tau}{\tau} \right) \right)\right] \mu^2 \rme^{-\frac{4\pi}{N\epsilon^2(\mu)}}
\end{align}
These formulae match the large-$\theta$ asymptotics obtained in Ref.~\cite{Sugeno:2025exv}, via the analytic continuation on the $\tau$ plane.
Again, we note that, although $M < 0$, the stability bound for the fluctuation operator $M +|F|>0$ is satisfied.

It is already interesting to substitute $    \tau = \left[ \left( \frac{1}{\epsilon^2} - \frac{1}{g^2} \right) \frac{1}{N}  - \frac{1}{2\pi} \log (g^2/\epsilon^2)\right] $ into these formulae by assuming $1/\epsilon^2 \gg 1/g^2$:
\begin{align}
    (F_* ,M_*) \simeq \left(\frac{4\pi \mu^2 \rme^{1-\gamma}}{N^2 g^4} \exp\left( -\frac{4\pi}{Ng^2} \right), - \frac{4\pi \mu^2 \rme^{1-\gamma}}{N^2 g^4} \exp\left( -\frac{4\pi}{Ng^2} \right) \right).  \label{eq:saddle_point_almost_selfdual}
\end{align}
We emphasize that the topological counterterm from the bosonic chiral anomaly $- \frac{1}{2\pi} \log (g^2/\epsilon^2) $ is essential to have a well-defined self-dual limit of $\epsilon \rightarrow +0$.
The dynamical scale in the self-dual limit is determined by
\begin{align}
    \Lambda_g^2 := \frac{\mu^2 }{N^2 g^4} \rme^{-\frac{4\pi}{Ng^2}}= \frac{\mu^2 }{N^2 } \rme^{-\frac{4\pi}{Ng_{\mathrm{sd}}^2}},
\end{align}
where we have used $ g^{-2}_{\text{sd}} = g^{-2} + \frac{N}{2\pi} \log (g^2)$.  
As a result, for the vacuum of the self-dual theory, we find the condensates:
\begin{align}
    (F_* ,M_*) \simeq 4\pi  \rme^{1-\gamma}  \left( \Lambda_g^2, -\Lambda_g^2 \right).  
    \label{eq:self-dual-sol}
\end{align}

We also note that, in the self-dual limit $\epsilon \rightarrow +0$, the lowest eigenvalue of the fluctuation operator converges to zero $M_* + |F_*| \rightarrow + 0$.
The lowest Landau levels can effectively play the role of the zeromodes in this limit within the semiclassical analysis of $(M,F)$. \footnote{ 
Note that, according to \eqref{eq:tau_sourcing}, the Euclidean $F_* \sim \Lambda^2_g$ condensate corresponds to an imaginary electric field condensate. This is natural because at this stage, only ${\rm Re}(\tau)$ is activated. Recall that  ${\rm Im}(\tau)$ sources real electric field in Minkowski formulation, hence  ${\rm Re}(\tau)$  must source an imaginary electric field. }

\subsection{Problem of unphysical saddles}

The large-$N$ saddle point method gives reasonable results so far.
We can easily recover the $\theta$ and $n$ dependence:
at large $\operatorname{Re} \tau \gg 1$, we have
\begin{align}
    F_*^{(n)} &\simeq \left[ 4 \pi \tau^2_n \rme^{4 \pi \tau_n + 1 - \gamma} \right]  \mu^2 \rme^{-\frac{4\pi}{N\epsilon^2(\mu)}} , 
    \label{saddlelarge} \\
    M_*^{(n)} &\simeq - \left[ 4 \pi \tau^2_n \rme^{4 \pi \tau_n + 1 - \gamma} \right] \mu^2 \rme^{-\frac{4\pi}{N\epsilon^2(\mu)}}   
\end{align}
with $\tau_n = \left[ \left( \frac{1}{\epsilon^2} - \frac{1}{g^2} \right) \frac{1}{N} + \im \frac{\theta + 2\pi n}{2\pi N} - \frac{1}{2\pi} \log \frac{g^2N}{\epsilon^2N}\right] $.
Since the vacuum energy density is $ L_{\rm eff}(M_*, F_*)  = \frac{1}{4\pi} (M_* - \mu^2)$, the total partition function would take the following form:
\begin{align}
    Z(\theta) ~``\simeq"~  \sum_{n \in \mathbb{Z}} \rme^{- N V_{\mathrm{2d}} L_{\rm eff}(M_*^{(n)}, F_*^{(n)}) }  = \rme^{\frac{1}{4\pi} N V_{\mathrm{2d}} \mu^2}\sum_{n \in \mathbb{Z}} \rme^{- \frac{1}{4\pi} N V_{\mathrm{2d}} M_*^{(n)}  },
\end{align}
if we accept a naive saddle-point prescription where Stokes multipliers of all branches are set to one.

When $\theta \sim O(1)$, we usually assume that the $n = 0$ saddle is dominant.
If this formula is taken literally, a wrong saddle at large $n$ with lower action than the correct one would contribute.
Even worse, the above formula is badly diverging.
As first pointed out in Ref.~\cite{Sugeno:2025exv}, this problem indeed exists in the standard $\mathbb{C}P^{N-1}$ model.

Let us inspect the value of $L_{\rm eff}(M_*^{(n)}, F_*^{(n)})$ in  two distinct limits:

\vspace{0.5em}
\noindent \textbf{1. The Self-Dual Limit ($\epsilon \rightarrow 0$):} 
In the strict self-dual limit, for finite $\bar \theta_n$,  $\tau_n \simeq 1/(N\epsilon^2)$.    The  $1/\epsilon^4$ divergence of  the prefactor  cancels exactly with the  log-factor that arises from anomaly in the  $\rme^{4 \pi \tau_n}$ factor.   Furthermore, $\rme^{-\frac{4\pi}{N\epsilon^2(\mu)}}$  combines with the  other terms in   $\rme^{4 \pi \tau_n}$ to generate  RG invariant strong scale of  the  coupling $g$ in the  self-dual limit, $\Lambda_g^2$.   Thus, we obtain a finite saddle value of the energy density in the self-dual limit. 
\begin{align}
L_{eff}^{(SD)} 
&\simeq - e^{1-\gamma} \left[ \frac{\mu^2}{N^2 g^4} \exp\left(-\frac{4\pi}{Ng^2}\right) \right] e^{2i\bar{\theta}_n} =  - e^{1-\gamma} \Lambda_g^2 e^{2i\bar{\theta}_n}
\end{align}
Note that if the correct anomaly factor were  not included,  one would obtain a divergent field strength condensate and vacuum energy density. In the subsequent section, we will determine the ranges of 
$\bar \theta_n$ where the Stokes multiplier of this saddle is zero and non-zero. 

\vspace{0.5em}
\noindent \textbf{2. The Physical Limit ($\epsilon = g$):} 
When the theory is continued back to the standard, symmetric $\mathbb{C}P^{N-1}$ model, the real part of $\tau_n$ identically vanishes. The topological  coupling becomes  purely imaginary, dictated entirely by the topological angle: $\tau_n = i \bar{\theta}_n / (2\pi)$. Substituting this yields:
\begin{align}
L_{eff}^{(Phys)} &\simeq - \left( i \frac{\bar{\theta}_n}{2\pi} \right)^2 e^{1-\gamma} \mu^2 \exp\left(-\frac{4\pi}{Ng^2}\right) e^{2i\bar{\theta}_n} \nonumber \\
&\simeq \frac{1}{4\pi^2} \bar{\theta}_n^2 e^{1-\gamma} \mu^2 \exp\left(-\frac{4\pi}{Ng^2}\right) e^{2i\bar{\theta}_n}  \approx e^{1-\gamma} \Lambda_g^2  \bar{\theta}_n^2 \exp(2i\bar{\theta}_n)
\label{disaster}
\end{align}
which indeed agrees precisely with Ref.~\cite{Sugeno:2025exv}. As pointed out there,  this is rather problematic.  Consider we have $|\theta | <\pi$, where the vacuum is certainly $n=0$. However,   \eqref{disaster} implies that when $n \sim N$, there are infinitely many values of $n$ such that the  energy density of the branch $n$, given approximately by  the real part of the above expression, 
$\propto \operatorname{Re}[ \Lambda_g^2 \bar{\theta}_n^2 \exp(2i\bar{\theta}_n) ] =\Lambda_g^2 \bar{\theta}_n^2 \cos 2 \bar \theta_n $ which becomes much lower than the stable vacuum state that is obtained via the weak field analysis \cite{Witten:1978bc, DAdda:1978vbw, DAdda:1978dle}. 

If the Stokes multiplier of these  saddles  remain  non-zero  all values of $\bar{\theta}_n$, this would yield unphysical and absurd vacuum branch structure for the physical $\mathbb CP^{N-1}$ model, and   this would be a disaster for the large-$N$ saddle point analysis.   It is one of  our goal to solve this problem in the next sections. 

Ref.~\cite{Sugeno:2025exv} discussed several possible resolutions to this paradox, for example, the breakdown of the saddle point approximation, contributions from nonzero momentum Fourier modes, Stokes phenomena with complicated analytic continuations, and delicate destructive cancellations over $n$ sectors. Despite these discussions, they pointed out that none of these possibilities offer an easy resolution.
However, we will show that the true resolution to this catastrophic growth requires a reassessment of the active/passive integration cycles, specifically through a lesser-known mechanism called ``the boundary Stokes phenomenon''. This mechanism mathematically forces the Stokes multiplier of the saddles which become unphysical to drop strictly to zero long before they become a disaster for the vacuum branch structure of the theory. 

\section{Stokes phenomenon with boundary: solution of the unphysical saddle problem}

\label{sec:Stokesph}

Although we applied the standard saddle-point method to the interpolating theory at large $N$, there is a problem of unphysical saddles.
This problem is rather severe. It would imply that either the large-$N$ weak field analysis \cite{DAdda:1978vbw, DAdda:1978dle, Witten:1978bc} is incorrect, or there is something wrong with the saddle point analysis. Perhaps,  the Stokes multipliers of these saddles vanish due to competition between different saddles, as is often the case in Lefschetz thimble analysis.  However, it turns out to have a more subtle solution.

In this section, we resolve this problem by the Stokes phenomenon with boundary.
Due to the singularity at $F=0$, this point will be regarded as a ``boundary'' of the contour.
We can see the Stokes phenomenon from (thimble from endpoint) $+$ (thimble from saddle) to (thimble from endpoint): 
the unphysical saddle becomes just irrelevant/inactive, but not through a competition with other saddles. Rather, the integral becomes saturated by the endpoint thimble, resolving the paradox pointed out in Ref.~\cite{Sugeno:2025exv}.




We first revisit the integral over $(F,M)$ in Section \ref{sec:revisit_integral} as a preparation.
An important point is that the effective action $S_{\mathrm{eff}}[F,M]$ has different analytic continuations from $F>0$ and $F<0$ due to the singularity at $F=0$. Thus, we cannot apply the usual thimble decomposition for a contour including $F=0$. 
Instead, we should apply the thimble analysis with the boundary.
For illustrations of Stokes and Anti-Stokes transition with an endpoint thimble, we consider a simple integral, which can be obtained at asymptotically large $\operatorname{Re} \tau$, in Sections \ref{sec:bdyStokes_toy} and \ref{sec:antiStokes_toy}.

After the analysis on the asymptotic model, we proceed to the numerical analysis on the actual integral in Section \ref{sec:bdyStokes_numerical}, and see that the boundary Stokes phenomenon indeed occurs.
A careful estimation of the endpoint integral is given in Section \ref{sec:endpoint_vs_negativeF}.
The vacuum energy as a function of $\theta/N$ is examined in Section \ref{sec:vacuum_theta}.

For small $\tau$, we can determine the Stokes line analytically by the weak-field analysis (Section \ref{sec:weak_field_maintext}).
In particular, we show that the saddle point is irrelevant for all $\operatorname{Im}\tau \neq 0$ in the physical case $\operatorname{Re}\tau = 0$.
This result only implies that the saddle becomes irrelevant for $\theta \sim O(N)$, and does not invalidate the standard large-$N$ calculation of the topological susceptibility (see Appendix \ref{sec:scaling_subtleties} for details).

\subsection{Reviewing the saddle-point calculation}
\label{sec:revisit_integral}

As a preparation for a careful analysis, we review the saddle point analysis.
To make the thimble analysis easier, we proceed with the iterated integral.

We begin with the following expression:
\begin{align}
    Z(\theta) &=  \sum_{n \in \mathbb{Z}}\tilde{Z}(\theta+2\pi n)  = \sum_{n \in \mathbb{Z}}\int dF dM ~\rme^{-S_{\mathrm{eff}}^{\theta + 2 \pi n}[F,M]},
\end{align}
with the uniform ansatz.

We here remember that the effective action $S_{\mathrm{eff}}[F,M]$ consists of two holomorphic functions of $\mathrm{Re}F > 0$ sector and $\mathrm{Re}F < 0$ sector, separately.
Thus, we should decompose in the following way:

\begin{align}
    \tilde{Z}(\theta+2\pi n) 
    &= \tilde{Z}_+(\theta+2\pi n) + \tilde{Z}_-(\theta+2\pi n) \notag \\
    \tilde{Z}_+(\theta+2\pi n) &= \int_{F>0} dF dM ~\rme^{-S_{\mathrm{eff}}^{\theta + 2 \pi n}[F,M]} \notag \\
    \tilde{Z}_-(\theta+2\pi n) &= \int_{F<0} dF dM ~\rme^{-S_{\mathrm{eff}}^{\theta + 2 \pi n}[F,M]} 
\end{align}

With the deformation towards the self-dual theory $(\operatorname{Re} \tau_n>0)$, we can focus on the positive-$F$ sector\footnote{
The negative-$F$ sector $\tilde{Z}_-(\theta)$ may affect the contribution near $F=0$, which we will call the thimble associated with the endpoint.
This point is discussed in Section \ref{sec:endpoint_vs_negativeF}.
Either way, we can see that the endpoint contribution does not vanish.
}, $\tilde{Z}_+(\theta) $. We proceed with the iterated integral:
\begin{align}
    \tilde{Z}_+(\theta+2\pi n) &= \int_{F>0} dF \int_{\im \mathbb{R}} dM ~\rme^{-S_{\mathrm{eff}}^{\theta + 2 \pi n}[F,M]} 
\end{align}
First, we can integrate out $M$ by assuming $F>0$, and then focus on the thimble analysis on complex $F$.
The subtle point is that the original contour of $F$ is the open line $F \in (0,\infty)$ (and not $F \in (-\infty,\infty)$ as one may naively extrapolate due to non-analyticity) , which makes the thimble analysis more nontrivial.

It is straightforward to integrate out $M$ when $F>0$ is assumed.
\begin{align}
    \int_{\im \mathbb{R}} dM ~\rme^{-S_{\mathrm{eff}}^{\theta + 2 \pi n}[F,M]} 
\end{align}
The result will be analytically continued after that.
We remember that the effective action $S_{\mathrm{eff}}[F,M]$ is given by,
\begin{align}
   \frac{1}{NV_{\mathrm{2d}}}S_{\mathrm{eff}}[F,M] = -\frac{1}{4\pi} \left[ - M\log\left(\frac{\Lambda_\epsilon^2}{2\rme^\gamma F}\right)  - F\log(2\pi) + 2F\log\Gamma\left(\frac{F+M}{2F}\right) \right] - F \tau_n,
\end{align}
where
\begin{align}
         \tau_n &= \frac{1}{N} \left(\frac{1}{\epsilon^2} - \frac{1}{g^2} \right) + \im \frac{\theta + 2 \pi n}{2\pi N} - \frac{1}{2\pi} \log \left( \frac{g^2N}{\epsilon^2 N} \right),\\
         \Lambda_\epsilon^2 &= \mu^2 \rme^{-\frac{4\pi}{N \epsilon^2}} .
\end{align}
The saddle-point equation of $M$ is given by,
\begin{align}
    \psi\left(\frac{1}{2}+\frac{M}{2F}\right) =\log\left(\frac{\Lambda_\epsilon^2}{2\rme^\gamma F}\right).
\end{align}
There are infinitely many real solutions of this equation:
one solution is located at $y:=\frac{1}{2}+\frac{M}{2F} = y_0 >0$, and the other solutions $y = y_k$ are labeled by $k \in \mathbb{Z}_{>0}$, which are between $-k < y_k  < -k+1$.

As a function of $M$, the action has singularities at $M = -F, -3F, -5F, \cdots$. 
In terms of $y$, the singularities are located at non-positive integers $y=0,-1,-2,\cdots$.
The singularities can be understood as the appearance of the zeromodes of the fluctuation operator: $\operatorname{Spec}(-D_\mu^2 + M) = \{ F (2k+1) + M  ~|~k\in \mathbb{Z}_{\geq 0}\}$, and the fluctuation determinant vanishes at these points $M = -F, -3F, -5F, \cdots$.

The dual thimbles of the saddles $\{ y_k\}_{k \in \mathbb{Z}_{\geq 0}}$ lie on the real axis, ending at either a singularity or $y=+\infty$.
The original contour intersects only the dual thimble of the positive saddle $y:= y_0 >0$.
On the complex $M$ plane, one can explicitly observe that the original contour $\im \mathbb{R}$ can be continuously deformed into the thimble of this saddle.

Thus, only the first saddle $y=\frac{1}{2}+\frac{M}{2F} >0$ is relevant.
Let $y(F)$ denote this solution, i.e., the principal branch of $\psi^{-1}(\log\left(\frac{\Lambda_\epsilon^2}{2\rme^\gamma F}\right))$.
Note that for complex $F$, $y(F)$ should be understood as the analytic continuation of the branch defined on $F>0$.
After the saddle-point approximation of $M$ integral, we have
\begin{align}
    \int_{\im \mathbb{R}} dM ~\rme^{-S_{\mathrm{eff}}[F,M]}& \xrightarrow{\text{saddle point}} \rme^{-S_{\mathrm{eff}}[F]}, \notag \\
    \frac{1}{NV_{\mathrm{2d}}}   S_{\mathrm{eff}}[F] &= \frac{F}{2\pi} \left[  \left( y(F) - \frac{1}{2}\right) \log\left(\frac{\Lambda_\epsilon^2}{2\rme^\gamma F}\right) - \log\Gamma\left(y(F)\right) \right] + \frac{F}{4\pi}\log(2\pi)  - F \tau_n. \label{eq:eff_action_F}
\end{align}

Then, our problem is reduced to the $F$ integral:
\begin{align}
    \tilde{Z}_+(\theta+2\pi n)  &= \int_{F>0} dF ~\rme^{-S_{\mathrm{eff}}[F]}
\end{align}

One can check that, as reported in \cite{Sugeno:2025exv} and explained briefly around \eqref{disaster}, the saddle point of $F=F_*$, which is continued to the standard saddle, shows oscillating exponential growth as $\bar \theta$ increases.
We thus face the problem pointed out above.

Below, we explore the thimble structure of this $F$ integral and see how this problem is resolved.
As a problem of this $F$ integral, we can take $n=0$ without loss of generality, and simply write $\tau_ n = \tau$.

\subsection{Stokes phenomenon with boundary: the asymptotic model (Large \texorpdfstring{$\operatorname{Re}\tau$}{Re(tau)}) }
\label{sec:bdyStokes_toy}

Below, we demonstrate that the saddle point becomes irrelevant at a certain $\bar{\theta} = \theta /N$.
This occurs due to the Stokes phenomenon with boundary.
In mathematics, the Stokes phenomenon with boundary is developed and used for hyperasymptotics \cite{Berry:1991but, howls1992hyperasymptotics, howls1997hyperasymptotics, Delabaere:2002}.

As a demonstration of this phenomenon, we first consider a simplified situation: an asymptotic model obtained by a formal limit $\operatorname{Re}\tau \rightarrow+\infty$ and $|F| \rightarrow +\infty$.
If we only keep the leading order of the effective action (\ref{eq:eff_action_F}) (with focusing on the scale $F \sim O(\rme^{4\pi \tau})$), the $F$-integral becomes very simple:
\begin{align}
    \int_{F>0} dF ~\rme^{-S_{\mathrm{asymp}}[F]} :=\int_{F>0} dF ~\rme^{-NV_{\mathrm{2d}}  F \left[ \frac{1}{4\pi} \log(F/\Lambda_\epsilon^2) - \tau \right]}. \label{eq:toy_model_integral}
\end{align}
To obtain this asymptotic action, we ignore the subleading $F \log(\log(F))$ term while keeping the scaling $F \sim O(\rme^{4\pi \tau})$.
Therefore, it serves as an instructive example for understanding the asymptotic structure of the original integral.

\begin{figure}[t]
\hspace{1cm}
    \centering
    \includegraphics[width=1 \linewidth]{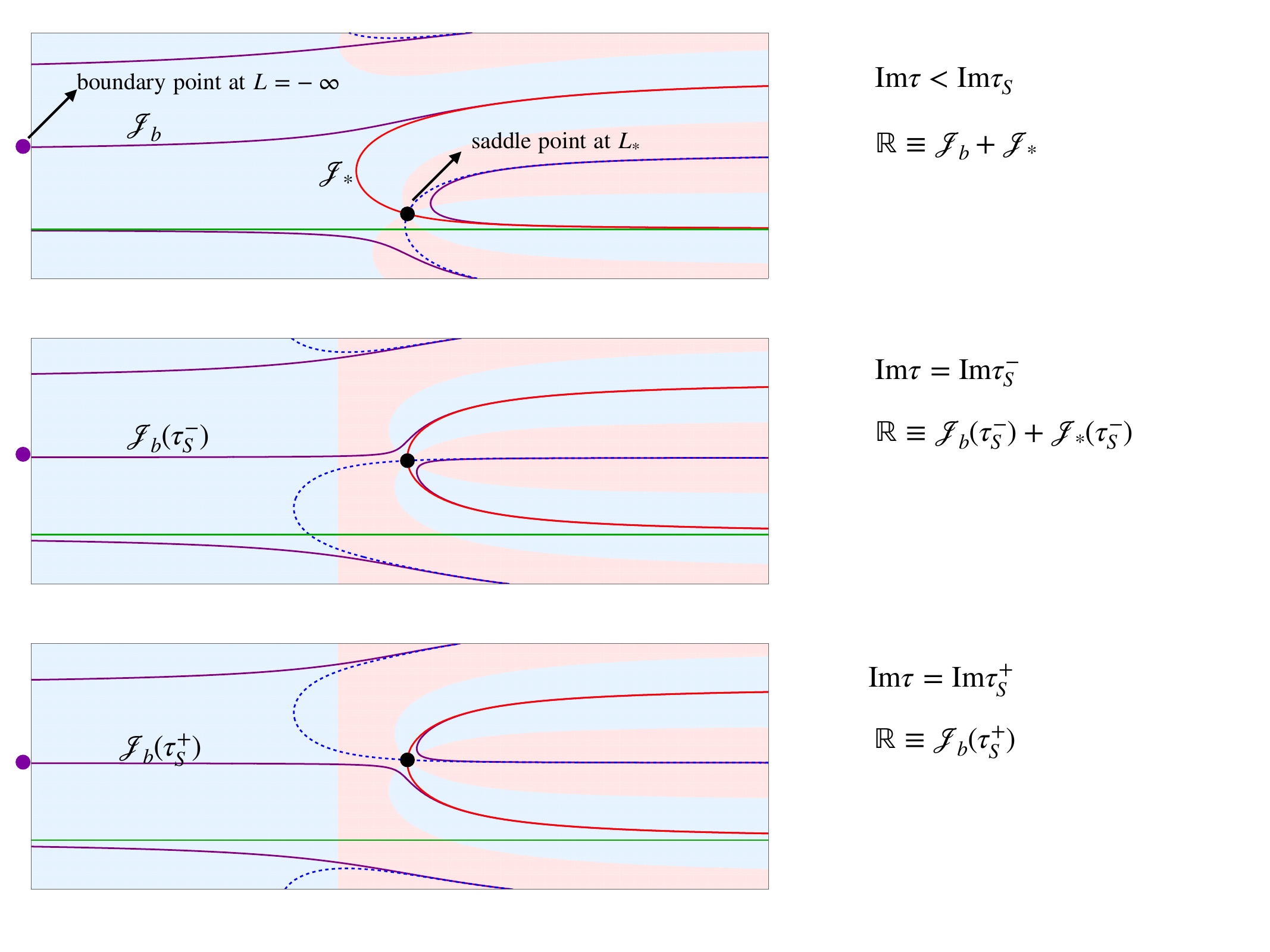}
    \vspace{-1cm}
    \caption{{\bf Boundary Stokes phenomenon:} Thimble structure of the asymptotic model integral (\ref{eq:asym_model_integral}) on the complex $L= \log (F/\Lambda_\epsilon^2)$ plane.  
We show the saddle point (black point), end point (star, $F=0, L=-\infty$), saddle thimble (red contour), dual thimble (blue dashed), boundary thimble (purple), and original contour $L: -\infty \to \infty$ (green). 
  For the right side $\operatorname{Re}L \rightarrow + \infty$, the integral is converging ($S \rightarrow + \infty$) in the blue shaded regions, and diverging ($S \rightarrow - \infty$) in the red shaded regions.}
  \label{fig:Boundarystokes}
\end{figure}

\begin{figure}[htbp]
  \centering
  \begin{minipage}[b]{0.32\textwidth}
    \centering
    \includegraphics[width=\linewidth]{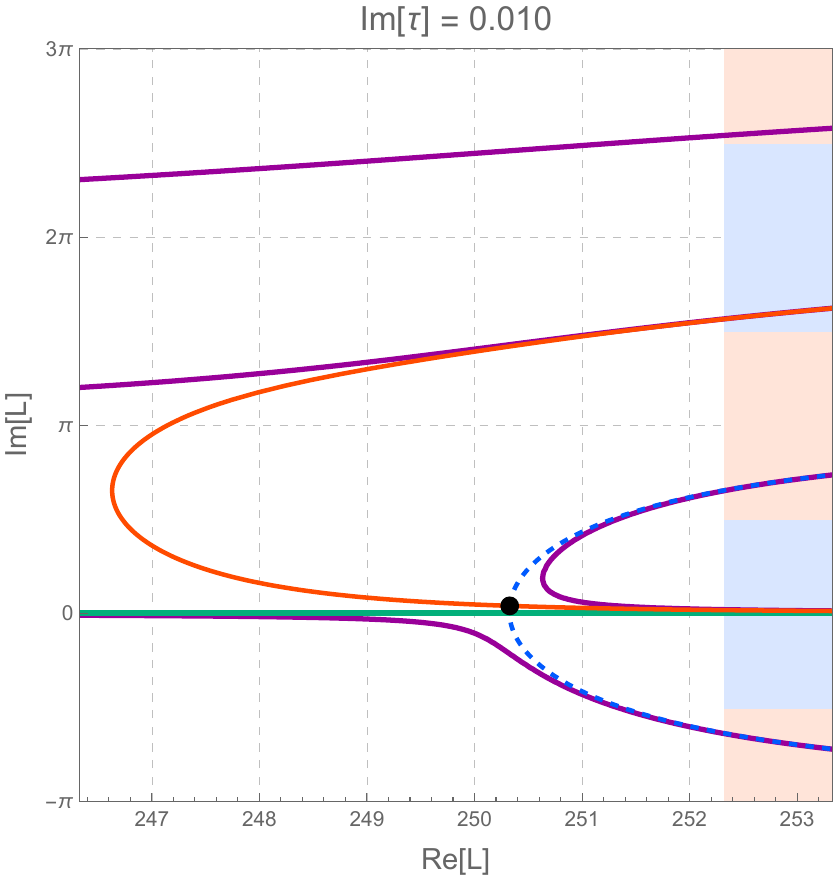}
  \end{minipage}
  \hfill
  \begin{minipage}[b]{0.32\textwidth}
    \centering
    \includegraphics[width=\linewidth]{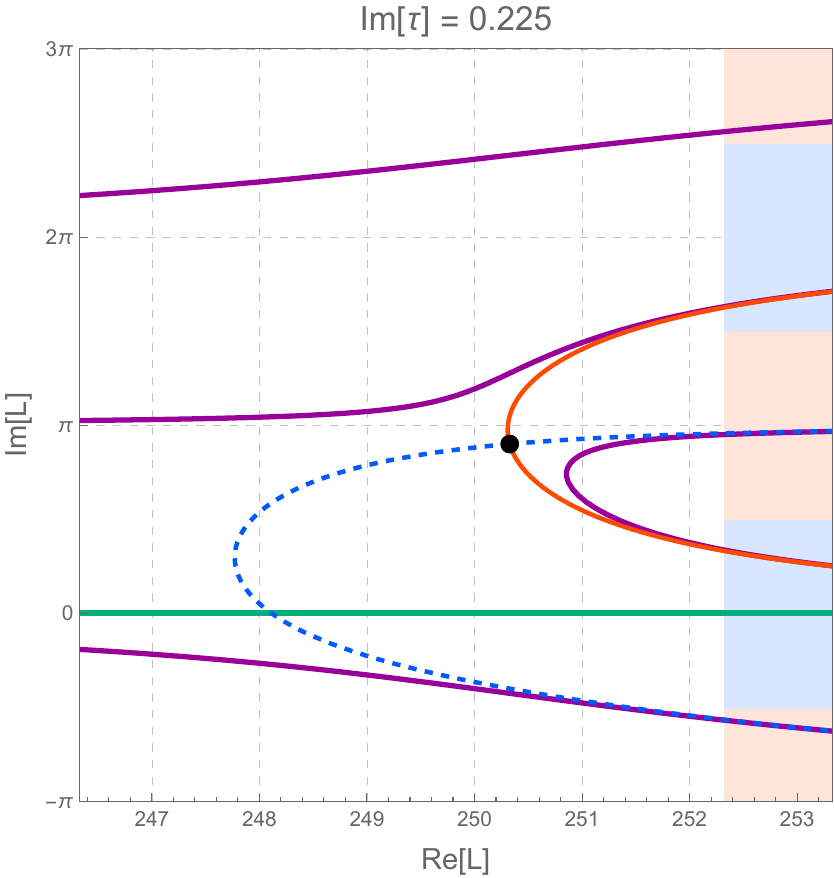}
  \end{minipage}
  \hfill
  \begin{minipage}[b]{0.32\textwidth}
    \centering
    \includegraphics[width=\linewidth]{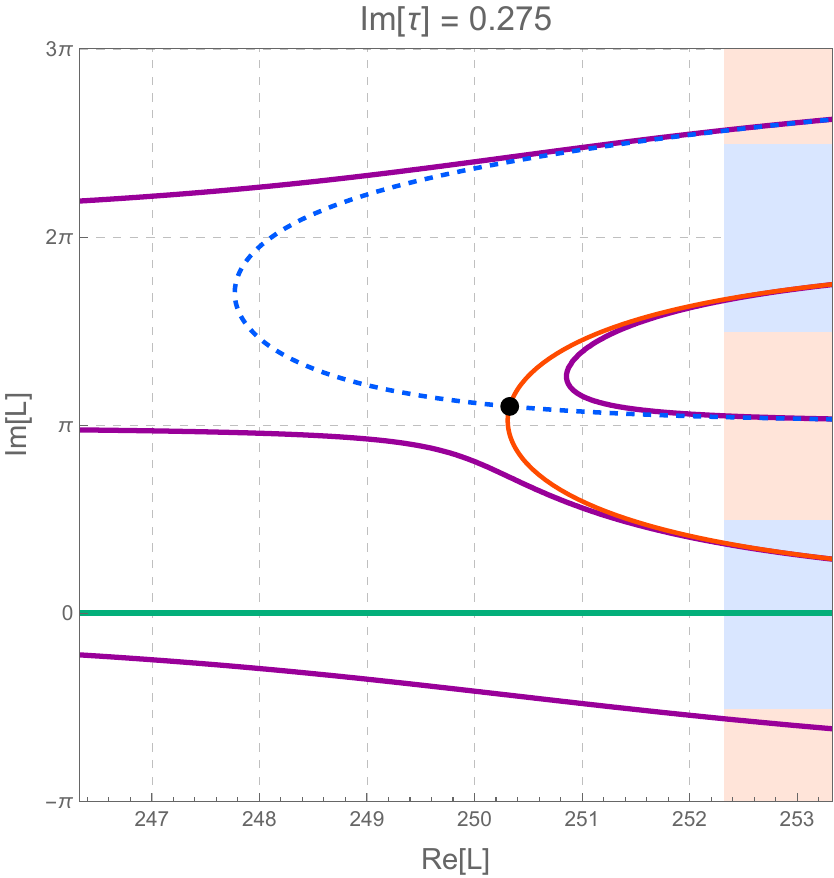}
  \end{minipage}
  \caption{Thimble structure of the toy model integral (\ref{eq:toy_model_integral}) on the complex $L= \log (F/\Lambda_\epsilon^2)$ plane at $\operatorname{Re}\tau =20$.
We show the saddle point (black point), thimble (red contour), dual thimble (blue dashed), $\operatorname{Im}S_{\mathrm{asymp}} = 0$ (purple), and original contour $L: -\infty \to \infty$ (green). 
  For the right side $\operatorname{Re}L \rightarrow + \infty$, the integral is converging ($S \rightarrow + \infty$) in the blue shaded regions, and diverging ($S \rightarrow - \infty$) in the red shaded regions. }
  \label{fig:toy_model_thimble}
\end{figure}

\noindent
{\bf $L$-plane analysis:} 
It is easy to visualize the thimble structure on the complex $L= \log (F/\Lambda_\epsilon^2)$ plane.
By regarding $L \in \mathbb{C}$, the whole log Riemann surface is well parametrized. 
\begin{align}
Z_{\mathrm{asymp}} = \Lambda_\epsilon^2 \int_{-\infty}^{\infty} dL \exp(L) \exp\left( -NV_{\mathrm{2d}}\Lambda_\epsilon^2 e^L \left[ \frac{L}{4\pi} - \tau \right] \right)
\label{eq:asym_model_integral}
\end{align}
where $ NV_{\mathrm{2d}}\Lambda_\epsilon^2$ is the large parameter in front of action\footnote{As a saddle-point analysis, the Jacobian factor does not contribute in the leading order in $1/NV_{\mathrm{2d}} $.}. 
The boundaries of integration transform accordingly: the origin $F \to 0$ is pushed to $L \to -\infty$, and $F \to \infty$ maps to $L \to \infty$.
The original integration contour is the real axis of $L$. The saddle point is located at   (ignoring prefactors in \eqref{saddlelarge} momentarily)
\begin{align}
F_* = \Lambda_\epsilon^2 \rme^{4\pi\tau-1}, \; \; {\rm equivalently,} \; \;      L_* = 4\pi\tau - 1
\end{align}
Figure~\ref{fig:Boundarystokes} depicts the qualitative features of the thimble structure of this integral.
We also present the plots of the thimble structure at $\operatorname{Re}\tau =20$ in Figure~\ref{fig:toy_model_thimble}.

The important fact is that, in addition to the thimble associated with the saddle point $\mathcal{J}_*$, a curve with $\operatorname{Im}S = \operatorname{Im}S[F=0]=0$ is also an important contour.
This contour $\operatorname{Im}S =0$ has the following properties:
\begin{enumerate}
    \item The integration over this contour does not have a sign problem: the phase $\operatorname{Im}S$ is constant.
    \item The real part $\operatorname{Re}S$ monotonically increases or decreases on such a constant-$\operatorname{Im}S$ curve (unless it hits a saddle point).
    \item The $\operatorname{Im}S =0$ contour passes through the endpoint of the original contour $L\rightarrow -\infty$ ($F=0$).
\end{enumerate}

From the properties 1. and 2., the integration over the constant-$\operatorname{Im}S$ contour can be well estimated in the limit $NV_{\mathrm{2d}} \rightarrow +\infty$.
An integration on such an open contour will be dominated by the endpoint contribution from these two properties.
Moreover, due to property 3., this contour is very useful for a deformation of the original open contour with the endpoint $L\rightarrow -\infty$ ($F=0$). This boundary thimble plays a role on the same footing with the standard thimbles in the problem.

Based on this strategy, we can decompose the original contour of the integral (\ref{eq:toy_model_integral}) along $\mathbb R^{+}$  in terms of saddle and boundary thimbles: 
\begin{align}
    \mathbb R^{+} = \sum_{\sigma} n_{\sigma} (\tau) {\cal J}_{\sigma}(\tau)
\end{align}
where $n_{\sigma}$ are integers  and possible thimbles are:
\begin{align}
{\cal J}(\tau) = \left\{   {\cal J}_{b}(\tau):{\rm boundary  \; thimble },   
     {\cal J}_{*}(\tau): {\rm saddle  \; thimble }  \right\}
\end{align}
as shown in  Figure \ref{fig:toy_model_thimble}. Let us now describe thimble decomposition in two complementary Stokes wedges, which will allow us to see the crucial Stokes phenomenon dictating branch structure 
of the large-$N$ sigma model. 
\begin{itemize}
    \item $-1/4 < \operatorname{Im}\tau < 1/4$: the saddle point is relevant.

    See left two panels of Figure \ref{fig:toy_model_thimble}. See also top two panels of Figure \ref{fig:Boundarystokes}.

    The black dot is the saddle point, the red curve is the thimble $\mathcal{J}_*$, and the blue dashed curve is the dual thimble.
    The original contour is the green line, which intersects with the dual thimble when the saddle point is relevant.

    On the right side, $\operatorname{Re}L \rightarrow + \infty$, there are diverging (red-shaded) and converging (blue-shaded) regions.
    Note that we cannot deform a contour through the diverging direction (red-shaded region) at the infinity $\operatorname{Re}L \rightarrow + \infty$.
    The original contour is from $L = -\infty$ to $L = + \infty~(-\pi/2 < \operatorname{Im}L < \pi/2)$. We can deform this contour into the linear combination of thimbles 
    \begin{align}
    \mathbb R \equiv      \mathcal{J}_b (\tau)+\mathcal{J}_* (\tau), \qquad 
    | \operatorname{Im}\tau | \leq \frac{1}{4}
    \label{Stokeswedge}
    \end{align}
    \begin{itemize}
        \item the thimble $\mathcal{J}_*$ (red curve): from $\operatorname{Re}L  = + \infty~(3\pi/2 < \operatorname{Im}L < 5\pi/2)$ to $\operatorname{Re}L  = + \infty~(-\pi/2 < \operatorname{Im}L < \pi/2)$
        \item the $\operatorname{Im}S_{\mathrm{asymp}}=0$ curve $\mathcal{J}_b$ (purple curve):  from $L = -\infty$ to $\operatorname{Re}L  = + \infty~(3\pi/2 < \operatorname{Im}L < 5\pi/2)$ 
    \end{itemize}

    \item $1/4 < \operatorname{Im}\tau $: the saddle point is irrelevant/inactive. 

    At $\operatorname{Im}\tau = 1/4$, the Stokes phenomenon occurs.
    See the transition from the center panel to the right panel of Figure \ref{fig:toy_model_thimble} (also center and bottom panels of Figure \ref{fig:Boundarystokes}).
We can see that, while the dual thimble intersects with the original contour at $\operatorname{Im}\tau = 0.225$, it does not intersect with the original contour at $\operatorname{Im}\tau = 0.275$. 

    At the point $\operatorname{Im}\tau = 1/4$, the imaginary part of the saddle point coincides with the imaginary part of the endpoint: 
    \begin{align} 
    {\rm Stokes:} \qquad \operatorname{Im}S_{\mathrm{asymp}}[F_*] = \operatorname{Im}S_{\mathrm{asymp}}[F=0] = 0
    \end{align}
    Above this point, the original contour no longer intersects with the dual thimble (See the right panel of Figure \ref{fig:toy_model_thimble}).

    Through this transition, one can observe the reconnection of the $\operatorname{Im}S_{\mathrm{asymp}} =0$ curve (purple) and the thimble (red).
    Due to this reconnection, a $\operatorname{Im}S_{\mathrm{asymp}} =0$ curve belongs to the same sector as the original contour (from $L = -\infty$ to $L = + \infty~(-\pi/2 < \operatorname{Im}L < \pi/2)$).

    Then, we can deform the original contour into just one $\operatorname{Im}S_{\mathrm{asymp}} =0$ curve $\mathcal{J}_b$: from $L = -\infty$ to $L = + \infty~(-\pi/2 < \operatorname{Im}L < \pi/2)$.

\end{itemize}
This is the Stokes phenomenon including the boundary thimble and saddle thimble, which renders the would-be dangerous saddles completely inactive, i.e, saddles disappear from the integration cycle completely. 
As a matter of fact, before one hits the Stokes line ${\rm Im} 
\tau_S= 1/4$, there is already an exchange of dominance at anti-Stokes line at ${\rm Im} \tau_{AS}= 1/8 $, as we discuss below. But first, let us recap the $F$-plane realization of the boundary Stokes phenomenon. 

\noindent
{\bf $F$-plane analysis:}
In the complex $F$-plane, this observation can be understood as follows. The original contour runs from $F=0$ to $F=(+\infty)_{1}$, where $(+\infty)_{1}$ denotes positive infinity on the first Riemann sheet. Accordingly, the contours $\mathcal{J}_b$ and $\mathcal{J}_*$ are mapped to:
\begin{itemize}
    \item When the saddle point is relevant ($|\operatorname{Im}\tau|<1/4$), the $\mathcal{J}_b$ is the $\operatorname{Im}S_{\mathrm{asymp}} =0$ curve from $F =0$ to $F=(+\infty)_{2}$, where $(+\infty)_{2}$ denotes the positive infinity on the second Riemann sheet.
    The thimble $\mathcal{J}_*$ is a path from $(+\infty)_{2}$ to  $(+\infty)_{1}$.
    \item When the dual thimble passes through the endpoint $F=0$, the reconnection of the $\operatorname{Im}S_{\mathrm{asymp}} =0$ curve and the thimble occurs.
    \item After this reconnection ($|\operatorname{Im}\tau| > 1/4$),  the $\operatorname{Im}S_{\mathrm{asymp}} =0$ curve $\mathcal{J}_b$ now becomes a path from $F=0$ to $F=(+\infty)_{1}$.
\end{itemize}

In summary, we can make the following decomposition of (\ref{eq:toy_model_integral}):
\begin{align}
    \int_{F>0} dF ~\rme^{-S_{\mathrm{asymp}}[F]} = \begin{cases}
         \int_{\mathcal{J}_b+\mathcal{J}_*} dF ~\rme^{-S_{\mathrm{asymp}}[F]} ~~~&(|\operatorname{Im}\tau|<1/4) \\
         \int_{\mathcal{J}_b} dF ~\rme^{-S_{\mathrm{asymp}}[F]} ~~~&(1/4 < |\operatorname{Im}\tau| ) ,
    \end{cases}
\label{StokesT}
\end{align}
and we can estimate these integrals by the endpoint or saddle point in the limit $NV_{\mathrm{2d}} \rightarrow \infty$:
\begin{align}
    \int_{\mathcal{J}_b} dF ~\rme^{-S_{\mathrm{asymp}}[F]} \simeq \rme^{-S_{\mathrm{asymp}}[F=0]},~~\int_{\mathcal{J}_*} dF ~\rme^{-S_{\mathrm{asymp}}[F]} \simeq \rme^{-S_{\mathrm{asymp}}[F_*]}.
\end{align} 
The first estimation is valid because $S_{\mathrm{asymp}}[F]$ increases monotonically on the path $\mathcal{J}_b$.

This mechanism proposes a solution of the unphysical saddle problem.
When $\theta/N$ is above a certain transition point, the saddle completely disappears from the integration cycle; it still exists as a critical point, but its Stokes multiplier drops to zero ($n_* = 0$). This topological truncation is vital. In the adjacent Stokes chamber (where $n_* = 0$), there are values of $\tau$ where the saddle contribution would naively dominate again if $n_*$ were non-zero. This is precisely the potential inconsistency pointed out by Sugeno, Yokokura, and Yonekura \cite{Sugeno:2025exv} but now realized in self-dual theory.  However, because the Stokes phenomenon renders the saddle topologically inactive, this dangerously unphysical contribution is completely projected out.

\subsection{Anti-Stokes transition with boundary}  
\label{sec:antiStokes_toy}

In the Stokes wedge $|\operatorname{Im}\tau| \leq 1/4$, the standard saddle point remains topologically active, meaning the original integration contour over the positive real axis, $F \in (0, \infty)$, cannot be deformed into a single steepest descent path. Instead, the contour decomposes into a sum of the boundary and saddle thimbles: $\mathbb{R} \equiv \mathcal{J}_b(\tau) + \mathcal{J}_{*}(\tau)$. 

However, topological relevance does not imply physical dominance. Midway through the Stokes wedge lies an \textit{anti-Stokes line} at $\tau_{\mathrm{AS}}$, which triggers a crucial exchange of dominance between the saddle and boundary contributions.

The total amplitude in the positive topological sector, $\tilde{Z}_{+}$, is given by the sum of the integrals over these two distinct paths. In the semi-classical, large-$NV_{\mathrm{2d}}$ limit, each integral is dominated by the configuration minimizing the real part of the action. For $\mathcal{J}_{*}$, this is the saddle point $F_*$; for $\mathcal{J}_{b}$, it is  the endpoint $F \to +0$. Thus, the partition function is well-approximated by the competition between these two discrete  weights as shown in \eqref{StokesT}, $(|\operatorname{Im}\tau|<1/4)$. 

The physical vacuum energy density, defined via $E_{\mathrm{vac}} = -\frac{1}{NV_{\mathrm{2d}}} \log(\tilde{Z}_{+})$, fundamentally relies on which configuration carries the lower real action:

\begin{align}
&    E_{\mathrm{vac}}(\theta) \approx \lim_{N \rightarrow \infty} -\frac{1}{NV_{\mathrm{2d}}} \log\left( n_* e^{-\operatorname{Re}(S_*)} +  n_b e^{-\operatorname{Re}(S_0)} \right) \cr  &= \frac{1}{NV_{\mathrm{2d}}}
    \begin{cases} 
        \operatorname{Re}(S_*)  &(n_{*}, n_b)=(1,1)  \quad  |\operatorname{Im} \tau| < |\operatorname{Im} \tau_{\mathrm{AS}}| \quad (\text{saddle dom.}) \\ 
        \operatorname{Re}(S_*)= \operatorname{Re}(S_0)   & (n_{*}, n_b)=(1,1) \quad |\operatorname{Im} \tau| = |\operatorname{Im} \tau_{\mathrm{AS}}| \quad (\text{co-dom.}) \\ 
        \operatorname{Re}(S_0) & (n_{*}, n_b)=(1,1)  \quad  |\operatorname{Im} \tau_{\mathrm{AS}}| < |\operatorname{Im} \tau| < |\operatorname{Im} \tau_{\mathrm{S}}| \quad (\text{boundary dom.})  \\
         \operatorname{Re}(S_0) & (n_{*}, n_b)=(0,1) \quad  |\operatorname{Im} \tau| > |\operatorname{Im} \tau_{\mathrm{S}}| \quad (\text{boundary only})
    \end{cases}
\end{align}

The anti-Stokes line $\operatorname{Re}(S_*) = \operatorname{Re}(S_0)$, situated at $\theta/N = \pi/4$ (or $\operatorname{Im} \tau = 1/8$), marks the exact point where the saddle and boundary thimbles contribute equally. As we increase $\operatorname{Im}\tau$, the system transitions sequentially through a region of saddle dominance, co-dominance at the anti-Stokes line, and finally boundary dominance in the second half of the Stokes wedge.  
Eventually, at the Stokes line ($|\operatorname{Im}\tau_{\mathrm{S}}| = 1/4$), the boundary contribution $S_0$ is maximally dominant\footnote{This stems from the Stokes line being horizontal in the $\tau$-plane ($\operatorname{Im}\tau = 1/4$) in this asymptotic model. See Appendix E for details.} over the saddle contribution $S_*$. 

In the strict large-$NV_{\mathrm{2d}}$ limit, the logarithmic sum acts as a hard minimum operator. Consequently, the vacuum energy exhibits a non-analytic kink and flattens completely to the boundary value:
\begin{equation}
    E_{\mathrm{vac}}(\theta) \sim 
    \begin{cases} 
        -\Lambda_g^2 \cos(2\theta/N) & \text{for } |\theta/N| < \pi/4 \,, \\ 
        0 & \text{for } |\theta/N| > \pi/4 \,. 
    \end{cases}
    \label{eq:vacuum_energy_asymptotic}
\end{equation}

\subsection{Stokes phenomenon with boundary: the numerical analysis (Small \texorpdfstring{$\operatorname{Re}\tau$}{Re(tau)})}
\label{sec:bdyStokes_numerical}

As the above toy model is obtained in the formal limit $\operatorname{Re}\tau \rightarrow+\infty$ and $|F| \rightarrow +\infty$, we expect that the actual integral at finite $\operatorname{Re}\tau$:
\begin{align}
    \tilde{Z}_+ = \int_{F>0} dF ~\rme^{-S_{\mathrm{eff}}[F]},
\end{align}
where $S_{\mathrm{eff}}[F]$ is given by (\ref{eq:eff_action_F}), has qualitatively the same structure.
Here, we numerically check that the Stokes phenomenon with the endpoint indeed occurs.

\begin{figure}[htbp]
  \centering
  \begin{minipage}[b]{0.32\textwidth}
    \centering
    \includegraphics[width=\linewidth]{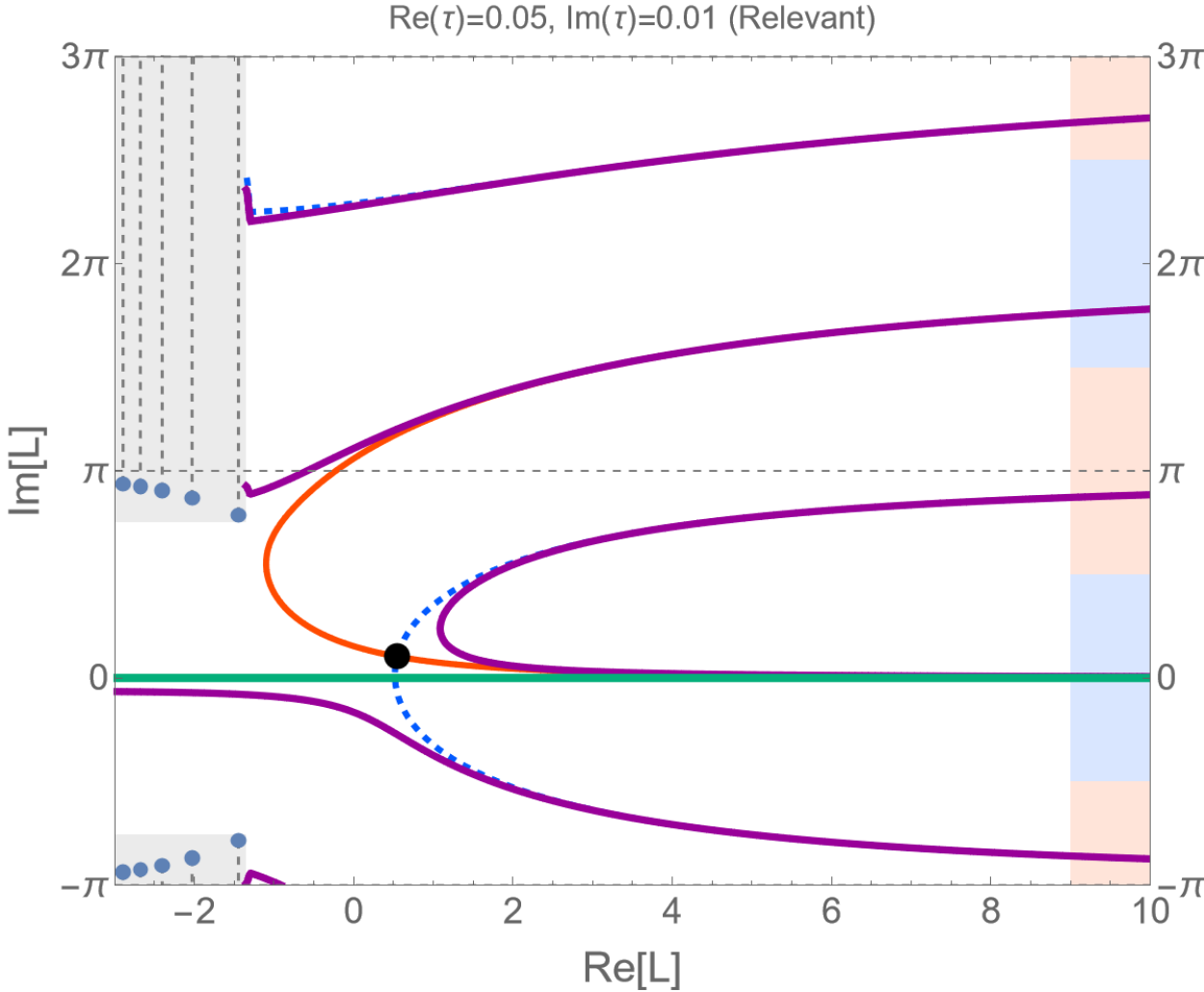}
  \end{minipage}
  \hfill
  \begin{minipage}[b]{0.32\textwidth}
    \centering
    \includegraphics[width=\linewidth]{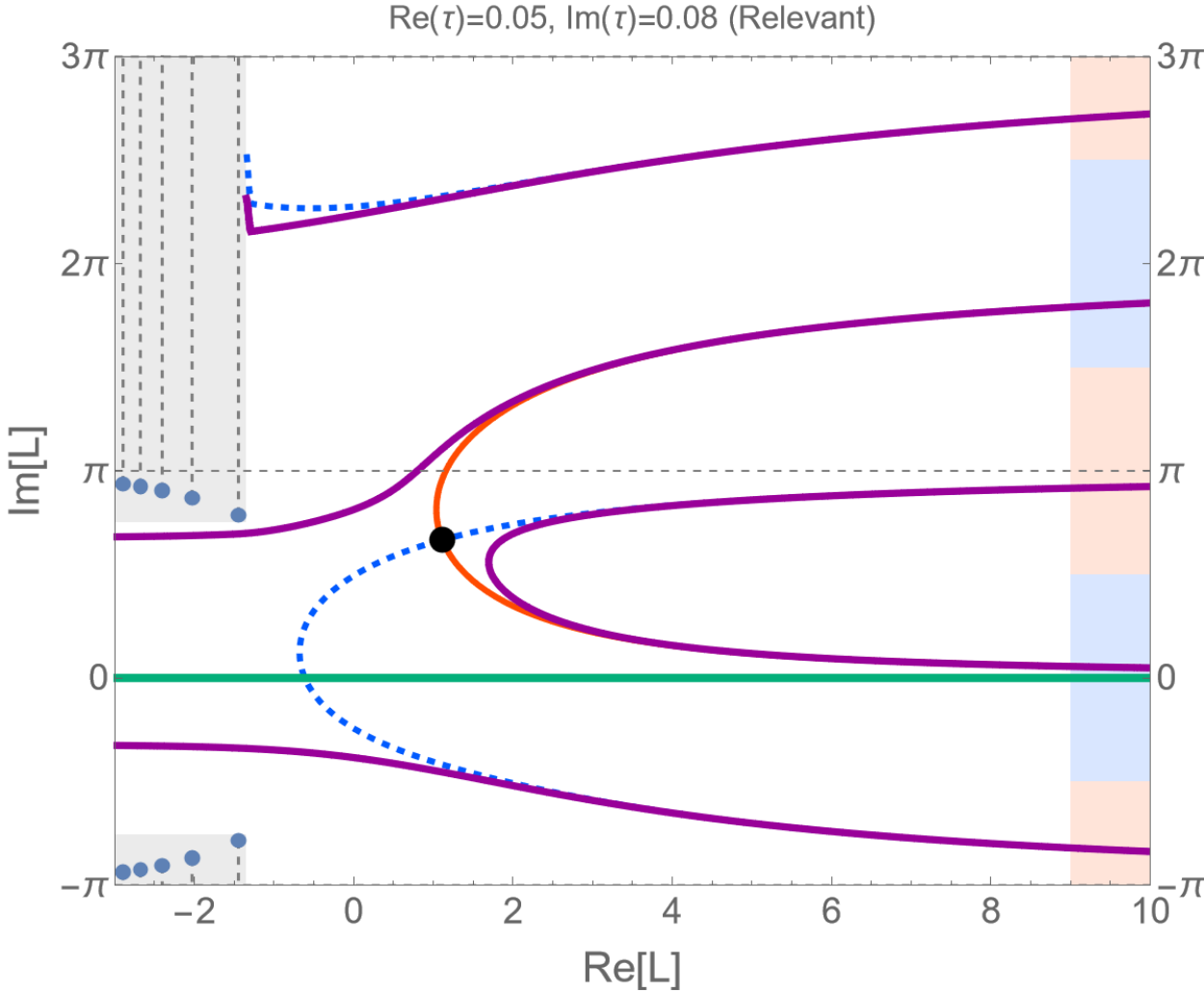}
  \end{minipage}
  \hfill
  \begin{minipage}[b]{0.32\textwidth}
    \centering
    \includegraphics[width=\linewidth]{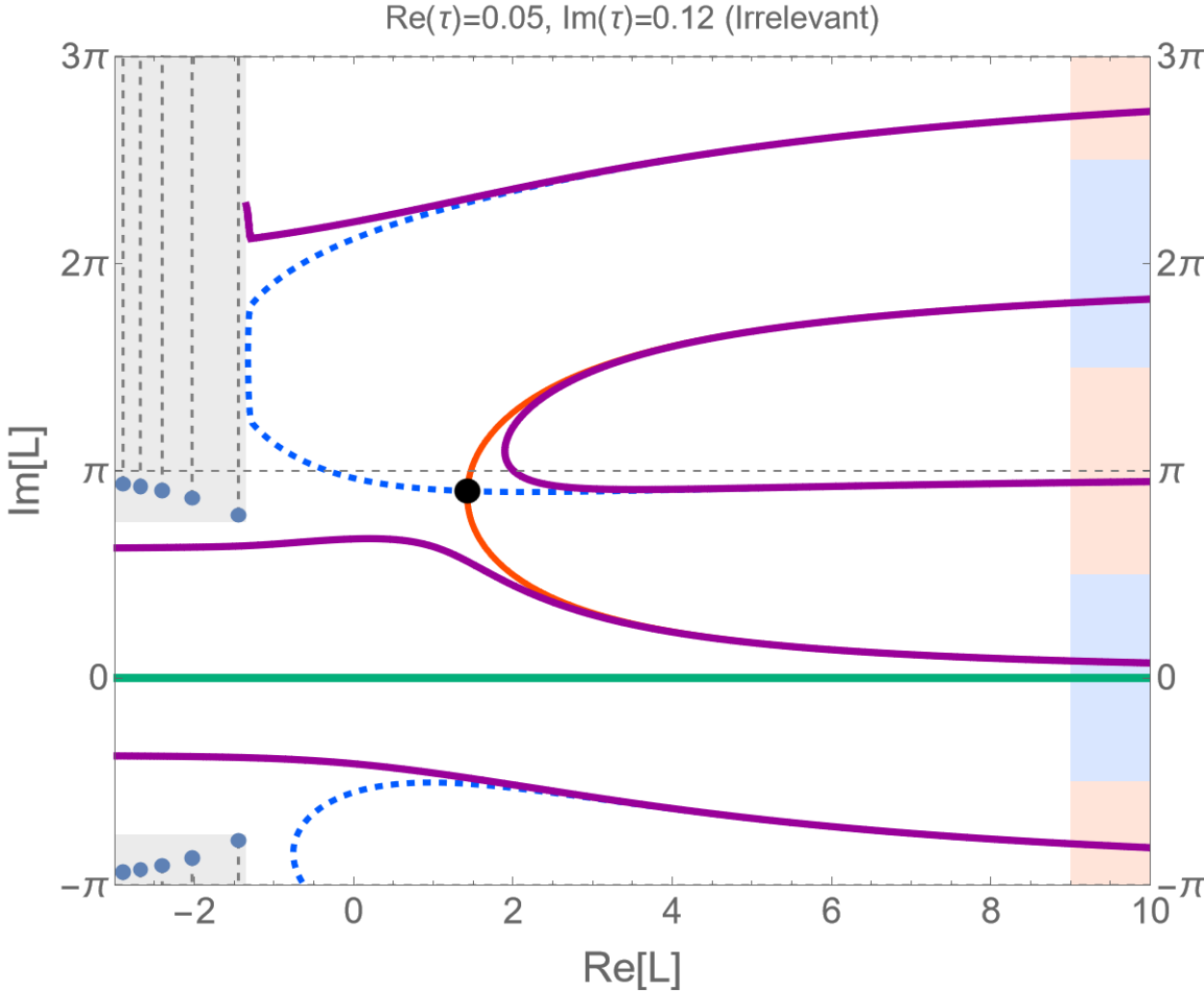}
  \end{minipage}
  \caption{Thimble structure of the original $F$ integral on the complex $L= \log (F/\Lambda_\epsilon^2)$ plane at $\operatorname{Re}\tau =0.05$.
   We plot for $\operatorname{Im}\tau =0.01$ (left panel), $\operatorname{Im}\tau =0.08$ (center panel), $\operatorname{Im}\tau =0.12$ (right panel).
We show the saddle point (black point), thimble (red contour), dual thimble (blue dashed), $\operatorname{Im}S_{\mathrm{eff}} = 0$ (purple), and original contour $L: -\infty \to \infty$ (green). 
  For the right side $\operatorname{Re}L \rightarrow + \infty$, the integral is converging ($S \rightarrow + \infty$) in the blue shaded regions, and diverging ($S \rightarrow - \infty$) in the red shaded regions.
As the action $S_{\mathrm{eff}}$ contains the inverse digamma function, there are infinitely many branch points shown as blue dots.
In these Figures, their branch cuts are parallel to the imaginary axis (gray dashed lines).
The contours on the gray-shaded regions are not shown because of these branch cuts.
Since the thimbles and dual thimbles are visualized using constant-$\operatorname{Im}S_{\mathrm{eff}}$ contours in this figure, any (blue dotted or red) curves not connected to the saddle point should be neglected.
  }
  \label{fig:numerical_thimble_tau005}
\end{figure}

For example, let us examine the case at $\operatorname{Re}\tau = 0.05$.
We show a few thimble structures in Figure \ref{fig:numerical_thimble_tau005}.
Most symbols and labels in this figure are the same as in the previous one.
Although a subtle point appears for small $\theta/N$, this plot indeed indicates that the boundary Stokes phenomenon happens in the same way.
Regarding the thimble structures, we have the following observation.

\begin{itemize}
    \item At small $\operatorname{Im} \tau$: $\tau = 0.05 + 0.01\im$ [left panel of Figure \ref{fig:numerical_thimble_tau005}].

    In this case, it is not clear how the original contour is decomposed into the thimbles, because the $\operatorname{Im} S_{\mathrm{eff}}=0$ curve enters another Riemann sheet. As shown in this figure, this curve encounters the branch cut of the inverse digamma function\footnote{While an analysis in the complex $y$-plane could in principle provide the full thimble decomposition, where $y=\psi^{-1}(L - \log(2\rme^\gamma))$, we will not go into those details here.}.
    
    Although a more intricate structure may appear in other Riemann sheets, it is fair to deduce that the dominant contribution arises from the saddle point.
 \begin{align}
    \int_{F>0} dF ~\rme^{-S_{\mathrm{eff}}[F]} = \int_{\mathcal{J}_b+\mathcal{J}_*+\cdots} dF ~\rme^{-S_{\mathrm{eff}}[F]} \simeq ~\rme^{-S_{\mathrm{eff}}[F_*]},
\end{align}   
where $\cdots$ denotes possible additional thimbles in higher Riemann sheets, and we have assumed that these contributions are subleading.

    \item At intermediate $\operatorname{Im} \tau$: $\tau = 0.05 + 0.08\im$ [center panel of Figure \ref{fig:numerical_thimble_tau005}].
    
At intermediate $\operatorname{Im} \tau$ (before the Stokes phenomenon), for example $\tau = 0.05 + 0.08\im$, we can see that the original contour is decomposed into the thimble associated with the saddle point and the  $\operatorname{Im} S_{\mathrm{eff}}=0$ curve:
 \begin{align}
    \int_{F>0} dF ~\rme^{-S_{\mathrm{eff}}[F]} = \int_{\mathcal{J}_b+\mathcal{J}_*} dF ~\rme^{-S_{\mathrm{eff}}[F]} \simeq ~\rme^{- \operatorname{min}(S_{\mathrm{eff}}[F_*],S_{\mathrm{eff}}[F=0])},
\end{align}  
where we have extracted the dominant term between $\mathcal{J}_b$ and $\mathcal{J}_*$ thimbles.

    \item The Stokes phenomenon: from $\tau = 0.05 + 0.08\im$ to $\tau = 0.05 + 0.12\im$ [center panel and right panel in Figure \ref{fig:numerical_thimble_tau005}] .

    Then, the structure is the same as above.
    The Stokes phenomenon happens when the dual thimble intersects with the endpoint: $\operatorname{Im} S[F_*] = 0$.
    Via some contour deformations, we obtain
\begin{align}
    \int_{F>0} dF ~\rme^{-S_{\mathrm{eff}}[F]} = \begin{cases}
         \int_{\mathcal{J}_b+\mathcal{J}_*} dF ~\rme^{-S_{\mathrm{eff}}[F]} ~~~&(\text{saddle point is active}) \\
         \int_{\mathcal{J}_b} dF ~\rme^{-S_{\mathrm{eff}}[F]} ~~~&(\text{saddle point is inactive}) ,
    \end{cases}
\end{align}
\end{itemize}

\begin{figure}[t]
    \centering
    \includegraphics[width=0.7 \linewidth]{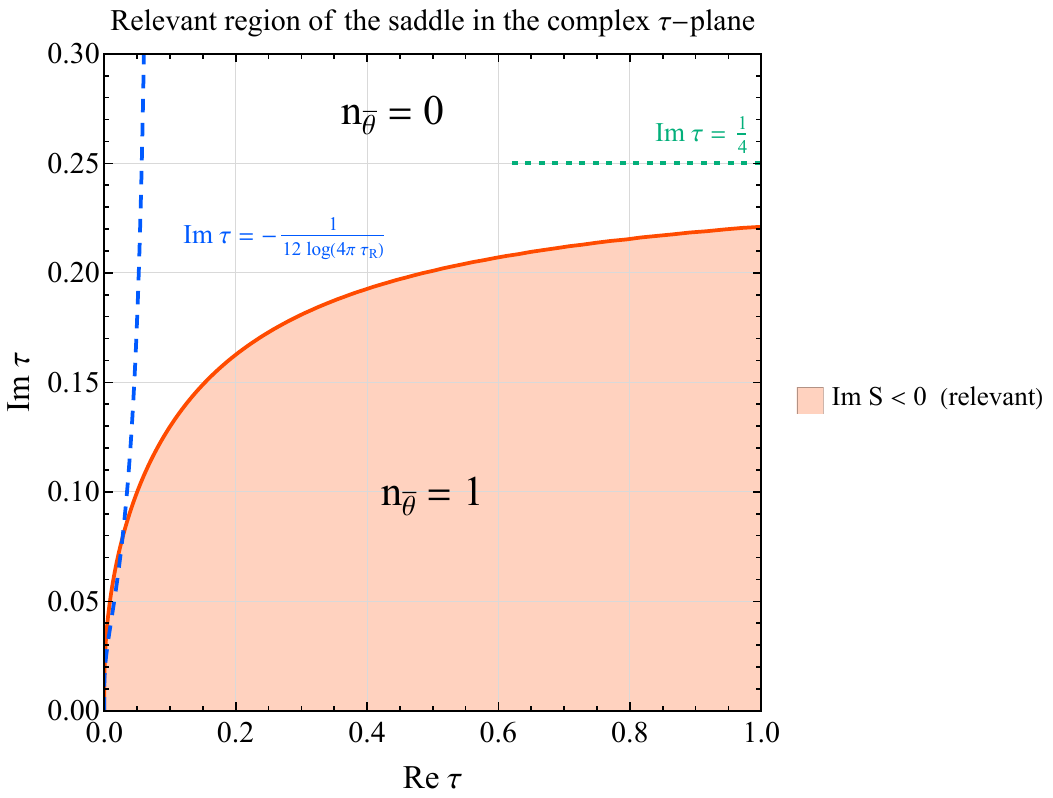}
    \caption{
    The shaded region indicates a parameter where the saddle point is relevant ($n_{\bar{\theta}} = 1$) on the complex $\tau$ plane.
    The asymptotic line ($\operatorname{Im}\tau = 1/4$) at large $\operatorname{Re}\tau$ is the Stokes transition point of the asymptotic model, and the asymptotic curve at small $\operatorname{Re}\tau$ can be obtained from the weak-field analysis of Section \ref{sec:weak_field_maintext}.
    }
    \label{fig:relevant_zone}
\end{figure}

Having clarified the thimble structure for some specific parameter setting, we now shift our focus to the overall behavior.
Figure \ref{fig:relevant_zone} plots the parameter region where the saddle point is relevant on the complex $\tau$-plane.
Since $\operatorname{Im} S[F] < 0$ on the original contour $F>0$, a necessary condition for the saddle to be relevant is that $\operatorname{Im} S[F_*] < 0$. 
Since the saddle contributes for real $\tau$, we deduce that the saddle point is relevant when $\tau$ lies in the region where $\operatorname{Im} S[F_*] < 0$ and which is connected to the real axis.
This region is plotted in Figure \ref{fig:relevant_zone}.

We can see that the Stokes line approaches the value of the asymptotic model ($\operatorname{Im} \tau = 1/4$) as $\operatorname{Re} \tau$ increases.
On the other hand, as $\operatorname{Re} \tau$ decreases, the relevant region of $\theta$ shrinks. 
Actually, as we will see later, the relevant region gradually vanishes in the physical limit ($\operatorname{Re} \tau \rightarrow 0$): the Stokes line on the complex $\tau$ plane converges to the origin at a nonperturbatively slow rate, see, e.g., (\ref{eq:Stokes_near_origin}).
We examine this point in detail later.

\subsection{Endpoint contribution and \texorpdfstring{$F<0$}{F<0} integral}
\label{sec:endpoint_vs_negativeF}

So far, we have estimated that the endpoint contribution has the order of $\sim \rme^{-S_{\mathrm{eff}}[F=0]}$.
One concern is that the negative-$F$ integral significantly affects the endpoint contribution near $F=0$.
Therefore, let us take into account the negative $F$ part and check that the endpoint contribution near $F=0$ indeed survives after the cancellation with the $F<0$ contribution.
This check is essential for justifying the above arguments on the anti-Stokes phenomenon.

Let us recall the original integration:
\begin{align}
    \tilde{Z}(\theta+2\pi n) 
    &=  \int_{F>0} dF dM \left(~\rme^{-S_{\mathrm{eff}}^{\theta + 2 \pi n}[F,M]} + ~\rme^{-S_{\mathrm{eff}}^{\theta + 2 \pi n}[-F,M]} \right)
\end{align}
and the effective action $S_{\mathrm{eff}}^{\theta + 2 \pi n}[-F,M]$ takes the almost same form:
\begin{align}
       \frac{1}{NV_{\mathrm{2d}}}S_{\mathrm{eff}}[-F,M] = -\frac{1}{4\pi} \left[ - M\log\left(\frac{\Lambda_\epsilon^2}{2\rme^\gamma F}\right)  - F\log(2\pi) + 2F\log\Gamma\left(\frac{F+M}{2F}\right) \right] + F \tau_n
\end{align}
for $F>0$, by noticing that the original integral (\ref{eq:integral_vac_energy}) is even in $F$.
The integral is now
\begin{align}
    \tilde{Z}(\theta+2\pi n) 
    &=  \int_{F>0} dF dM ~ \left(\rme^{-S_{\mathrm{eff}}^{\theta + 2 \pi n}[F,M]} + \rme^{-S_{\mathrm{eff}}^{-(\theta + 2 \pi n)}[F,M]} \right) \notag \\
    &\simeq  \int_{F>0} dF ~ \left(\rme^{-S_{\mathrm{eff}}^{(\tau_n)}[F]} + \rme^{-S_{\mathrm{eff}}^{(-\tau_n)}[F]} \right),
\end{align}
where we add the label of $\tau_n$ to $S_{\mathrm{eff}}[F]$.

The crucial point is that the effective action $S_{\mathrm{eff}}^{\theta + 2 \pi n}[F]$ is perturbatively an even function but is not an even function nonperturbatively due to the nonanalyticity.
If the action is analytic at $F=0$ and even, we can directly consider the contour $F \in (-\infty,+\infty)$, and the endpoint contribution must vanish.
Hence, the nonanalyticity is essential for the appearance of the endpoint contribution.
We shall see that the endpoint contribution near $F=0$ indeed survives after the cancellation with the $F<0$ contribution:
\begin{align}
    I_{F\approx 0} = \int_{\mathcal{J}_0^{(\tau_n)}} dF ~ \rme^{-S_{\mathrm{eff}}^{(\tau_n)}[F]} +  \int_{\mathcal{J}_0^{(-\tau_n)}} dF ~ \rme^{-S_{\mathrm{eff}}^{(-\tau_n)}[F]},
\end{align}
where we also add the label of $\tau_n$ to the endpoint thimble.

The dominant contribution arises near $F=0$.
However, if we use the small-$F$ expansion within the perturbative level, the integral is completely canceled due to the above argument.
Therefore, we need to extract a leading-order nonperturbative correction, and we have
\begin{align}
    I_{F\approx 0} \sim \exp\!\Big[-\,\frac{NV_{\mathrm{2d}}\Lambda_\epsilon^2}{4\pi\rme^\gamma}
    \;-\;2\sqrt{\pi\rme^{-\gamma} (-\im \tau_n) NV_{\mathrm{2d}} \Lambda_\epsilon^2}\;\Big]\ , \label{eq:endpoint_behavior_F0}
\end{align}
where we assumed $\operatorname{Im}\tau_n > 0$ without loss of generality.
We present its derivation in Appendix \ref{sec:endpoint_vs_negativeF_appendix}.
The above estimation is also valid for the physical limit $\operatorname{Re}\tau_n =0$.

Although there is a new nonperturbative suppression, this suppression does not change the energy density $\frac{N\Lambda_\epsilon^2}{4\pi\rme^\gamma}$.
Hence, compared to other saddles with higher actions, the endpoint contribution is still dominant even after the $F<0$ contribution is included.

Let us add a remark concerning the endpoint contribution.
Recall that we have discussed one sector $\tilde{Z}(\theta+2\pi n) $, which is obtained from the decomposition via the Poisson summation formula (\ref{eq:Poisson_summation}):
$Z(\theta) =  \sum_{n \in \mathbb{Z}}\tilde{Z}(\theta+2\pi n)$.
However, in the presence of dynamical charged matter, this decomposition is not a kinematical requirement.
Consequently, although the fixed-$n$ partition function $\tilde{Z}(\theta+2\pi n)$ is conventionally interpreted as describing the $n$-th metastable vacuum, the appearance of an endpoint contribution in the sector $\tilde{Z}(\theta+2\pi n)$ does not pose an inconsistency.
On the contrary, due to the non-analyticity\footnote{The non-analyticity at $F=0$ can be observed in the context of the Schwinger effect. Its decay rate is nonperturbative in $F$, which implies that the weak-field expansion does not converge. Indeed, in our derivation of (\ref{eq:endpoint_behavior_F0}), the nonperturbative part (\ref{eq:odd_action_nonpert}) is essential, and it shares the same structure as the decay rate associated with the Schwinger effect. See eg. \cite{Dunne:2004nc}.} at $F=0$, the fixed-$n$ partition function must necessarily contain the endpoint contribution.

\subsection{\texorpdfstring{$\theta$}{theta} dependence of the vacuum energy}
\label{sec:vacuum_theta}

With the above knowledge, we can discuss $\theta$ dependence of the vacuum energy of the $\epsilon$-deformed $\mathbb{C}P^{N-1}$ model.
Below, we provide three illustrative examples.
(For the location of the Stokes transition, see also Figure \ref{fig:relevant_zone}.)

\begin{figure}[H]
  \centering
  \begin{minipage}[b]{0.75 \textwidth}
    \centering
    \includegraphics[width=\linewidth]{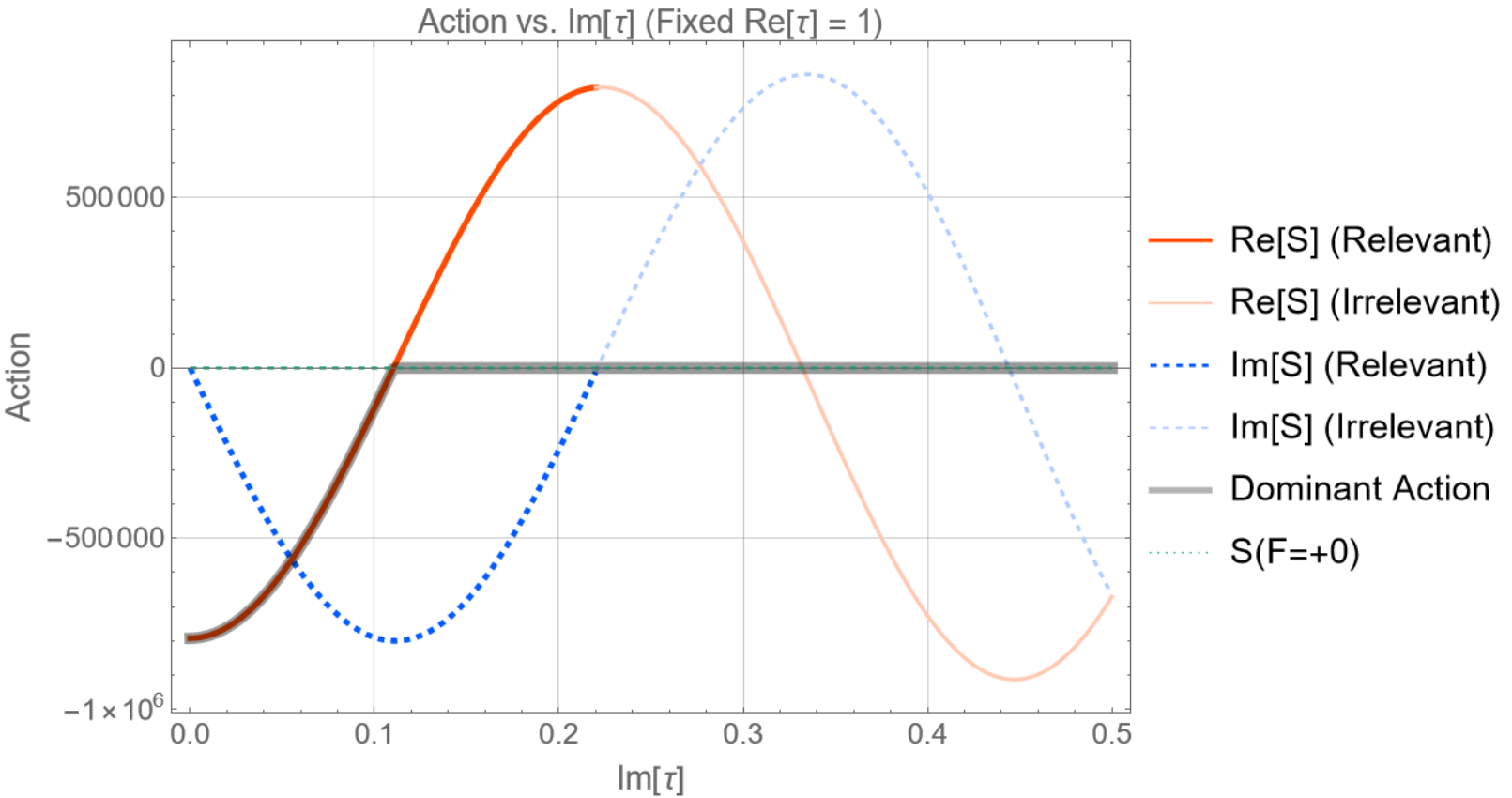}
  \end{minipage}
  \hfill
  \begin{minipage}[b]{0.75 \textwidth}
    \centering
    \includegraphics[width=\linewidth]{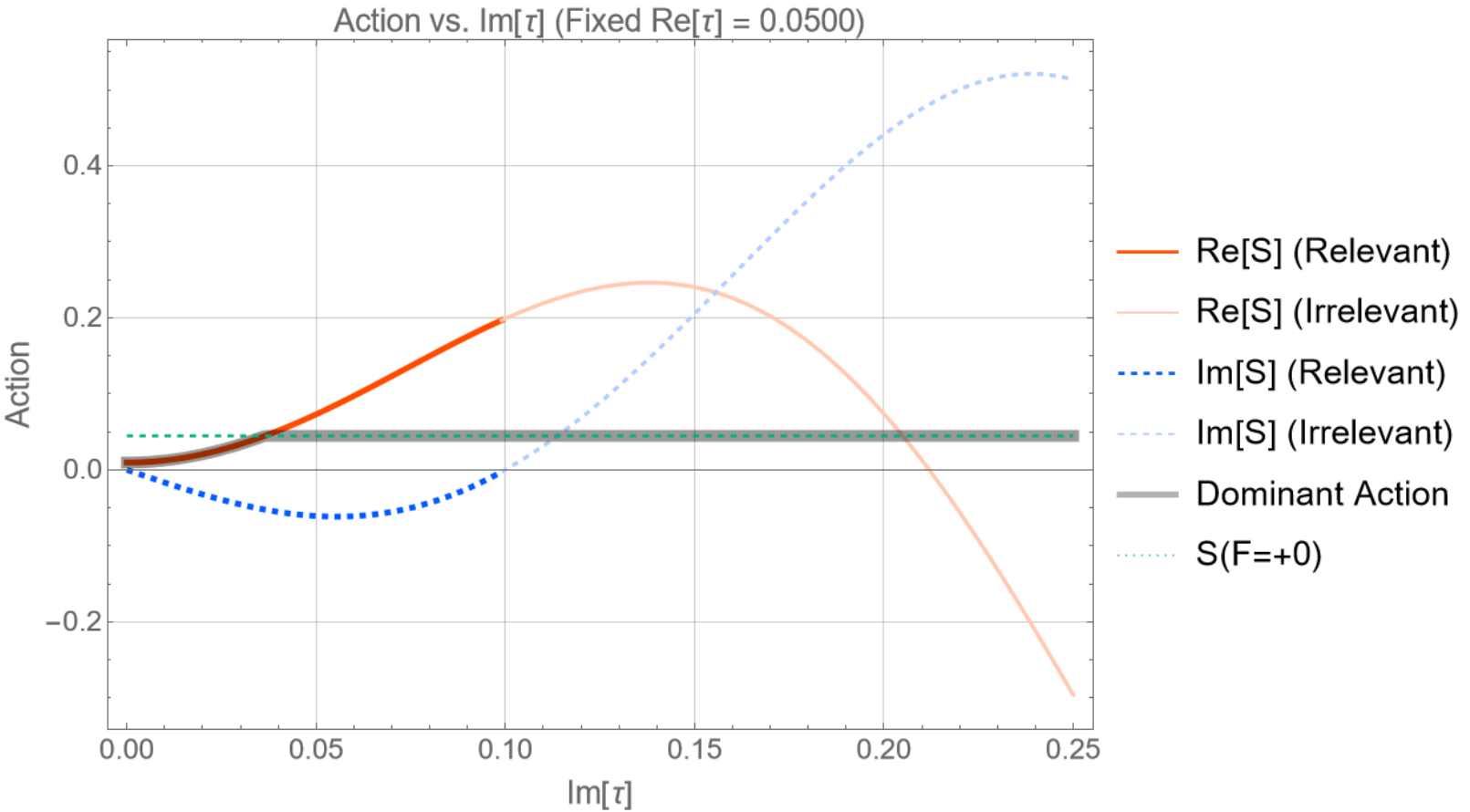}
  \end{minipage}
    \hfill
    
    \begin{minipage}[b]{0.75 \textwidth}
    \centering
    \includegraphics[width=\linewidth]{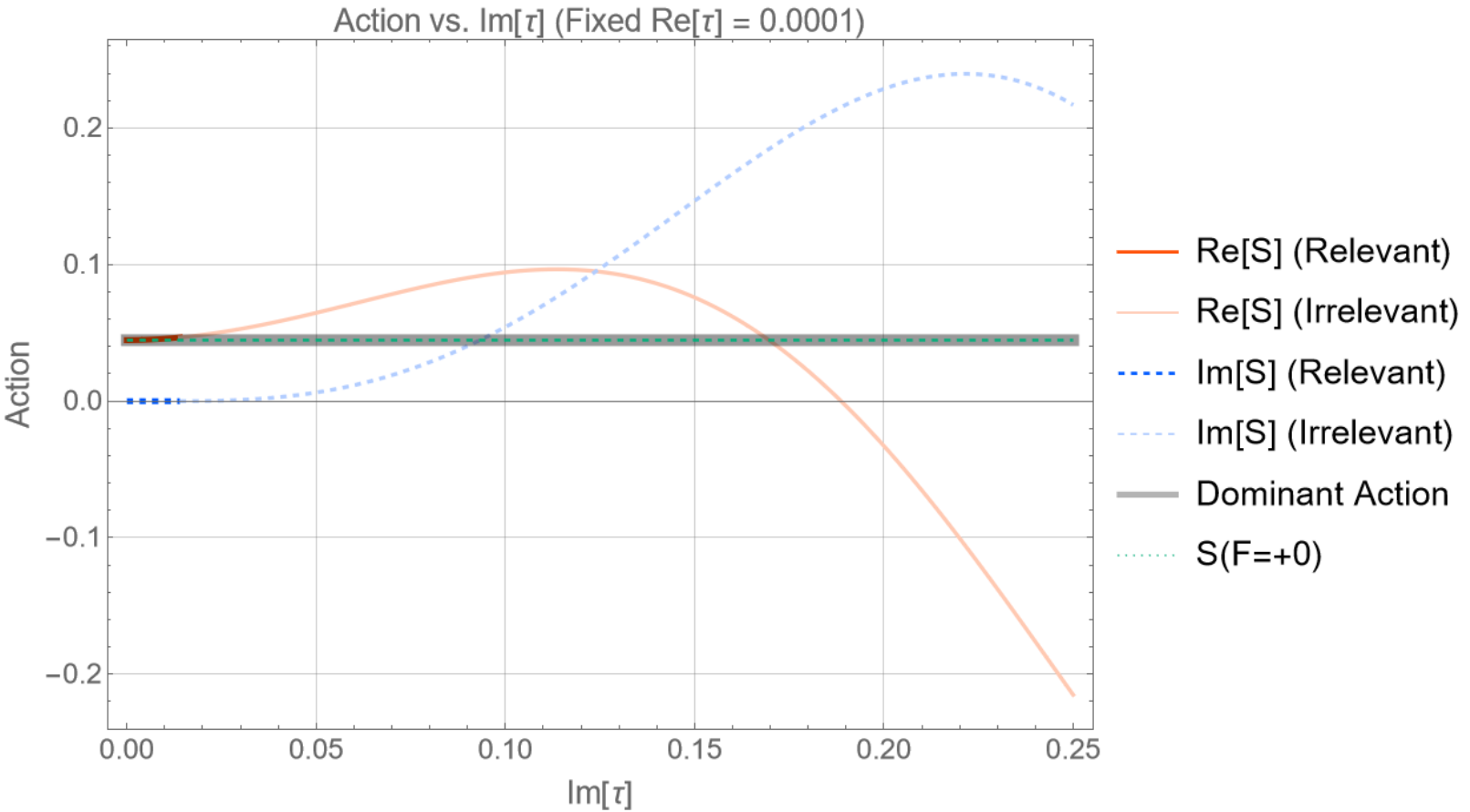}
  \end{minipage}
  \caption{ Plot of the saddle-point action at $\operatorname{Re}\tau = 1.00$ (top panel), $\operatorname{Re} \tau = 0.05$ (center panel), and $\operatorname{Re} \tau = 0.0001$ (bottom panel) in the unit of $\Lambda_\epsilon = 1$. The action represents the vacuum energy density.
  The red solid and blue dotted curves represent $\operatorname{Re} S$ and $\operatorname{Im} S$, respectively.
  The anti-Stokes phenomenon occurs at the level-crossing point $\operatorname{Re} S=S[F=+0]$.
  After that, the Stokes phenomenon happens at $\operatorname{Im} S = 0$.
  The curves are plotted with thin lines after the saddle becomes irrelevant.
  We can observe that the $\theta$ dependence approaches the asymptotic form (\ref{eq:vacuum_energy_asymptotic}) as $\operatorname{Re} \tau$ increases, and that the relevant region shrinks as $\operatorname{Re} \tau$ decreases.
  }
  \label{fig:vac_ene}
\end{figure}

\begin{itemize}
    \item Vacuum energy at relatively large $\operatorname{Re} \tau$ (top panel of Figure \ref{fig:vac_ene}: $\operatorname{Re} \tau = 1$).

    Already at $\operatorname{Re} \tau = 1$, the behavior is qualitatively identical to the vacuum energy of the asymptotic model (\ref{eq:vacuum_energy_asymptotic}): the anti-Stokes phenomenon occurs at $\operatorname{Im} \tau = 1/8$ and the Stokes phenomenon at $\operatorname{Im} \tau = 1/4$.

    \item Vacuum energy at relatively small $\operatorname{Re} \tau$ (center panel of Figure \ref{fig:vac_ene}: $\operatorname{Re} \tau = 0.05$).

    The $\operatorname{Im} \tau = \theta/2 \pi N$ dependence of the saddle-point action ($\operatorname{Re} S$ and $\operatorname{Im} S$) is approaching to that of the standard $\mathbb{C}P^{N-1}$ model (see Ref.~\cite{Sugeno:2025exv}) as $\operatorname{Re} \tau$ decreases.

    \item Vacuum energy near the physical limit $\operatorname{Re} \tau \rightarrow + 0$ (bottom panel of Figure \ref{fig:vac_ene}: $\operatorname{Re} \tau = 0.0001$)

    Although the curves look similar to those of $\operatorname{Re} \tau = 0.05$, the locations of the Stokes and anti-Stokes transition points are noteworthy.
    
    The endpoint action $S[F=+0] = \frac{N\Lambda_\epsilon^2}{4\pi\rme^\gamma}$ is identical to the saddle-point action at the physical point (\ref{eq:physical_saddle_location}) at $\operatorname{Im} \tau = 0$.
    Thus, in the physical limit, we can see that the anti-Stokes phenomenon occurs at $\operatorname{Im} \tau = +0$.

    Not only the anti-Stokes transition, but also the Stokes transition point approaches the origin $\operatorname{Im} \tau = 0$.
    The region where the saddle point is relevant shrinks.
    As we see in the next section, this region indeed collapses to the origin in the physical limit.
\end{itemize}

A notable feature of the top panel is that the maximal dominance of $\operatorname{Re} S_*$ and the Stokes transition seem to occur simultaneously, much like the conventional Stokes phenomenon.
As noted in Appendix \ref{app:boundary_stokes}, the maximal dominance over the subdominant saddle occurs at the Stokes transition point when the parameter direction ($\operatorname{Im} \tau$) is perpendicular to the Stokes line, and this is true also for the bulk-boundary Stokes transition. 
Figure \ref{fig:relevant_zone} shows that the Stokes line is almost horizontal on the complex $\tau$ plane for large $\operatorname{Re} \tau$.
This can also be seen from the asymptotic model.
The holomorphy in $\tau$ and the perpendicularity guarantee that the Stokes phenomenon and maximal dominance happen simultaneously.


\subsection{Physical limit: weak-field analysis}
\label{sec:weak_field_maintext}

In the previous section, we observed that the relevant region of the saddle point shrinks as $\operatorname{Re} \tau$ decreases.
Because the shrinking is quite slow (see also Figure \ref{fig:relevant_zone}), we need to determine the relevant region analytically near $\operatorname{Re} \tau=0$.

In the physical case $\operatorname{Re} \tau=0$ with small $\bar{\theta} = 2 \pi \operatorname{Im}\tau$, the saddle point $F_*$ is located near the origin $F=0$.
Therefore, we can analytically determine $\operatorname{Im}S[F_*]$ in this setup.
The detailed calculations are provided in Appendix \ref{app:weak_field}.
The imaginary part of the saddle-point action can be calculated as
\begin{align}
    \operatorname{Im} \left( \frac{1}{NV_{\mathrm{2d}}} S_{\mathrm{eff}}[F_*] \right) \simeq \frac{3 \bar{\theta} \Lambda_\epsilon^2}{2\pi \rme^\gamma} \exp\left(-\frac{\pi}{6 |\bar{\theta}|}\right). \label{eq:weak-field_imaginarypart}
\end{align}
This quantity is nonperturbatively small in $\bar{\theta}$, and this factor cannot be seen within the perturbative expansion in $F$.

The important implication is that the saddle point does not contribute for any $\operatorname{Im} \tau \neq 0$ at the physical point $\operatorname{Re} \tau = 0$.
Indeed, the integral we consider is
\begin{align}
    \tilde{Z}(\theta+2\pi n) 
    &= \int_{0}^\infty dF ~\rme^{-S_{\mathrm{eff}}^{\theta + 2 \pi n}[F]} +  \int_{0}^\infty dF ~\rme^{-S_{\mathrm{eff}}^{-(\theta + 2 \pi n)}[F]},
\end{align}
where the integral over negative $F$ becomes the second term.
We can choose $\theta+2\pi n>0$, and then the original contour ($0<F<\infty$) in the first term satisfies $\operatorname{Im} S_{\mathrm{eff}}[F] <0$, and that of the second term satisfies $\operatorname{Im} S_{\mathrm{eff}}[F] >0$.
However, the above result indicates that the imaginary part of the saddle-point action $\operatorname{Im} S_{\mathrm{eff}}[F_*]$ has the opposite sign, which implies that the dual thimble cannot intersect the original contour.
Therefore, the saddle is always irrelevant for any $\operatorname{Im} \tau \neq 0$ at the physical point $\operatorname{Re} \tau = 0$.

As seen in Figures \ref{fig:relevant_zone} and \ref{fig:vac_ene}, the relevant region of the saddle point shrinks as $\operatorname{Re} \tau$ decreases.
The above result directly shows that, in the physical limit, the saddle point is relevant only at $\operatorname{Im} \tau = 0$. This implies, rather dramatically, that no saddles with $\theta + 2 \pi n \sim \mathcal{O}(N) $ contribute to the large-$N$ partition function of the physical $\mathbb CP^{N-1}$.

Here, it is important to note that there is a subtlety in resolution at the large-$N$ leading order. 
Our result implies that the saddle becomes irrelevant as long as $(\theta + 2 \pi n)/N \sim O(1)$. 
Since the saddle is relevant at $\theta + 2 \pi n =0$, it is still relevant for $(\theta + 2 \pi n) \sim O(1)$, where the topological term does not affect the saddle-point equation. 
Hence, our result does not contradict the classic calculations of, e.g., the topological susceptibility. 
Indeed, although the saddle $F_*$ itself does not contribute at nonzero $\operatorname{Im} \tau$, the $O(\theta^2)$ term of its action gives the same topological susceptibility, since it is fixed by the weak-field expansion.
See Appendix \ref{sec:scaling_subtleties} for details.

We can also determine how the relevant region shrinks near the physical limit $\operatorname{Re} \tau \approx 0$.
By adding $\tau_R$ as a perturbation, we can obtain the Stokes transition point, which is located at
\begin{align}
 \tau_R \simeq \frac{1}{4\pi } \rme^{-\frac{\pi}{6 \bar{\theta}}}~~~\Leftrightarrow ~~~ \bar{\theta} \simeq - \frac{\pi}{6 \log(4\pi \tau_R)}.  \label{eq:Stokes_near_origin}
\end{align}
Its derivation is also presented in Appendix \ref{app:weak_field}.
This also shows why the physical theory  ($ \tau_R  =0 $) is so special and more difficult. 
Turning on some small $\tau_R$ gives a small Stokes wedge where $\theta + 2 \pi n \sim \mathcal{O}(N) $ saddles become relevant as shown in Fig.~\ref{fig:relevant_zone}.

\section{Towards the self-dual theory: amplitude at fixed topological charge}
\label{sec:amplitude_fixedQ}

In the exact self-dual theory, the partition function should be written as the sum over the self-dual configuration with the weight of the fluctuation determinant:
\begin{align}
    Z_{\mathrm{sd}}(\theta) &= \sum_{Q \geq 0} Z_Q\rme^{\im Q \theta}, \\
     Z_Q &= \rme^{-\frac{2\pi}{g^2_{\mathrm{sd}}}Q}  \int_{\mathcal{M}_{Q}} d\mu_{Q} ~(\text{fluctuation det.})^{-1}.
\end{align}
It is highly interesting to see how the large-$N$ analysis predicts the ``$Q$-instanton amplitude'' $Z_Q$.
To this end, let us investigate the ``$Q$-instanton amplitude'' by Fourier transformation of our result in the almost self-dual case\footnote{
In the large-$N$ analysis, there is a known subtlety regarding the order of the large-volume and large-$N$ limits \cite{Aguado:2010ex}. Since our approach additionally involves the self-dual limit, there can be a potential issue about the order of these limits.
In this paper, we simply assume that the self-dual limit is smooth.
}, $(\operatorname{Re} \tau \gg 1)$.
We return to the original representation of the partition function.
\begin{align}
    Z(\theta) &=  \sum_{n \in \mathbb{Z}}\tilde{Z}(\theta+2\pi n) 
\end{align}
We now define the amplitude of fixed topological charge $Q$ as,
\begin{align}
    Z(\theta) &=  \sum_{Q \in \mathbb{Z}} Z_Q \rme^{\im Q \theta} ~~ \notag \\
    &\Leftrightarrow~~Z_Q = \int_{-\pi}^{\pi}\frac{d\theta}{2\pi} Z(\theta)  \rme^{-\im Q \theta} = \int_{-\infty}^{+\infty} \frac{d\theta}{2\pi} \tilde{Z}_+(\theta) \rme^{-\im Q \theta}
\end{align}
By rescaling $\theta = N\tilde{\theta}$, we have
\begin{align}
    Z_Q = N\int \frac{d\tilde{\theta}}{2\pi} \tilde{Z}(N \tilde{\theta}) \rme^{-\im N Q \tilde{\theta}}.
\end{align}

From the discussion so far, $ \tilde{Z}(N \tilde{\theta})$ consists of two constituents:
\begin{align}
    \tilde{Z}(\theta+2\pi n) &=  \tilde{Z}_{\text{endpoint}}(\theta+2\pi n) + \tilde{Z}_{\text{saddle}} (\theta+2\pi n), 
\end{align}
where the leading order of each contribution is
\begin{align}
\tilde{Z}_{\text{endpoint}}(\theta+2\pi n) &\sim \rme^{-\,\frac{NV_{\mathrm{2d}}\Lambda_\epsilon^2}{4\pi\rme^\gamma}
    \;-\;2\sqrt{\pi\rme^{-\gamma} (-\im \tau_n) NV_{\mathrm{2d}} \Lambda_\epsilon^2}} \\
\tilde{Z}_{\text{saddle}} (\theta+2\pi n) &\sim \rme^{-S_{\mathrm{eff}} [M_*(\tau_n),F_*(\tau_n)]}~~~(\text{if the saddle is active}).
\end{align}

In the self-dual limit ($\operatorname{Re} \tau \rightarrow \infty$ with fixed $\Lambda_g$), the endpoint contribution would become a constant: $\tilde{Z}_{\text{endpoint}}(\theta+2\pi n) \sim 1$ because $(-\im \tau_n)  \Lambda_\epsilon^2 \rightarrow 0$.
Therefore, the endpoint contribution produces:
\begin{align}
    Z_Q^{\text{(endpoint)}} \sim \delta(Q),
\end{align}
which contributes only at $Q=0$.
It is reasonable that the $F=0$ endpoint contributes only at $Q=0$.
For $Q>0$, we can neglect this endpoint contribution, and we can now focus on the saddle-point contribution.

In the limit $\epsilon^2 \rightarrow +0$, the vacuum energy density is given by
\begin{align}
    L_{\mathrm{eff}}(M_*, F_*)  = \frac{1}{4\pi} (M_* - \mu^2),~~M_*\simeq - 4\pi \rme^{1-\gamma}\Lambda_g^2 \rme^{2 \im \tilde{\theta}}
\end{align}
where we have recalled $\Lambda_g^2 = \frac{\mu^2 }{N^2 g^4} \exp\left( -\frac{4\pi}{Ng^2} \right) = \frac{\mu^2 }{N^2 } \rme^{-\frac{4\pi}{Ng_{\mathrm{sd}}^2}}$.
Because the Stokes phenomenon occurs at $\tilde{\theta} = \pm \pi /2$, we can write
\begin{align}
    \tilde{Z}_{\text{saddle}} (N \tilde{\theta}) =
    \begin{cases}
        \rme^{+NV_{\mathrm{2d}} \rme^{1-\gamma}\Lambda_g^2 \rme^{2 \im \tilde{\theta}}}~~~&(-\pi/2 < \tilde{\theta} < \pi/2)\\
        0 &(\text{otherwise}),
    \end{cases}
\end{align}
up to some $O(1)$ factor.
Thus, the integration is
\begin{align}
    Z_Q &= \frac{N}{2\pi} \int_{|\tilde{\theta} | < \pi/2} d\tilde{\theta} ~\rme^{+NV_{\mathrm{2d}} \rme^{1-\gamma}\Lambda_g^2 \rme^{2 \im \tilde{\theta}} -\im N Q \tilde{\theta}} 
\end{align}
For $NQ/2 \in \mathbb{Z}$, this integral can be exactly computed.
Indeed, by expanding $\rme^{+NV_{\mathrm{2d}} \rme^{1-\gamma}\Lambda_g^2 \rme^{2 \im \tilde{\theta}}}$, we have
\begin{align}
    Z_Q &= \begin{cases}
        \frac{N}{2}  \frac{(NV_{\mathrm{2d}} \rme^{1-\gamma}\Lambda_g^2)^{NQ/2} }{\left(\frac{N Q }{2} \right)!}~~~&(\text{for} ~NQ \in 2 \mathbb{Z}_{> 0}) \\
        0 &(\text{for} ~NQ \in 2 \mathbb{Z}_{<0}) 
    \end{cases}
\end{align}
Since we consider the large-$N$ limit, this result can be understood as an approximation for general $NQ \in \mathbb{Z}$.

This amplitude indeed has an instanton factor:
\begin{align}
    Z_Q \propto \Lambda_g^{NQ} \propto \rme^{-\frac{2\pi}{g^2_{\mathrm{sd}}}Q}
\end{align}
The $\rme^{-\frac{2\pi}{g^2_{\mathrm{sd}}}Q}$ is the Boltzmann factor of $Q$ instanton in the exact self-dual theory.
Also from this formula, one can understand the necessity of the topological counterterm.
In the $\epsilon$-deformed theory, the normalization of fluctuation zeromodes gives the factor $\epsilon^{-2NQ}$ under the instanton background.
We have canceled this singularity by the $\log$ correction of the topological term, then we get $g^{-2NQ}$.
Via the coupling redefinition $g^{-2}_{\text{sd}} = g^{-2} + \frac{N}{2\pi} \log (g^2)$, we can correctly reproduce the instanton factor in the exact self-dual theory.

Hence, within the uniform ansatz and large-$N$ saddle point approximation, we predict the partition function of the exact self-dual theory at large-$N$:\footnote{ Note the similarity between \eqref{ZQ} and its counterpart in a particle on a circle with $N$ minima  quantum mechanical system \cite{Unsal:2020yeh} where $Z_Q \sim (V_{ \rm 1d} e^{-S_I/N})^{NQ}/(NQ)! \sim V_{ \rm 1d}^{NQ}(e^{-S_I})^{Q}/(NQ)! $, where the instanton factor appears as a  composite  of $N$ fractional instantons. In particular, instanton with winding number $Q=1$ appears not with naive extensive volume factor $V_{ \rm 1d} =\beta$, but rather $V_{ \rm 1d}^N$ reflecting the moduli space structure of  $N$ fractional instantons building up an instanton.  Furthermore, $1/(NQ)!$ is the standard symmetry factor arising from integrating over the positions of $N|Q|$ indistinguishable particles (fractional instantons) in a dilute gas. 
Although the resemblance of \eqref{ZQ} to a charge-$2/N$ fractional instanton gas is intriguing, the dilute gas picture is invalid because the system is not in the weak-coupling regime. Furthermore, in purely bosonic theories like the present one, the weight factor arising from the fluctuation determinant is non-trivial. Therefore, information about the moduli space alone is insufficient, making a semiclassical interpretation difficult.  A more thorough study of these features is desirable perhaps along the lines of \cite{Dunne:2012ae, Fujimori:2016ljw}.}
\begin{align}
    Z_{\mathrm{sd}}(\theta) &= \sum_{Q \geq 0} Z_Q\rme^{\im Q \theta}, \\
     Z_Q &=  \rme^{-\frac{2\pi}{g^2_{\mathrm{sd}}}Q}  \int_{\mathcal{M}_{Q}} d\mu_{Q} ~(\text{fluctuation det.})^{-1} \sim \frac{(NV_{\mathrm{2d}} \rme^{1-\gamma}\Lambda_g^2)^{NQ/2} }{\left(\frac{N Q }{2} \right)!},
     \label{ZQ}
\end{align}
where $\int_{\mathcal{M}_{Q}} d\mu_{Q} $ denotes the moduli integration of topological charge $Q$.

In the large-$N$ limit, we can evaluate the amplitude by casting it into its asymptotic exponential form. Applying Stirling's approximation to the factorial, $Z_Q$ can be written as
\begin{equation}
    Z_Q \approx \exp\left[ \frac{NQ}{2} \left( \ln \frac{2 V_{2d} e^{1-\gamma} \Lambda_g^2}{Q} + 1 \right) \right] .
\end{equation}
The vacuum dynamically selects the sector that maximizes this amplitude. By evaluating the saddle point of the exponent with respect to $Q$ at $\theta=0$, we find the dominant macroscopic sector $Q_*$ is
\begin{equation}
    Q_* \simeq 2 V_{2d} e^{1-\gamma} \Lambda_g^2 = \frac{F_* V_{2d}}{2\pi} ,
\end{equation}
where $F_*$ is the saddle point in the $n=0$ sector (3.29). Thus, we can understand that the dynamical magnetic field is determined by the competition between the action suppression $e^{-\frac{2\pi}{g_{\mathrm{sd}}^2}Q}$ and the ``entropic factor'' from the moduli integral (with the weight of fluctuation determinant): $\int_{\mathcal{M}_{Q}} d\mu_{Q} ~(\text{fluctuation det.})^{-1} \sim \frac{(N^{-1} \rme^{1-\gamma}\mu^2V_{\mathrm{2d}})^{NQ/2} }{\left(\frac{N Q }{2} \right)!}$. 
Note that, unlike the standard $\mathbb{C}P^{N-1}$ model at $\operatorname{Re}\tau = 0$, this entropic factor favors higher topological charge.
This can be interpreted as an origin of the condensation of the field strength.

It is important to note that the topological charge supported by the vacuum is equal to  density of states  of lowest  Landau level $\frac{F_*}{2\pi} $ times 
area of the 2d space $V_{\mathrm{2d}}$, i.e., the  total  number of states in the lowest Landau level with magnetic field $F_* $.  
However, unlike the standard Landau level problem where $F_*$ is external, in the present problem, it is generated dynamically, and is dictated  by  the strong scale 
$F_* \simeq \Lambda_g^2$.

\section{Discussion}
\label{sec:discussion}

Before concluding our paper, let us highlight a few implications and observations of our findings, and discuss various potential avenues for future work.


\paragraph{Implications of bosonic chiral anomaly:}
Our derivation of the anomalous functional measure (Section \ref{sec:bosonic_chiral}) has important implications for the broader literature utilizing first-order deformations. In the study of perturbative scattering amplitudes---such as in twistor string theory or  deformations of self-dual Yang-Mills (SDYM) theory---it is standard practice to recover the full physical theory via a classical $g^2 |h|^2$ (or equivalent $g^2 B^2$) deformation of the first-order action 
\cite{Chalmers:1996rq,Bittleston:2024efo,Bittleston:2025jmk,Witten:2003nn}. Because standard perturbative expansions are performed around a topologically trivial vacuum, the topological measure anomaly derived in this work evaluates strictly to zero. Consequently, the classical equivalence between the first-order and second-order formulations is perfectly sufficient for determining perturbative Feynman rules and scattering matrices. 

However, our analysis demonstrates that this classical and perturbative  equivalence fundamentally breaks down at the non-perturbative quantum level. When evaluating the thermodynamics of the instanton gas, the sum over all topological sectors activates the chiral measure anomaly. Therefore, a strict quantum equivalence---which is explicitly necessary for determining the true non-perturbative vacuum structure---cannot be achieved by a classical deformation alone, but rigorously requires the inclusion of the compensating topological counterterm.

\paragraph{Adiabatic continuity under \texorpdfstring{$\epsilon$}{epsilon} deformation:}
Analytic nonperturbative methods are generally based on identifying a tractable/simplified setup that preserves the essential aspects of the strongly-coupled dynamics. 
This underlying philosophy applies to the large-$N$ expansion, supersymmetric theories, semiclassical approaches via compactification, and so on. 
The findings in this paper strongly motivate the addition of a novel simplification, deformation toward a self-dual theory, to these theoretical toolkits. 
Indeed, our analysis of the $\mathbb{C}P^{N-1}$ model in the large-$N$ limit demonstrates that this self-dual deformation can be achieved without encountering a phase transition, implying a new form of adiabatic continuity  directly on infinite space $\mathbb R^2$ (Section \ref{sec:continuity_realtau}).
Exploring the interplay between deformations toward self-dual theories and other nonperturbative methods, particularly $\mathbb{R} \times S^1$ semiclassics \cite{Dunne:2012ae, Dunne:2012zk, Misumi:2014jua, Sulejmanpasic:2016llc, Fujimori:2016ljw, Fujimori:2017oab}, presents a highly promising avenue for future research.

\paragraph{Endpoint Stokes phenomenon:}
In the present case, the endpoint also possesses ``a thimble'' as if it were a saddle point, and the competition between the endpoint and the saddle point played a crucial role in avoiding the unphysical saddle problem (Section \ref{sec:Stokesph}). 
However, it is still an open problem to give the physical interpretation of the endpoint contribution due to the singularity at $F=0$.

Meanwhile, this result could have broad implications. 
For example, in some cases of quasi-moduli integrals, one may encounter integrals with boundaries or singularities. In such cases, it would be worthwhile to pay attention to the endpoint contribution.

\paragraph{Lowest Landau levels as an effective theory of the large-$N$ self-dual theory:}

The low-energy effective theory of the supersymmetric self-dual theory is often supposed to be a logarithmic CFT \cite{Frenkel:2006fy, Frenkel:2008vz, Frenkel:2007ux}.
The bosonic self-dual theory at large-$N$  exhibits a different behaviour. 
We obtain a dynamically generated field strength $F_* \simeq \Lambda_g^2$ with the multiplier field $M \simeq - \Lambda_g^2$, and the massless fluctuations on this uniform $F_*$ are the lowest Landau levels whose gap vanishes in the self-dual limit $(M + |F|) \rightarrow 0$.
On the other hand, the partition function of the self-dual theory takes the following form (Section \ref{sec:amplitude_fixedQ}):
\begin{align}
    Z_{\mathrm{sd}}(\theta) &= \sum_{Q \geq 0} Z_Q\rme^{\im Q \theta} \simeq Z_{Q_* = \frac{F_* V_{\mathrm{2d}}}{2\pi}} \rme^{\im Q_* \theta}, \\
     Z_Q &= \rme^{-\frac{2\pi}{g^2_{\mathrm{sd}}}Q} \int_{\mathcal{M}_{Q}} d\mu_{Q} ~(\text{fluctuation det.})^{-1}.
\end{align}
Because the lowest Landau levels are gapless modes under the uniform field-strength ansatz $F_*$, they will represent the moduli $\mathcal{M}_{Q}$ near the uniform-$F$ configuration.
Therefore, the low-energy effective theory of the large-$N$ self-dual theory would consist of the lowest Landau levels, which seemingly form a non-commutative theory.
This would be a 2d analog of the non-commutative effective theory under the self-dual background of 4d Yang-Mills theory \cite{Unsal:2026aax}, but it is important to note that the nontrivial self-dual field is dynamically generated.

\paragraph{Other future prospects:}

Related to the prediction on the self-dual limit of the partition function, it would be interesting to consider the interpretation of $Z_Q$ obtained in our large-$N$ calculation.
If we can control the fluctuation determinant well, the relationship between the $F$ condensate and the moduli space structure of the BPS solutions \cite{Aguado:2001xg, Hayashi:2025odr} will be able to be discussed.
To this end, it would be useful to extend our analysis to the supersymmetric theories \cite{DAdda:1978dle}.
We speculate that the fractional-instanton-like structure of $Z_Q$ might be understood from the structure of the moduli space at large-$N$.

The deformation towards the self-dual limit is equivalent to introducing an imaginary $\theta$ angle, and we found various nonperturbative aspects from the large-$N$ analysis.
It is a common approach to conduct numerical simulations at an imaginary $\theta$ towards understanding the $\theta$-dependence to avoid the sign problem \cite{Bhanot:1984rx, Azcoiti:2003qe, Imachi:2006gq, Bonanno:2018xtd} (see also \cite{Panagopoulos:2011rb, Bonati:2015sqt, Bonati:2016tvi, Hirasawa:2024fjt} for Yang-Mills theories).
Analytic structure on the complex $\theta$ plane ($\tau$ parameter in this paper) would be relevant to the ``analytic continuation'' from the imaginary to real $\theta$ angles.
Also, the imaginary electric field has been discussed in the phenomenological context \cite{Endrodi:2024cqn}, and our 2d Euclidean $F$-condensate has a potential relevance.

\acknowledgments
The authors are especially indebted to Yuya Tanizaki for collaboration at an early stage of this work.
This work was partially supported by Japan Society for the Promotion of Science (JSPS) Research Fellowship for Young Scientists Grant No. 23KJ1161 (Y.H.).  M. \"U. is supported by U.S. Department of Energy, Office 
of Science, Office of Nuclear Physics under Award Number 
DE-FG02-03ER41260 and  by the Simons Foundation Grant 
(Simons Collaboration on Confinement and QCD Strings). 


\appendix

\section{On zeromodes and topological term}
\label{sec:zeromodes}

Here, we provide a detailed derivation of the topological counterterm from the zeromode structure, which is partially presented in Section \ref{sec:bosonic_chiral}.

We consider the original $\epsilon$-deformed action of $(\phi,\phi^\dagger,h,h^\dagger)$,
\begin{align}
    &S_{def}[\phi,\phi^\dagger,h,h^\dagger, \alpha, A] \notag \\
    &= \int d^2 x \left[ + \epsilon^2 |h|^2 - \im h^\dagger D_+ \bm{\phi} - \im (D_+ \bm{\phi})^\dagger h \right] + \im \alpha (|\bm{\phi}|^2 -1)+  \left(\frac{1}{g^2_{\text{sd}}}   - \im \frac{\theta}{2 \pi}  \right) \int {d^2 x} F
\end{align}
with the Lagrange multiplier $\alpha \in \mathbb{R}$, which will be identified as $M/\epsilon^2 = \im \alpha$ in the saddle-point approximation in the main text.
In this appendix, we verify that this $\epsilon$-deformed theory reproduces (i) the self-dual theory in the $\epsilon \to 0$ limit, and (ii) the physical $\mathbb{C}P^{N-1}$ model in the $\epsilon \to g$ limit, paying careful attention to the zeromodes of the fluctuation operator.
In particular, the physical theory is reproduced when $\epsilon^2 = g^2$ where $g^2$ is defined by $\frac{1}{g^2_{\text{sd}} }   = \frac{1}{g^2 }+ \frac{N}{2\pi} \log (g^2)$.
A careful analysis on the zeromodes leads to the topological counterterm $+ \frac{N}{2\pi} \log (g^2)$ when moving from the first-order to the second-order formalism.


Naively, the ($\phi,\phi^\dagger,h,h^\dagger$) Gaussian integration leads to the fluctuation determinant
\begin{align}
    \left( \det \begin{pmatrix}
        \im \alpha & +\im D_-\\
        -\im D_+ & \epsilon^2
    \end{pmatrix} \right)^{-1} .
\end{align}
To arrive at the exact self-dual limit, we need a careful analysis of the zeromodes.

The fluctuation operator can be rewritten as,
\begin{align}
    \begin{pmatrix}
  \multicolumn{2}{c}{\multirow{2}{*}{\large $\im \alpha$}} & 0 & 0  \\
 \multicolumn{2}{c}{} & 0 & +\im  D_- \\
        0 & 0 & \epsilon^2  &0\\
        0 &  -\im  D_+ & 0 & \epsilon^2
\end{pmatrix}
    \begin{pmatrix}
        \phi_0 \in \ker D_+ \\
        \phi_\perp \in (\ker D_+)^\perp \\
        h_0 \in \ker D_- \\
        h_\perp \in (\ker D_-)^\perp
    \end{pmatrix} 
    \label{eq:total_fluct_oper}
\end{align}
where we have used $\operatorname{im}D_- \subset (\ker D_+)^\perp $.
As $\alpha$ is an arbitrary function, the operator $\im \alpha$ can mix $\ker D_+$ and $(\ker D_+)^\perp$.
In order to interpret the limit $\epsilon^2 \rightarrow +0$ as the self-dual theory, we have to address the following subtle but critical points.

\subsubsection*{About \texorpdfstring{$D_-$ }{D-} zeromodes}

The first subtlety arises in the $(h,h^\dagger)$ integration over $\ker D_-$.

We notice that the $\ker D_-$ sector decouples, which yields $\epsilon^{-2 \dim \ker D_-}$.
To remove this extra factor, it seems necessary to include an additional factor of $\epsilon^{2 \dim \ker D_-}$ in the integration measure $\mathcal{D}h \mathcal{D} h^{\dagger}$.
However, we actually do not have to attach this factor in order to take the self-dual limit, because $\left. (\dim \ker D_-) \right|_{\text{self dual}} = 0$. Indeed, if there were a nontrivial $h \in \ker D_-$, it satisfies
\begin{align}
    0 = \int d^2x~|D_-h|^2 = \int d^2x~(|D_\mu h|^2 + F(x)|h|^2).
\end{align}
On the other hand, for any self-dual configuration, the field strength is non-negative $F(x) \geq 0$ because of the BPS equation.
Thus, there are no zeromodes with wrong chirality under the self-dual $A$.
Therefore, it is not necessary to attach the factor $\epsilon^{2 \dim \ker D_-}$ by hand, whereas it is possible\footnote{
In the procedure of the main text, this prescription is expressed as follows.
The $(h,h^\dagger)$ integration can be written as,
 \begin{align}
   \int \mathcal{D}h \mathcal{D} h^{\dagger} &\;  \exp\left[ -\int \left( \epsilon^2 |h|^2-i h^\dagger D_+ \bm{\phi} - i (D_+ \bm{\phi})^\dagger h \right) \right] \notag \\
   &= \epsilon^{-2 \dim \ker D_-} \epsilon^{-2 \dim (\ker D_-)^\perp} \rme^{-\frac{1}{\epsilon^2}|D_+\phi|^2},
\end{align}
and $\epsilon^{-2 \dim (\ker D_-)^\perp} $ will be canceled by the further $\phi$-integration.
In this expression, we can see that the non-self-dual configuration is non-perturbatively suppressed $\rme^{-\sharp / \epsilon^2}$, whereas $\epsilon^{-2 \dim \ker D_-}$ is just a power in $\epsilon$.
As the self-dual configuration satisfies $\dim \ker D_-=0$, we can simply ignore $\epsilon^{-2 \dim \ker D_-}$ in the self-dual limit $\epsilon \rightarrow +0$.
}.
As an $\epsilon$-deformed theory, we choose to employ the untouched fluctuation determinant (\ref{eq:total_fluct_oper}).

\subsubsection*{Recovering self-dual theory}

Let us observe how the self-dual limit is obtained.
Because of $\left. (\dim \ker D_-) \right|_{\text{self dual}} = 0$ as presented above, we can omit $(\ker D_-)$ modes from the fluctuation determinant in the self-dual limit $\epsilon^2 \rightarrow +0$: we can concentrate on the following sector:
\begin{align}
    \begin{pmatrix}
  \multicolumn{2}{c}{\multirow{2}{*}{\large $\im \alpha$}}& 0  \\
 \multicolumn{2}{c}{}  & + \im D_- \\
        0 & - \im D_+  & \epsilon^2
\end{pmatrix}
    \begin{pmatrix}
        \phi_0 \in \ker D_+ \\
        \phi_\perp \in (\ker D_+)^\perp \\
        h_\perp \in (\ker D_-)^\perp
    \end{pmatrix} 
\end{align}
This determinant can be calculated as,
\begin{align}
  \det  \begin{pmatrix}
  \multicolumn{2}{c}{\multirow{2}{*}{\large $\im \alpha$}}& 0  \\
 \multicolumn{2}{c}{}  & + \im D_- \\
        0 & - \im D_+  & 0
\end{pmatrix} = \left( \operatorname{det}_{\ker D_+}(\im \alpha) \right) \left( \operatorname{det}_{(\ker D_+)^\perp} (- D_- D_+)\right)
\end{align}
It is crucial to note that the UV regularization satisfies the pairing $\dim (\ker D_+)^\perp = \dim (\ker D_-)^\perp$ to obtain the well-behaved fluctuation determinant $\left( \operatorname{det}_{(\ker D_+)^\perp} (- D_- D_+)\right)^{-1}$ from the $(\ker D_+)^\perp \oplus (\ker D_-)^\perp$ integration.
Otherwise, the ($\phi,\phi^\dagger,h,h^\dagger$) integration measure would produce a $U(1)$ gauge anomaly.

The first factor leads to the moduli integration:
\begin{align}
    \int \mathcal{D}\alpha \frac{\rme^{- \im \int d^2 x\, \alpha}}{ \operatorname{det}_{\ker D_+}(\im \alpha) } &= \int \mathcal{D}\alpha \int_{\ker D_+} \mathcal{D}\phi_0~\rme^{\im \int d^2 x\, \alpha(x) (|\phi_0(x)|^2 - 1)} \notag \\
    &= \int_{\ker D_+} \mathcal{D}\phi_0~\delta(|\phi_0|^2 - 1).
\end{align}
Once the path integral $\int \mathcal{D}A$ is included, the expression becomes equivalent to the integration over the self-dual $\mathbb{C}P^{N-1}$ fields.
In the limit $\epsilon^2 \rightarrow +0$, up to the topological term, the partition function becomes 
\begin{align}
    Z &= \int \mathcal{D}A\left( \int_{\ker D_+} \mathcal{D}\phi_0~\delta(|\phi_0|^2 - 1) \right) \left( \operatorname{det}_{(\ker D_+)^\perp} (- D_- D_+)\right)^{-1},
\end{align}
which indeed represents the self-dual theory, that is, the moduli integral with the fluctuation determinant.

\subsubsection*{Recovering the standard \texorpdfstring{$\mathbb{C}P^{N-1}$}{CP(N-1)} model}

The subtlety appears when we recover the standard $\mathbb{C}P^{N-1}$ model.
In the standard $\mathbb{C}P^{N-1}$ model, the partition function should be,
\begin{align}
    Z_{\text{standard }\mathbb{C}P^{N-1}}= \int \mathcal{D}A \int \mathcal{D}\alpha ~\rme^{-\frac{\im }{g^2} \int d^2x\, \alpha} \left[ \det (-D_\mu^2 +\im \alpha )\right]^{-N}.
\end{align}
where $D_\mu^2$ is an operator for one component.
Although we may have $\det(g^2)$ depending on the normalization of the kinetic term, this factor only yields an irrelevant cosmological constant. This rescaling is ``vector-like'' in the terminology discussed below.

Let us take $\epsilon^2 = g^2$ in our fluctuation determinant (\ref{eq:total_fluct_oper}).
We can simply integrate out $(h,h^\dagger)$ components, which is equivalent to,
\begin{align}
  & \det \begin{pmatrix}
  \multicolumn{2}{c}{\multirow{2}{*}{\large $\im \alpha$}} & 0 & 0  \\
 \multicolumn{2}{c}{} & 0 & + \im D_- \\
        0 & 0 & g^2  &0\\
        0 & - \im D_+ & 0 & g^2
\end{pmatrix} \notag \\
&= (g^2)^{\dim \ker D_- + \dim (\ker D_-)^\perp}  \times \left( \operatorname{det}_{ \ker D_+ \oplus (\ker D_+)^\perp} \left[
\im \alpha + \frac{1}{g^2} (-D_- D_+)\right] \right) \notag \\
&= (g^2)^{\dim \ker D_- - \dim \ker D_+}  \times \left( \operatorname{det}_{ \ker D_+ \oplus (\ker D_+)^\perp} \left[
\im g^2 \alpha +  (-D_- D_+)\right] \right)
\end{align}
where $(-D_- D_+)$ is defined on $ \ker D_+ \oplus (\ker D_+)^\perp$ in the second and third lines, and we have used $\dim (\ker D_+)^\perp = \dim (\ker D_-)^\perp$.

As $-D_- D_+ = -D_\mu^2 -F$, by properly redefining the Lagrange multiplier $\alpha$ (which gives an irrelevant constant and $1/g^2$ topological term), we arrive at
\begin{align}
    Z_{\epsilon^2=g^2} = \int \mathcal{D}A \int \mathcal{D}\alpha ~(g^2)^{ \frac{N}{2\pi} \int d^2x F} \rme^{-\frac{\im }{g^2} \int d^2x\, \alpha} \left[ \det (-D_\mu^2 +\im \alpha )\right]^{-N},
\end{align}
where we have used the index theorem (for $N$ components)
\begin{align}
    \dim \ker D_+ - \dim \ker D_- = \frac{N}{2\pi} \int d^2x \,F.
\end{align}
Hence, the fluctuation operator (\ref{eq:total_fluct_oper}) at $\epsilon^2 = g^2$ gives the standard $\mathbb{C}P^{N-1}$ model with the extra anomalous topological term $\rme^{ \frac{N}{2\pi} \log g^2 \int d^2x F}$.
Therefore, let us add a counterterm into the original topological term:
in the first-order $(\phi, h)$ formalism, we employ the following topological term
\begin{align}
    S_{\mathrm{top}}^{(\phi, h)} = \frac{1}{g^2} - \im \frac{\theta}{2\pi} + \frac{N}{2\pi} \log (g^2) = \frac{1}{g^2_{\text{sd}}} - \im \frac{\theta}{2\pi} \label{eq:phi-h-top} 
\end{align}
together with the fluctuation operator (\ref{eq:total_fluct_oper}).
This dictates how the self-dual coupling $g^2_{\text{sd}}$ and standard coupling $g^2$ are related.

\subsubsection*{The \texorpdfstring{$\epsilon$}{epsilon}-deformed theory in the second-order formalism}

Then, let us consider the $\epsilon$-deformed theory in the second-order formalism, through integrating out the auxiliary field $h$.

With a similar manipulation, we can integrate out $(h,h^\dagger)$ in the $\epsilon$-deformed theory:
\begin{align}
  & \det \begin{pmatrix}
  \multicolumn{2}{c}{\multirow{2}{*}{\large $\im \alpha$}} & 0 & 0  \\
 \multicolumn{2}{c}{} & 0 & + \im D_- \\
        0 & 0 & \epsilon^2  &0\\
        0 & - \im D_+ & 0 & \epsilon^2
\end{pmatrix} \notag \\
&= (\epsilon^2)^{\dim \ker D_- - \dim \ker D_+}  \times \left( \operatorname{det}_{ \ker D_+ \oplus (\ker D_+)^\perp} \left[
\im \epsilon^2 \alpha +  (-D_- D_+)\right] \right)
\end{align}
From this fluctuation determinant, with the topological term (\ref{eq:phi-h-top}), we obtain the correct $\epsilon$-deformed theory in terms of the rescaled $\phi$:
\begin{align}
    S[\phi, \phi^\dagger, M] &= \int d^2x \left[ |D_\mu \phi|^2 + M |\phi|^2 - \frac{M}{\epsilon^2} \right] \notag \\
    &~~~~~~~- \left[\frac{1}{\epsilon^2} - \frac{1}{g^2} + \im \frac{\theta}{2\pi} - \frac{N}{2\pi} \log (g^2/\epsilon^2) \right] \int d^2x~F \label{eq:rescaled_phi_action}
\end{align}
We need not only $\frac{1}{\epsilon^2} - \frac{1}{g^2}$, but also the topological term $- \frac{N}{2\pi} \log (g^2/\epsilon^2)$ to reproduce the correct self-dual limit.
Indeed, in the main text, we see that the large-$N$ calculation naturally encodes this anomaly, see Section \ref{sec:self-dual_limit_analytic}.

\subsubsection*{Chiral anomaly in bosonic spinor}

We have encountered the topological term generated by the difference of zeromodes.
One may understand this topological term as the chiral anomaly of the bosonic spinor field.

Indeed, we can represent $(\phi, h)$ as a bosonic Dirac spinor $\Psi=(\phi,h)^T$, and the Lagrangian can be written as
\begin{align}
    \bar{\Psi} \left(\gamma^\mu D_\mu + \begin{pmatrix}
        \im \alpha/\epsilon^2 &0 \\
        0 &\epsilon^2
    \end{pmatrix} \right) \Psi
\end{align}
after the rescaling of $\im \alpha$.

The $\epsilon$ dependence can be absorbed by the ``chiral'' transformation:
\begin{align}
\begin{pmatrix}
    \phi \\
    h
\end{pmatrix} \mapsto 
\begin{pmatrix}
    \epsilon \phi \\
    \epsilon^{-1} h
\end{pmatrix}.
\end{align}
However, under this transformation, the integration measure produces the topological term $2 \frac{N}{2\pi} \log (\epsilon) \int d^2 x~F$, through, e.g., heat kernel regularization, exactly like the standard chiral anomaly.

\section{Calculation on endpoint contribution and \texorpdfstring{$F<0$}{F<0} integral}
\label{sec:endpoint_vs_negativeF_appendix}

Here, we shall show: for $\operatorname{Im} \tau_n > 0$,
\begin{align}
    I_{F\approx 0} &= \int_{\mathcal{J}_0^{(\tau_n)}} dF ~ \rme^{-S_{\mathrm{eff}}^{(\tau_n)}[F]} +  \int_{\mathcal{J}_0^{(-\tau_n)}} dF ~ \rme^{-S_{\mathrm{eff}}^{(-\tau_n)}[F]} \notag \\
    &\sim \exp\!\Big[-\,\frac{NV_{\mathrm{2d}}\Lambda_\epsilon^2}{4\pi\rme^\gamma}
    \;-\;2\sqrt{\pi\rme^{-\gamma} (-\im \tau_n) NV_{\mathrm{2d}} \Lambda_\epsilon^2}\;\Big]\ ,
\end{align}
whose derivation was postponed in Section \ref{sec:endpoint_vs_negativeF}.

\subsection*{Setup}
For convenience, let us repeat the effective action
\begin{align}
        \frac{1}{NV_{\mathrm{2d}}}   S_{\mathrm{eff}}[F] &= \frac{F}{2\pi} \left[  \left( y(F) - \frac{1}{2}\right) \log\left(\frac{\Lambda_\epsilon^2}{2\rme^\gamma F}\right) - \log\Gamma\left(y(F)\right) \right] + \frac{F}{4\pi}\log(2\pi)  - F \tau_n. \tag{\ref{eq:eff_action_F}}
\end{align}
where $y(F) = \psi^{-1}(\log \left( \frac{\Lambda_\epsilon^2}{2 \rme^\gamma F} \right))$ is defined by a principal branch: $y(F) >0$ for $F>0$.
We define $C:=\Lambda_\epsilon^2/(2\rme^\gamma)$ for brevity: $y(F) = \psi^{-1}(\log \left( \frac{C}{F} \right))$.
We also define
\begin{align}
    \tilde{S}(F) &:= \frac{F}{2\pi}\,W\big(y(F)\big), \\
    W(y) &:= \Big(y-\frac12 \Big)\psi(y)-\log\Gamma(y)+\frac12 \log(2\pi).
\end{align}
so that  $\frac{1}{NV_{\mathrm{2d}}}   S_{\mathrm{eff}}[F] = \tilde{S}(F)  - \tau_n F$.

As we focus on the $F\approx0$ contribution, we can just take convenient contours due to the Cauchy theorem.
Thus, let us just choose the steepest descent direction in $O(F)$.
Because the linear term is only $- \tau_n F$, we can take the following path
\begin{align}
    F = \mp \frac{\rho}{\tau_n}~~~(\rho>0)~~~\text{for}~S_{\mathrm{eff}}^{(\pm \tau_n)}[F]. \label{eq:linear_segment}
\end{align}
Then, the problem is now to extract the asymptotic behavior from $F\approx 0$ in the following integral
\begin{align}
     I_{F\approx 0} =\int_0^{\rho_{\mathrm{max}}} \frac{d\rho}{\tau_n} \,\rme^{-NV_{\mathrm{2d}}\rho} \Big[ \rme^{-NV_{\mathrm{2d}}\tilde{S}(\rho/\tau_n)} - \rme^{-NV_{\mathrm{2d}}\tilde{S}(-\rho/\tau_n)} \Big], \label{eq:integral_nearF0}
\end{align}
where the upper bound $\rho_{\mathrm{max}}>0$ is not specified because we focus on the contribution near $F=0$.
Without loss of generality, we can choose $\operatorname{Im} \tau_n >0$.

We also note the small-$F$ expansion of $y(F)$:
\begin{align}
    y(F) = \frac{C}{F} + \frac{1}{2} - \frac{1}{24}\left( \frac{F}{C} \right) + O(F^3). \label{eq:yofF_perturbative}
\end{align}

\subsection*{Odd part of the action}
From this representation, the odd part of $\tilde{S}(\rho/\tau_n)$ is essential.
Actually, $\tilde{S}(F)$ is a perturbatively even function to all orders, and we need to extract the nonperturbative correction.

Let us evaluate the odd part 
\begin{align}
    \tilde{S}(F) - \tilde{S}(-F) = \frac{F}{2\pi}\big[ W(y_+) + W(y_-) \big],
\end{align}
where $y_+$ and $y_-$ are the exact roots for $+F$ and $-F$. 
Let us choose that $F$ lies in the lower half-plane ($\operatorname{Im}F<0$), then $-F$ lies in the upper half-plane ($\operatorname{Im}(-F)>0$).
As we define $y(F)$ as the principal branch, which is constructed from an analytic continuation from positive real $F$, we have
\begin{align}
    \psi(y_-) = \log\Big(\frac{C}{-F}\Big) = \log\Big(\frac{C}{F}\Big) - \im\pi = \psi(y_+) - \im\pi,
\end{align}
since we chose $\operatorname{Im} C/F > 0$.
Note that $y_- \simeq C/(-F)$ lies in the lower half-plane ($\operatorname{Im}y_- < 0$) as $-F$ is chosen to be in the upper half-plane.

We can evaluate the combination $W(y_+) + W(y_-)$ using the exact reflection formula $W(y_-) + W(1-y_-)$. 
First, we apply the digamma reflection formula $\psi(1-y_-) = \psi(y_-) + \pi\cot(\pi y_-)$ to find the exact relation between the roots. Because $\operatorname{Im}y_-<0$, we have $\cot(\pi y_-) \simeq \im(1+2\rme^{-2\pi\im y_-})$, which gives:
\begin{align}
    \psi(1-y_-) \simeq (\psi(y_+) - \im\pi) + \im\pi(1+2\rme^{-2\pi\im y_-}) = \psi(y_+) + 2\pi\im\rme^{-2\pi\im y_-}.
\end{align}
As $y_- \simeq C/(-F)$, the residual difference $\delta y = y_+ - (1-y_-)$ is nonperturbatively small $O(\rme^{-2\pi\im y_-})$.
We can concretely determine $\delta y$ by expanding $\psi(1-y_-+\delta y) = \psi(y_+)$, which gives
\begin{align}
    \delta y \simeq -\frac{2\pi\im}{\psi'(1-y_-)}\rme^{-2\pi\im y_-}
\end{align}
in the leading order.

Now we evaluate the sum using the exact $W$ reflection formula:
\begin{align}
    W(y_+) + W(y_-) &= W(1-y_- + \delta y) + W(y_-) \notag \\
    &\simeq \Big[ W(1-y_-) + W(y_-) \Big] + W'(1-y_-)\delta y.
\end{align}
The bracket part can be evaluated via the reflection formula for $W(y)$: 
\begin{align}
    W(y_-) + W(1-y_-) &= \log(2\sin(\pi y_-)) - \pi \left( y_- - \frac{1}{2} \right) \cot(\pi y_-) \notag \\
    &\simeq (-1 - 2\pi\im(y_- -1/2))\rme^{-2\pi\im y_-}
\end{align}
for $\operatorname{Im}y_-<0$.
On the other hand, since $W'(y) = (y-1/2)\psi'(y)$, the last term $W'(1-y_-)\delta y$ becomes $+2\pi\im(y_- -1/2)\rme^{-2\pi\im y_-}$.

Adding them together, we arrive at 
\begin{align}
    W(y_+) + W(y_-) = -\rme^{-2\pi\im y_-} = \rme^{2\pi\im C/F}.
\end{align}
where we used $y_- \simeq -C/F + 1/2 + O(F)$, which leads to the odd part:
\begin{align}
    \tilde{S}(F) - \tilde{S}(-F) = \frac{F}{2\pi}\rme^{2\pi\im C/F} =: 2 \tilde{S}_{\mathrm{np}}(F). \label{eq:odd_action_nonpert}
\end{align}
Crucially, all perturbative terms cancel together, and only the nonperturbative term remains.
We recall that this formula is valid for $\operatorname{Im} F < 0$.

\subsection*{Estimation of the integral}

Returning to the integral (\ref{eq:integral_nearF0}), we can factor out the purely even perturbative part $\tilde{S}_{\mathrm{pert}}(F)$:
\begin{align}
    \rme^{-NV_{\mathrm{2d}}\tilde{S}(F)} - \rme^{-NV_{\mathrm{2d}}\tilde{S}(-F)}
    &= \rme^{-NV_{\mathrm{2d}}\tilde{S}_{\mathrm{pert}}(F)}\Big[ \rme^{-NV_{\mathrm{2d}}\tilde{S}_{\mathrm{np}}(F)} - \rme^{+NV_{\mathrm{2d}}\tilde{S}_{\mathrm{np}}(F)} \Big] \notag \\
    &\simeq  -2NV_{\mathrm{2d}}\,\rme^{-NV_{\mathrm{2d}}\tilde{S}_{\mathrm{pert}}(F)} \tilde{S}_{\mathrm{np}}(F),
\end{align}
where we set $F = \rho/\tau$, where $\operatorname{Im} F < 0$ so that (\ref{eq:odd_action_nonpert}) can be applied, as we chose $\operatorname{Im }\tau >0$.
We can find that the higher nonperturbative corrections such as $ (NV_{\mathrm{2d}} \tilde{S}_{\mathrm{np}})^3$ are strongly suppressed by doing the same operation as described below.
With this operation, the integral (\ref{eq:integral_nearF0}) becomes
\begin{align}
    I_{F\approx 0} \simeq  -NV_{\mathrm{2d}} \int_0^{\rho_{\mathrm{max}}} \frac{d\rho}{\tau}\,\frac{\rho/\tau}{2\pi}\,\rme^{2\pi\im C\tau/\rho} \, \rme^{-NV_{\mathrm{2d}}(\rho + \tilde{S}_{\mathrm{pert}}(\rho/\tau))}.
\end{align}
This integral is the competition between the highly suppressed odd part $\tilde{S}_{\mathrm{np}}(F)$ and the action cost $NV_{\mathrm{2d}}(\rho + \tilde{S}_{\mathrm{pert}}(\rho/\tau))$.

By using $\tilde{S}_{\mathrm{pert}}(F) = \frac{\Lambda_\epsilon^2}{4\pi\rme^\gamma}+O(F^2)$ and by changing the variable $\tilde{\rho} = \sqrt{NV_{\mathrm{2d}}}\rho$,
we have
\begin{align}
    I_{F\approx 0} \sim  \rme^{-NV_{\mathrm{2d}}\,\frac{\Lambda_\epsilon^2}{4\pi\rme^\gamma}}\int_0^{\sqrt{NV_{\mathrm{2d}}}\rho_{\mathrm{max}}} d\tilde{\rho}\,\tilde{\rho} \,  \rme^{-\sqrt{NV_{\mathrm{2d}}}(-2\pi\im C\tau/\tilde{\rho}+\tilde{\rho})},
\end{align}
where irrelevant prefactors are omitted.

 More explicitly, we drop the $O(F^2)$ and higher-order terms from the perturbative action $\tilde{S}_{\text{pert}}(\rho/\tau)$. To see why, observe how each term scales under the change of variables $\tilde{\rho} = \sqrt{N V_{2d}} \rho$. 
 Both the linear action cost and the essential singularity term scale as $\sqrt{N V_{2d}}$ and they compete  to dictate the location of the boundary saddle point. In stark contrast, the quadratic perturbative term contributes $N V_{2d} O(F^2) \sim N V_{2d} (\tilde{\rho} / (\tau \sqrt{N V_{2d}}))^2 = O(1)$ to the exponent. Because the steepest descent is strictly governed by the large parameter $\sqrt{N V_{2d}}$ in this analysis, the $O(F^2)$ and higher-order terms provide strictly subleading corrections that are washed out of the saddle-point approximation. 

Consequently, we arrive at
\begin{align}
    I_{F\approx 0}\;\sim\;
    \exp\!\Big[-NV_{\mathrm{2d}}\,\frac{\Lambda_\epsilon^2}{4\pi\rme^\gamma}
    \;-\;2\sqrt{-2\pi\im\,C\,\tau\,NV_{\mathrm{2d}}}\;\Big]\ , 
    \label{no-nonzero-saddle}
\end{align}
where we recall that $C=\frac{\Lambda_\epsilon^2}{2\rme^\gamma}$ and $\operatorname{Im} \tau > 0$ is chosen.
This completes the derivation of (\ref{eq:endpoint_behavior_F0}).


\section{Large-\texorpdfstring{$N$}{N} scaling limits of the topological term} 
\label{sec:scaling_subtleties}

A crucial subtlety in the large-$N$ analysis of the $\mathbb{C}P^{N-1}$ model concerns the treatment of the topological phase factor, specifically whether the imaginary part of the chemical potential, $i\text{Im}(\tau) F \propto i (\theta + 2\pi n) F$, is treated as part of the macroscopic effective action or strictly as a subleading weight factor in the path integrand. The correct analytical framework depends on the scaling of $\theta + 2 \pi n$ relative to $N$. There are two distinct regimes, which lead to profoundly different thimble geometries and results:

\begin{align}
    \tilde{Z}_{\pm} (\theta+2\pi n) &= \int_{\pm}  dF dM \, e^{- N ( \tilde{S}[F,M] - \tau F )},   \\
   & \qquad  \text{Im}(\tau_n)  \sim \mathcal{O}(N^0)  \qquad \text{i.e.,} \;  (\theta + 2 \pi n)  \sim \mathcal{O}(N) \\ 
   \notag \\
    \tilde{Z}_{\pm} (\theta+2\pi n) &= \int_{\pm}  dF dM \, e^{- N ( \tilde{S}[F,M] - \text{Re}(\tau_n) F) }  \, e^{+ i  \frac{(\theta + 2 \pi n)}{2 \pi}  F },  \\ 
    & \qquad \text{Im}(\tau_n)   \sim \mathcal{O}(N^{-1})   \qquad \text{i.e.,} \;  (\theta + 2 \pi n)  \sim \mathcal{O}(N^0)  
\end{align} 

 Let us explain these two cases in slightly more detail.  

\noindent
\textbf{Standard large-$N$ Limit $\bar{\theta}_n \sim \mathcal{O}(N^0)$ or $\theta + 2\pi n \sim \mathcal{O}(N)$):} \\
If the topological sector scales extensively with $N$, the parameter $\bar{\theta}_n = \frac{\theta + 2\pi n}{N}$ remains finite in the large-$N$ limit. In this regime,  $\text{Im}(\tau) \sim \mathcal{O}(N^0)$  and must be included directly within the macroscopic effective action: 
\begin{align}
   S_{\text{eff}} [F,M]  =  \tilde{S}[F,M] -  \text{Re}(\tau)F  - i\text{Im}(\tau) F  
\end{align}
The classical saddle points are determined by varying this full, complexified action. 

As demonstrated in Section \ref{sec:Stokesph}, the inclusion of the macroscopic imaginary phase shifts the steepest descent contour into the complex plane. In the undeformed physical theory ($\text{Re}(\tau) = 0$), this complex shift immediately triggers a boundary Stokes phenomenon at $F=0$. The Stokes multiplier of this would-be metastable saddle evaluates strictly to zero for any non-zero $\bar{\theta}_n$. Consequently, the saddle is completely projected out of the physical spectrum, and the path integral is dominated entirely by the non-analytic endpoint singularity \eqref{no-nonzero-saddle}.

\noindent
\textbf{Fixed $(\theta + 2\pi n)$ Limit:} \\
Conversely, if we consider branches where $\theta + 2\pi n \sim \mathcal{O}(1)$\footnote{We can relax the fixed $\mathcal{O}(N^0)$ condition to a sufficient suppression condition, in particular, to $\mathcal{O}(N^p)$ with $p < 1$, such that the quantity $\bar{\theta}_n$ vanishes as $N \to \infty$.}, the topological parameter $\text{Im}(\tau)$ is $\mathcal{O}(N^{-1})$. Therefore, it is \textit{not} part of the large-$N$ effective action, but remains simply as a phase factor in the integrand. In this regime, the effective action is
\begin{align}
 S_{\text{eff}} [F,M]  =  \tilde{S}[F,M] -  \text{Re}(\tau)F 
 \end{align}
Because the physical theory also satisfies $\text{Re}(\tau) = 0$, the macroscopic effective action used to find the saddle point is completely stripped of the $\tau F$ term. The classical saddle point remains trivially at $F=0$.  By performing a Gaussian fluctuation analysis around this zero-field endpoint and evaluating the phase factor against these fluctuations, we recover the classic quadratic energy spectrum \cite{Witten:1978bc, DAdda:1978dle}:
\begin{align}
    E_n \approx N \frac{\Lambda_g^2}{4\pi e^\gamma} + \frac{3 \Lambda_g^2}{2\pi e^\gamma N} (\theta + 2\pi n)^2
    \label{fixedtheta}
\end{align}
Note that the first term here is $\mathcal{O}(N)$ and the latter is $\mathcal{O}(N^{-1})$ because $(\theta + 2\pi n)$ is fixed. 

This establishes a profound non-commutativity of limits.  If we first take the large-$N$ limit holding $\bar{\theta}_n$ fixed (yielding the endpoint singularity result in \eqref{no-nonzero-saddle}), and subsequently take the limit  
$\bar{\theta}_n \sim O(N^{-1}) \rightarrow 0$, Eq.~\eqref{no-nonzero-saddle} fails to recover the classic multi-branch spectrum of Witten and D'Adda et al.  
This failure occurs because the dominant contribution for $\theta + 2 \pi n \sim \mathcal{O}(N)$ \eqref{no-nonzero-saddle} differs from the leading one for $\theta + 2 \pi n \sim \mathcal{O}(1)$, even though the endpoint thimble itself may capture the correct behavior as a whole.
Indeed, the leading contribution for $\theta + 2 \pi n \sim \mathcal{O}(N)$~\eqref{no-nonzero-saddle} arises solely from the linear segment (\ref{eq:linear_segment}) near $F=0$, where $F \propto 1/\tau$.
This estimation based on the segment is not valid for $\theta + 2 \pi n \sim \mathcal{O}(1)$, causing the naive extrapolation to break down.
Such a behavior is expected because the saddle point and the endpoint coalesce at $\tau = 0$.


\noindent
\textbf{The Relevance of the $\epsilon$-Deformation:} \\
These scaling observations underscore the unique analytical power of the $\epsilon$-deformation. In the undeformed physical theory, examining the finite $\bar{\theta}$ regime leads to a vanishing Stokes multiplier, rendering the classical saddle irrelevant. 

By introducing the kinetic deformation $\epsilon < g$, we turn on a finite real part of the topological parameter, $\text{Re}(\tau) > 0$. This real deformation naturally  alters the thimble geometry.  This creates a finite, shrinking domain in $\bar{\theta} < \bar{\theta}_{\rm c}$ where the Stokes multiplier is strictly non-zero. Within this domain, the macroscopic saddle point becomes topologically active and physically relevant. In this sense, the $\epsilon$-deformation acts as an analytical regulator, providing a rigorous mathematical window into metastable branches that the undeformed physical theory simply projects out. 


\section{Details on weak-field analysis at physical limit}
\label{app:weak_field}

In this Appendix, we present detailed calculations on the weak-field analysis (small-$\theta$ or small-$F$), which gives the small-$\theta$ asymptotic behavior of the vacuum energy at the physical point.
In particular, we focus on the imaginary part $\operatorname{Im} S_*$, which determines the relevance or irrelevance of the saddle.

Let us start with recalling the $F$-integral, including the negative-$F$ contribution:
\begin{align}
    \tilde{Z}(\theta+2\pi n) 
    &= \int_{0}^\infty dF ~\rme^{-S_{\mathrm{eff}}^{\theta + 2 \pi n}[F]} +  \int_{0}^\infty dF ~\rme^{-S_{\mathrm{eff}}^{-(\theta + 2 \pi n)}[F]},
\end{align}
where we inverted the negative-$F$ integral to the positive-$F$ integral via flipping the sign of the topological term.

We focus on the small-$(\theta+2\pi n)$ case so that the saddle point is located near $F=0$: the thimble structure can be investigated via the weak-field analysis.
Without loss of generality, it is sufficient to consider
\begin{align}
    \int_{0}^\infty dF ~\rme^{-S_{\mathrm{eff}}^{\theta}[F]}
\end{align}
for $\theta \in \mathbb{R}$.

Let us recall the notations:
\begin{align}
    \frac{1}{NV_{\mathrm{2d}}}   S_{\mathrm{eff}}[F] &= \tilde{S}(F)  - \tau F \notag \\
    \tilde{S}(F) &= \frac{F}{2\pi}\,W\big(y(F)\big), \notag \\
    W(y) &= \Big(y-\frac12 \Big)\psi(y)-\log\Gamma(y)+\frac12 \log(2\pi),
\end{align}
where $y(F) = \psi^{-1}(\log \left( \frac{C}{F} \right))$ with $C:=\Lambda_\epsilon^2/(2\rme^\gamma)$.
For convenience, we introduce $\bar{\theta} := \theta/ N = -  2 \pi \im \tau $.

\subsection*{Saddle-point equation at small \texorpdfstring{$\theta$}{theta}}

Let us solve the saddle-point condition $\partial_F S_{\mathrm{eff}} = 0$.

From the definition $\psi(y(F)) = \log\left(\frac{C}{F}\right)$, we have $y'(F) \psi'(y) = -\frac{1}{F}$.
By using this relation, we differentiate $W(y)$ with respect to $F$: $\partial_F W(y) =-\frac{1}{F}\left(y - \frac{1}{2}\right)$.
Therefore, the derivative of the entire action is,
\begin{align}
    \partial_F \left( \frac{1}{NV_{\mathrm{2d}}} S_{\mathrm{eff}}[F] \right) 
    &= \partial_F \tilde{S}(F) - \tau \notag \\
    &= \frac{1}{2\pi} \left[ W(y) - \left(y - \frac{1}{2}\right) \right] - \tau.
\end{align}
The saddle-point condition requires this derivative to be zero, which simplifies to:
\begin{align}
    W(y_*) - \left(y_* - \frac{1}{2}\right)  = 2\pi\tau = \im \bar{\theta}.
\end{align}

At the saddle point $F_*$ (with corresponding $y_*$), the action value is
\begin{align}
    S_* 
    &= \frac{F_*}{2\pi} W(y_*) - F_* \tau = \frac{F_*}{2\pi} \left( y_* - \frac{1}{2} \right).
\end{align}

At small $\theta$, the saddle point $F_*$ should be weak, and the value of $|y_*|$ becomes large.
The leading order of $2\pi \partial_F \tilde{S}(F(y)) = \left(y - \frac{1}{2}\right)\psi(y) - \log\Gamma(y) + \frac{1}{2}\log(2\pi) - y + \frac{1}{2}$ can be evaluated as:\footnote{Recall the asymptotic expansions:
\begin{align}
    \log\Gamma(y) &\sim \left(y - \frac{1}{2}\right)\log y - y + \frac{1}{2}\log(2\pi) + \frac{1}{12y} - \dots, \\
    \psi(y) &\sim \log y - \frac{1}{2y} - \frac{1}{12y^2} + \dots,
\end{align}}
\begin{align}
    2\pi \partial_F \tilde{S}(F(y)) &\sim \left[ \left(y - \frac{1}{2}\right)\log y - \frac{1}{2} + \frac{1}{6y} \right] - \left[ \left(y - \frac{1}{2}\right)\log y - y + \frac{1}{12y} \right] - y + \frac{1}{2} \notag \\
    &= \frac{1}{12y}.
\end{align}
From the saddle-point equation $2\pi \partial_F \tilde{S}(F_*) = \im\bar{\theta}$, we have
\begin{align}
    y_* \simeq -\im/(12\bar{\theta}).
\end{align}

Using $F(y) = C \rme^{-\psi(y)} \simeq C/y$, the saddle point in $F$ is located at:
\begin{align}
    F_* \simeq \frac{C}{-\im/(12\bar{\theta})} = 12 C \im \bar{\theta}  .
\end{align}
We observe that the perturbative saddle point $F_*$ is purely imaginary, which is often interpreted as the real electric field in the Minkowski spacetime.

\subsection*{Evaluation of imaginary part}

We evaluate the imaginary part of the saddle-point action $\operatorname{Im} S_*$.
From the previous subsection, the saddle point is located at $F_* \simeq \im 12 C \bar{\theta}$ to leading order.
Since both $F_*$ and $\tau = \im \bar{\theta}/(2\pi)$ are purely imaginary at this order, the topological term $F_*\tau$ is real, and thus:
\begin{align}
    \operatorname{Im} S_* \simeq \operatorname{Im}\, \tilde{S}(F_*).
\end{align}
Since $\tilde{S}(F)$ takes real values on the positive real axis, the Schwarz reflection principle gives $\tilde{S}(F^*) = [\tilde{S}(F)]^*$ for all $F$. Therefore, we have
\begin{align}
    \operatorname{Im}\, \tilde{S}(F_*) = \frac{\tilde{S}(F_*) - \tilde{S}(-F_*)}{2\im}.
\end{align}

In Appendix \ref{sec:endpoint_vs_negativeF_appendix}, we evaluated this odd part non-perturbatively using the digamma and gamma function reflection formulas. As established in \eqref{eq:odd_action_nonpert}, for $\operatorname{Im} F < 0$:
\begin{align}
    \tilde{S}(F) - \tilde{S}(-F) = \frac{F}{2\pi} \rme^{2\pi\im C/F}.
\end{align}

For $\bar{\theta} > 0$, we apply \eqref{eq:odd_action_nonpert} to $-F_*$ (which lies in the lower half-plane) since $F_* \simeq \im 12 C \bar{\theta}$:
\begin{align}
    \tilde{S}(-F_*) - \tilde{S}(F_*) = \frac{-F_*}{2\pi} \rme^{2\pi\im C/(-F_*)}.
\end{align}
Substituting $F_* \simeq \im 12 C \bar{\theta}$, and recalling $C=\Lambda_\epsilon^2/(2\rme^\gamma)$, we have\footnote{
Here, we assume that the higher-order corrections to $F_*$ do not change the leading order of $\operatorname{Im} S_*$.
For imaginary-$F$ corrections, the odd-part formula guarantees that the corrections do not change the leading order.
For the correction to $\operatorname{Re} F_*$, we can in fact show $\operatorname{Re} F_* \sim O(\rme^{-\pi/(6\bar{\theta})})$ by the same computation in Appendix \ref{sec:endpoint_vs_negativeF_appendix}.
This nonperturbatively small $\operatorname{Re} F_*$ does not contribute to the leading order.
Indeed, when we write $F_* = \im F_I + \delta F_R$ with $\im F_I$ the purely
imaginary solution of the perturbative saddle-point equation
$\partial_F \tilde{S}_{\mathrm{pert}}(\im F_I) = \tau$, the action expands as
\begin{align}
\operatorname{Im} \left( \frac{1}{NV_{\mathrm{2d}}} S_{\mathrm{eff}}[\im F_I + \delta F_R] \right)
= \operatorname{Im} \left( \frac{1}{NV_{\mathrm{2d}}} S_{\mathrm{eff}}[\im F_I] \right)
+ \operatorname{Im}\Big[ \big( \partial_F \tilde{S}(\im F_I) - \tau \big)\, \delta F_R \Big]
+ O(\delta F_R^2).
\end{align}
In the second term, the perturbative part cancels
against $\tau$ by the saddle-point equation, and only the nonperturbative
gradient survives:
$\partial_F \tilde{S}(\im F_I) - \tau = \partial_F \tilde{S}_{\mathrm{np}}(\im F_I)
= O(\rme^{-\pi/(6\bar{\theta})})$.
Hence the corrections from $\delta F_R$ start at
$O(\rme^{-\pi/(3\bar{\theta})})$ and are doubly suppressed.
}
\begin{align}
    \operatorname{Im} \left( \frac{1}{NV_{\mathrm{2d}}} S_{\mathrm{eff}}[F_*] \right) \simeq +\frac{3 \bar{\theta} \Lambda_\epsilon^2}{2\pi \rme^\gamma} \exp\left(-\frac{\pi}{6 \bar{\theta}}\right) > 0~~~~(\text{for}~\bar{\theta} > 0).
\end{align}
For $\bar{\theta} < 0$, we can apply \eqref{eq:odd_action_nonpert} to $F_*$ and carry out the same calculation, we finally obtain
\begin{align}
    \operatorname{Im} \left( \frac{1}{NV_{\mathrm{2d}}} S_{\mathrm{eff}}[F_*] \right) \simeq \frac{3 \bar{\theta} \Lambda_\epsilon^2}{2\pi \rme^\gamma} \exp\left(-\frac{\pi}{6 |\bar{\theta}|}\right), \tag{\ref{eq:weak-field_imaginarypart}}
\end{align}
which is positive for $\bar{\theta} > 0$ and is negative for $\bar{\theta} < 0$.

This result indicates that the saddle point is always irrelevant for $\bar{\theta} \neq 0$.
Indeed, for $\bar{\theta} > 0$, the original contour $F>0$ has a negative imaginary part $\operatorname{Im} S(F) < 0$.
For $\bar{\theta} < 0$, the original contour $F>0$ has a positive imaginary part $\operatorname{Im} S(F) > 0$.
Thus, the above calculation shows that the dual thimble of the saddle at small $\bar{\theta}$ never intersects with the original contour: the saddle becomes irrelevant immediately at $\bar{\theta} \neq 0$.

\subsection*{Stokes-transition point at small $\operatorname{Re}\tau$}

At the physical limit, we have seen that the saddle point is always irrelevant for $\bar{\theta} \neq 0$.
From this calculation, we can analytically determine the point of the Stokes phenomenon $\bar{\theta}$ for very small $\operatorname{Re}\tau$.
For simplicity, we restrict ourselves to the case $\tau_R:=\operatorname{Re}\tau >0$ and $\bar{\theta} >0$.

For small $\tau$, the saddle point is located at 
\begin{align}
    F_* = F_{*,R} + \im F_{*,I} \simeq 24 \pi C \tau.
\end{align}
from the same calculation.
To determine the Stokes-transition point, it is sufficient to see the region with the hierarchy: $\tau_R\ll \bar{\theta}$, because it turns out that the Stokes phenomenon occurs at $\tau_R \sim O(e^{-\pi/(6 \bar{\theta})})$ below.

In the case $\tau_R = O(e^{-\pi/(6 \bar{\theta})})$, we have $F_{*,R} \sim O(e^{-\pi/(6 \bar{\theta})})$ (including the nonperturbative correction at $\tau_R =0$).
This real part does not contribute, because 
\begin{align}
\operatorname{Im} \left( \frac{1}{NV_{\mathrm{2d}}} S_{\mathrm{eff}}[F_{*,R} + \im F_{*,I}] \right)
&\simeq \operatorname{Im} \left( \frac{1}{NV_{\mathrm{2d}}} S_{\mathrm{eff}}[\im F_{*,I}] \right) - \tau_R F_{*,I} 
\notag \\
&~~~~+ \operatorname{Im}\Big[ \big( \partial_F \tilde{S}(\im F_{*,I}) - \im \frac{\bar{\theta}}{2\pi} \big)\, F_{*,R}  \Big],
\end{align}
and the last term vanishes in the leading order due to the saddle point equation at $\tau_R = 0$.
Therefore, in the leading order, we obtain
\begin{align}
\operatorname{Im} \left( \frac{1}{NV_{\mathrm{2d}}} S_{\mathrm{eff}}[F_{*,R} + \im F_{*,I}] \right)
&\simeq \frac{3 \bar{\theta} C}{\pi } \rme^{-\frac{\pi}{6 \bar{\theta}}} - (12 C  \bar{\theta} ) \tau_R.
\end{align}
Hence, for very small $\tau_R$, the Stokes phenomenon occurs at
\begin{align}
 \tau_R \simeq \frac{1}{4\pi } \rme^{-\frac{\pi}{6 \bar{\theta}}}~~~\Leftrightarrow ~~~ \bar{\theta} \simeq - \frac{\pi}{6 \log(4\pi \tau_R)}, \tag{\ref{eq:Stokes_near_origin}}
\end{align}
This confirms the result presented in the main text.

\section{General formalism of the boundary Stokes phenomenon}
\label{app:boundary_stokes}

In the semiclassical analysis of finite-dimensional contour integrals or path integrals on manifolds with boundaries, the asymptotic expansion is governed not only by the stationary points of the action, but also by the boundary points themselves. 
While standard Picard--Lefschetz theory decomposes an integration cycle into steepest-descent paths originating from bulk saddles  
\cite{Pham:1983,Witten:2010cx,Behtash:2015zha} (see also \cite{Dersy:2026jat,Ai:2019fri,Garbrecht:2025tos,Loges:2022nuw,Feldbrugge:2017kzv} for recent applications) , the presence of a boundary introduces a new class of integration cycles attached directly to the boundary points \cite{howls1992hyperasymptotics, Delabaere:2002}, giving rise to the \emph{boundary Stokes phenomenon}. 
There are fundamental differences between the standard Stokes phenomenon that occurs between two saddles and the boundary Stokes phenomenon between a boundary and a saddle. We describe them below.

Consider an exponential integral with a general parameter-dependent action:
\begin{equation}
    I(t) = \int_{\Gamma} dz \, g(z) e^{-S(z, t)}, \qquad t = \rho e^{i\theta} \in \mathbb{C},
\end{equation}
where $t$ is an external complex parameter ($\rho > 0$), and the integration contour $\Gamma$ spans between two endpoints, $\partial \Gamma = \{a, b\}$, one or both of which may reside at fixed points in the finite complex plane. To obtain a well-defined asymptotic expansion, $\Gamma$ is deformed along the gradient flow of the real action with respect to a real flow time $u$: 
\begin{equation}
    \frac{dz}{du} = \overline{\left(\frac{\partial S(z)}{\partial z}\right)}.
\end{equation}
By construction, this flow preserves the imaginary part of the action, $\frac{d}{du}\text{Im}[S(z, t)] = 0$, while monotonically increasing the real part, $\frac{d}{du}\text{Re}[S(z)] > 0$.

To establish the homological decomposition of the integration contour as an element of the relative homology group $H_1(\mathbb{C}, X_\infty)$ (more precisely, $H_1(\mathbb{C}, X_\infty \cup \{a, b\})$), we classify the asymptotic regions at infinity into \textbf{good domains} $X_\infty \subset \mathbb{C}$---where $\mathrm{Re}[S(z)] \to +\infty$ and the integrand $e^{-S(z)}$ is exponentially damped---and \textbf{bad domains}, where $\mathrm{Re}[S(z)] \to -\infty$ and the integrand diverges. The steepest-descent cycles associated with the action $S(z)$ fall into two distinct classes:
\begin{itemize}
\item \textbf{Saddle Thimbles ($\mathcal{J}_i(t)$):} A bulk Lefschetz thimble $\mathcal{J}_i$ attached to a critical point $S'(z_i) = 0$ is a two-sided trajectory of the gradient flow that interpolates between two distinct good domains in $X_\infty$.  
\item \textbf{Boundary Thimbles ($\mathcal{J}_a(t), \mathcal{J}_b(t)$):} These cycles are attached to the boundary points $a$ and $b$, where generically $\partial_z S \neq 0$. The point $a$ partitions the continuous gradient flow line into two distinct segments: the forward steepest-descent segment that lands in a good domain $X_\infty$, and its complementary steepest-ascent segment that terminates in a bad domain.
\end{itemize}
The original integration contour is homologous to an integer linear combination of both saddle and boundary thimbles:
\begin{equation}
    \Gamma \equiv \sum_i n_i \mathcal{J}_i(t) + n_a \mathcal{J}_a(t) + n_b \mathcal{J}_b(t),
\end{equation}
where the Stokes multipliers $n_i(t) \in \mathbb{Z}$ are given by the intersection pairing between $\Gamma$ and the dual (steepest-ascent) thimbles $\mathcal{K}_i(t)$:\footnote{For a boundary thimble $\mathcal{J}_a$, the intersection number $n_a = \langle \Gamma, \mathcal{K}_a \rangle$ is defined via a transverse intersection at the endpoint $a$, which is regularized by an infinitesimal push-off of the integration contour $\Gamma$ along the steepest-descent direction into the bulk. 
The intersection number of a boundary thimble is naturally fixed by the original contour: $n_a = +1$ and $n_b = -1$ when $\Gamma$ is a path from $a$ to $b$.}
\begin{equation}
    n_i(t) = \langle \Gamma, \mathcal{K}_i(t) \rangle.
\end{equation}

\noindent 
{\bf Bulk-bulk Stokes lines:}
In the ordinary saddle--saddle Stokes phenomenon, a Stokes line is determined by setting the imaginary part of the action difference between two saddles $\Delta S_{ij}(t) = S(z_i(t), t) - S(z_j(t), t)$ to zero. 
\begin{align}
\text{Im}[\Delta S_{ij}(t)] = 0 
\label{saddle-saddle}
\end{align}

This defines the Stokes lines for the bulk. We will have extra Stokes lines in the presence of boundaries. 

\noindent 
{\bf Boundary-bulk Stokes lines:}
A boundary-bulk Stokes line in parameter space is defined by the alignment of the imaginary actions of a stationary point $z_i(t)$ and a boundary point $a$:
\begin{equation}
    \text{Im}\left[\Delta S_{ai}(t)\right] \equiv \text{Im}\left[S(z_i(t), t) - S(a, t)\right] = 0.
    \label{eq:phase_alignment}
\end{equation}
This defines the locus in parameter space where boundary-to-saddle Stokes transitions occur.

\noindent 
{\bf Phase alignment and maximal dominance:}

In a standard Stokes transition, the phase alignment ($\text{Im}[\Delta S_{ij}(t)] = 0 $) identically coincides with the locus where the real action difference is at an extremum with respect to the angular parameter $\theta = \arg(t)$, representing a state of maximal exponential dominance: $\frac{\partial}{\partial \theta} \left(\text{Re}\,\Delta S_{ij}(t)\right) = 0.$

In general, this coincidence is guaranteed by the two properties: (i) $\Delta S_{ij}(t)$ is holomorphic in $t$, (ii) the Stokes line is perpendicular to the direction of the parameter variation.
Indeed, a holomorphic function $\Delta S_{ij}(t)$ has real and imaginary parts with orthogonal gradients, satisfying $(\nabla (\operatorname{Re} \Delta S_{ij})) \cdot (\nabla (\operatorname{Im} \Delta S_{ij})) = 0,$ where $\nabla$ is the two-dimensional gradient with respect to the real and imaginary parts of $t$.
Therefore, if the direction of the varying parameter is perpendicular to the Stokes line ($\text{Im}[\Delta S_{ij}(t)] = 0$), the real part has an extremum at the Stokes transition point, $\frac{\partial}{\partial \theta} \left(\text{Re}\,\Delta S_{ij}(t)\right) = 0$ where $\theta$ is the varying parameter.

We remark that this property is a consequence of only the holomorphy and perpendicularity.
This conclusion remains valid in both bulk-bulk and boundary-bulk Stokes lines.
When the parameter is the angular one $\theta$, the perpendicularity condition corresponds to the property that the Stokes line extends straight from the origin.

In the Airy example examined below, the bulk saddle actions respond homogeneously to variations of the parameter $t$, and the Stokes lines are the straight ones from the origin. 
In contrast, because $\Delta S_{ai}(t)$ is inhomogeneous in $\theta$, the bulk-boundary Stokes lines do not coincide with the maximal dominance when the angular parameter is varied. The phase matching condition Eq.~\eqref{eq:phase_alignment} that triggers the Stokes jump in $\langle \Gamma, \mathcal{K}_i(t) \rangle$ is no longer the same as the extremum of $\text{Re}[\Delta S_{ai}(t)]$.

We note again that this coincidence depends on the variation parameter.
For the bulk-boundary case as well, the Stokes transition coincides with the maximal dominance when the Stokes line is perpendicular to the direction of the parameter variation.
An example of this is presented in the main text (see Figures \ref{fig:relevant_zone} and \ref{fig:vac_ene} at large $\operatorname{Re} \tau$).




As $\arg(t)$ is varied across the Stokes line, the Stokes multiplier $n_*$ of the saddle may discontinuously drop to zero:
\begin{equation}
    \Gamma \equiv \mathcal{J}_{\text{bdy}} + \mathcal{J}_* \xrightarrow{\quad \text{Stokes Line} \quad} \Gamma \equiv \mathcal{J}_{\text{bdy}}.
\end{equation}
The integration cycle becomes saturated entirely by the boundary thimble. 
This boundary Stokes phenomenon is precisely what takes place in the large-$N$ limit of the generalized and in particular, the physical $\mathbb{C}P^{N-1}$ model, and this mechanism makes the problematic saddle inactive.

\begin{figure}[t]
    \centering
    \includegraphics[width=0.8\textwidth]{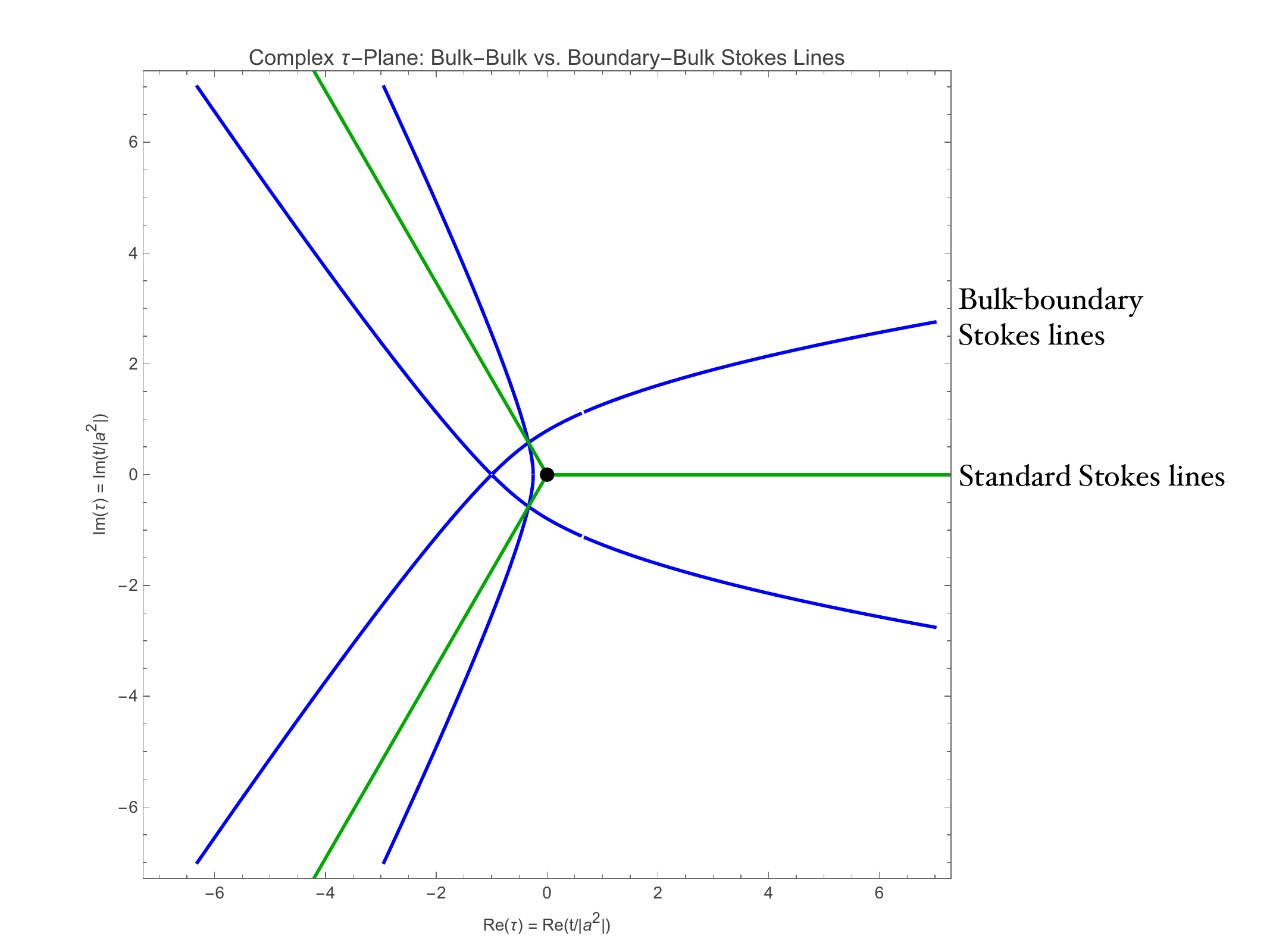}
    \vspace{-0cm}
    \caption{ Standard bulk-bulk Stokes lines (green lines) and extra bulk-boundary Stokes lines of the Airy example. 
    The bulk-boundary Stokes lines are plotted for purely imaginary $a$ ($\alpha = \pm \pi/2$).}
    \label{fig:BoundaryAiry}
\end{figure}

\subsection{Boundary Stokes phenomenon in the Airy example}
\label{app:cubic_example}

In this section, we illustrate the boundary Stokes phenomenon using the cubic action (the Airy example) on a domain with a finite lower boundary. The action is given by:
\begin{equation}
    S(z, t) = \frac{z^3}{3} - t z,
    \label{eq:cubic_action}
\end{equation}
where $t = \rho e^{i\theta}$ is a complex parameter ($\rho > 0$). To clearly distinguish standard saddle--saddle transitions from boundary--saddle transitions, we first examine the bulk--bulk Stokes lines of the boundary-free integral before incorporating a generic finite boundary.

\paragraph{1. Bulk--bulk Stokes lines (the symmetric or boundary-free case):}
Consider the standard Airy integral without finite boundaries,
\begin{equation}
    I(t) = \int_{\Gamma_0} dz \, e^{-S(z, t)}, \qquad t \in \mathbb{C}, 
    \label{eq:airy}
\end{equation}
where $\Gamma_0 \equiv (-i \infty, i \infty)$ starts and ends in asymptotic good domains (once it is infinitesimally tilted at both ends). When the boundary is absent, 
the asymptotic behavior is governed entirely by the stationary points of Eq.~\eqref{eq:cubic_action}, located at $z_{\pm} = \pm \sqrt{t}$. Their critical actions are:
\begin{equation}
    S_{\pm} = \mp \frac{2}{3} t^{3/2} = \mp \frac{2}{3} \rho^{3/2} e^{i 3\theta/2}.
\end{equation}
In this symmetric configuration, the action difference between the two bulk saddles ($\Delta S_{+-} \equiv S_+ - S_-$) scales homogeneously with $t^{3/2}$. This uniform power-law scaling is a consequence of the exact scaling symmetry:
\begin{equation}
    z \to \lambda z, \quad t \to \lambda^2 t \implies S \to \lambda^3 S,
\end{equation}
under which the origin $z=0$ is a fixed point. 

Consequently, the bulk--bulk Stokes lines---defined by the algebraic phase alignment condition between the two saddles---manifest as scale-invariant rays in the complex $t$-plane:
\begin{align}
    \text{Im}[\Delta S_{+-}] \propto \sin\left(\frac{3\theta}{2}\right) = 0 \implies \theta_{\text{S}} \in \left\{0, \,\frac{2\pi}{3}, \,\frac{4\pi}{3}\right\}.
\end{align}
The real action difference $\text{Re}[\Delta S_{+-}] \propto \cos(3\theta/2)$ reaches an absolute extremum with respect to $\theta$ along these exact same rays (see Figure \ref{fig:BoundaryAiry}):
\begin{equation}
    \frac{\partial}{\partial \theta} \left(\text{Re}\,\Delta S_{+-}\right) \propto -\sin\left(\frac{3\theta}{2}\right) = 0.
\end{equation}
Therefore, for bulk--bulk transitions, the equal-phase Stokes condition and the locus of maximal exponential dominance identically coincide:
\begin{align}
    \text{Im}[\Delta S_{+-}] = 0 \iff \frac{\partial}{\partial \theta} \left(\text{Re}\,\Delta S_{+-}\right) = 0.
\end{align}
As we show below, this coincidence is broken as soon as a non-trivial finite boundary is introduced.

\begin{figure}[t]
    \centering
    \includegraphics[width=\textwidth]{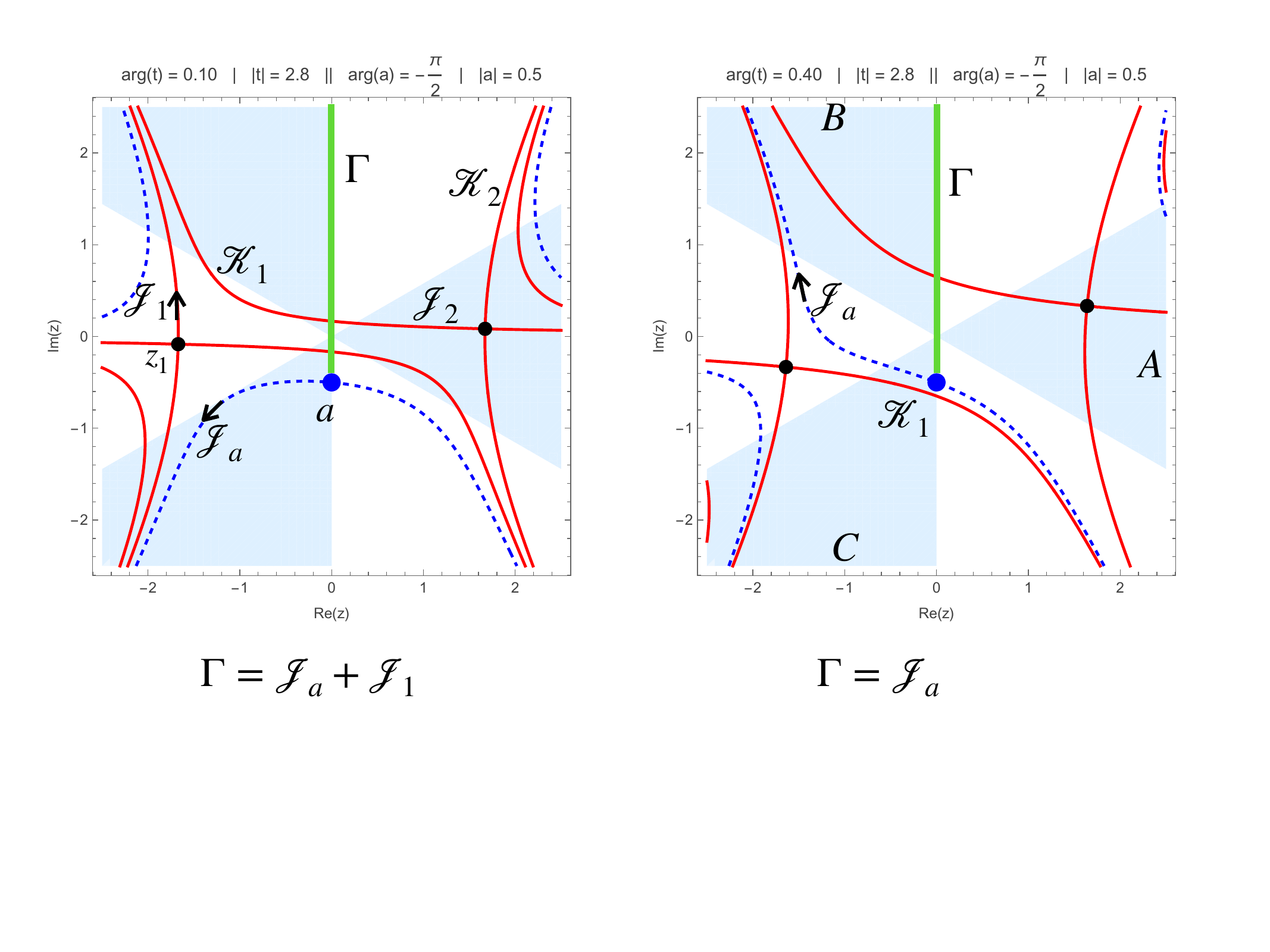}
    \vspace{-3cm}
    \caption{Boundary Stokes phenomenon. 
    \textbf{Left:} At $\arg(t) =0.10$, the integration contour $\Gamma = [a, i\infty)$ decomposes into the boundary thimble and the bulk saddle thimble, $\Gamma \equiv \mathcal{J}_a + \mathcal{J}_1$. 
    \textbf{Right:} After crossing the bulk--boundary Stokes line to $\arg(t) = 0.40$, the saddle multiplier drops to zero ($n_1 = 0$), leaving the cycle saturated entirely by the boundary thimble, $\Gamma \equiv \mathcal{J}_a$. The Stokes jump occurs away from the locus of maximal exponential dominance.}
    \label{fig:BoundaryAiry2}
\end{figure}

\paragraph{2. Bulk--boundary Stokes lines (\texorpdfstring{$a \neq 0$}{a != 0}) and decoupling:}
We now examine the integral where the domain is truncated at a fixed, $t$-independent complex boundary $z = a = |a|e^{i\alpha} \in \mathbb{C}$ (see Fig.~\ref{fig:BoundaryAiry2}):
\begin{equation}
    I(t,a) = \int_{\Gamma} dz \, e^{-S(z, t)}, \qquad \Gamma = [a, i \infty).
    \label{eq:airywb}
\end{equation}
When $a \neq 0$, the scaling symmetry is explicitly broken. The action evaluated at the boundary point is:
\begin{equation}
    S(a) = \frac{a^3}{3} - t a = \frac{|a|^3}{3} e^{3i\alpha} - |a| \rho e^{i(\theta+\alpha)}.
\end{equation}
Let us analyze the transition between the dominant bulk saddle $z_+ \equiv z_2$ and the boundary $z=a$. The action difference $\Delta S_{a+}$ takes the form:
\begin{equation}
    \Delta S_{a+}(\rho, \theta; a) \equiv S_+ - S(a) = -\frac{2}{3} \rho^{3/2} e^{i 3\theta/2} - \frac{|a|^3}{3} e^{3i\alpha} + |a| \rho e^{i(\theta+\alpha)}.
\end{equation}
This expression mixes the $\mathcal{O}(\rho^{3/2})$ scaling of the saddle with the $\mathcal{O}(\rho^1)$ linear boundary correction, while also introducing a constant complex boundary phase. We now extract the exact algebraic conditions for phase alignment and maximal dominance:

Setting $\text{Im}[\Delta S_{a+}] = 0$ yields the  
equation for the bulk-boundary Stokes lines, where new Stokes transitions occur:
    \begin{equation}
        \text{Im}[\Delta S_{a+}] = -\frac{2}{3} \rho^{3/2} \sin\left(\frac{3\theta}{2}\right) - \frac{|a|^3}{3}\sin(3\alpha) + |a| \rho \sin(\theta + \alpha) = 0.
        \label{eq:cubic_stokes_cond}
    \end{equation}
See Figure \ref{fig:BoundaryAiry} for the bulk-boundary Stokes lines for purely imaginary $a \neq 0$.
On the other hand, requiring the real action difference to be an extremum with respect to the angular parameter yields:
    \begin{equation}
        \frac{\partial}{\partial \theta} \left(\text{Re}\,\Delta S_{a+}\right) = \rho^{3/2} \sin\left(\frac{3\theta}{2}\right) - |a| \rho \sin(\theta + \alpha) = 0.
        \label{eq:cubic_dom_cond}
    \end{equation}
A comparison of Eq.~\eqref{eq:cubic_stokes_cond} and Eq.~\eqref{eq:cubic_dom_cond} reveals that
\begin{align}
    \text{Im}[\Delta S_{a+}]  \not\propto \frac{\partial}{\partial \theta} \left(\text{Re}\,\Delta S_{a+}\right).
\end{align}
The phase alignment and extremum loci are decoupled; the bulk--boundary Stokes line is defined solely through $\text{Im}[\Delta S_{ai}] = 0$.  Consequently, for any generic boundary $a \neq 0$:
\begin{equation}
    \theta_{\text{S}}(\rho; a) \neq \theta_{\text{MaxDom}}(\rho; a).
\end{equation}
Unlike the straight scale-invariant rays of the bulk--bulk case, a bulk--boundary Stokes line $\text{Im}[\Delta S_{ai}] = 0$ forms a curved trajectory in the complex $t$-plane.

In the full angular domain $\theta \in [0, 2\pi)$ (with sufficiently large $\rho$),  
the parameter space contains exactly nine Stokes lines: three scale-invariant bulk--bulk rays and six parameter-dependent bulk--boundary curves (see Figure \ref{fig:BoundaryAiry}).
In the asymptotic limit ($\rho \to \infty$), the linear boundary action $\mathcal{O}(\rho^1)$ becomes subleading to the saddle action $\mathcal{O}(\rho^{3/2})$, causing the nine Stokes lines to cluster into three triplets centered around the unperturbed symmetry angles $\theta_k = \frac{2\pi k}{3}$ for $k \in \{0, 1, 2\}$.

By expanding the phase alignment condition $\mathrm{Im}[\Delta S_{a\pm}(\rho, \theta)] = 0$ in powers of $\rho^{-1/2}$ around each bulk root $\theta_k$, we obtain an explicit large-$\rho$ perturbative series for the angular positions of the six bulk--boundary Stokes lines:
\begin{equation}
    \theta_{\text{S}}^{(\pm)}(\rho; \theta_k) = \theta_k \pm (-1)^k \frac{|a| \sin(\theta_k + \alpha)}{\sqrt{\rho}} + \mathcal{O}\left(\frac{1}{\rho}\right), \qquad k \in \{0, 1, 2\},
    \label{eq:stokes_perturbative}
\end{equation}
where the superscripts $(\pm)$ denote the Stokes transitions associated with the saddles $z_\pm$, respectively, and we assume the non-degenerate boundary orientation $\sin(\theta_k + \alpha) \neq 0$. Equation~\eqref{eq:stokes_perturbative} demonstrates that at large $|\rho|$, each unperturbed bulk--bulk ray at $\theta_k$ is flanked by a pair of bulk--boundary Stokes lines separated by an angular splitting $\Delta \theta = \mathcal{O}(\rho^{-1/2})$ that vanishes asymptotically as $\rho \to \infty$.

Figure~\ref{fig:BoundaryAiry2} shows a simple example of boundary-saddle Stokes phenomenon as ${\rm Arg}(t)$  is varied. 
Before crossing the bulk--boundary Stokes line (eg. at $\arg(t) = 0.10$), the integration contour decomposes into the boundary thimble and the bulk saddle thimble, $\Gamma \equiv \mathcal{J}_a + \mathcal{J}_1$, since  $\langle \Gamma, \mathcal{K}_1(t) \rangle=1$.
   As $\arg(t)$ is increased across the Stokes line,  the intersection number of the saddle drops discontinuously to zero ($ \langle \Gamma, \mathcal{K}_1(t) \rangle= 0$), leaving the cycle saturated entirely by the boundary thimble, $\Gamma \equiv \mathcal{J}_a$ (eg. at $\arg(t) = 0.40$).   
  \begin{equation}
    \Gamma \equiv \mathcal{J}_{{a}} + \mathcal{J}_1 \xrightarrow{\quad \text{Bulk-boundary Stokes Line} \quad} \Gamma \equiv \mathcal{J}_{{a}}.
\end{equation}
Crucially, this Stokes  jump occurs without requiring destructive competition between saddles and far from the locus of maximal exponential dominance similar to our field theory example.

\bibliographystyle{utphys}
\bibliography{./QFT.bib, ./refs.bib}

\end{document}